\documentclass[a4paper,11pt]{article}
\usepackage{jcappub} 
\usepackage{lineno}

\usepackage{dcolumn}
\usepackage{verbatim}

\usepackage{breakurl}

\usepackage{lmodern}

\pdfoutput=1
\usepackage{graphicx}
\usepackage{bm}
\usepackage{amsmath}
\usepackage{amsfonts}
\usepackage{amssymb}
\usepackage{multirow}
\usepackage[capitalise]{cleveref}
\usepackage{acro2}
\usepackage[utf8]{inputenc}

\usepackage{tikz}
\usepackage{pgfplots}
\usepackage{subcaption}

\crefname{figure}{Fig.}{Figs.}
\def\({\left(}
\def\){\right)}
\def\[{\left[}
\def\]{\right]}

\newcommand{\be}{{\begin{eqnarray}}}
\newcommand{\ee}{{\end{eqnarray}}}

\newcommand{\overbar}[1]{\mkern 1.5mu\overline{\mkern-1.5mu#1\mkern-1.5mu}\mkern 1.5mu}

\newcommand{\mpl}{m_\mathrm{Pl}}

\newcommand{\fnl}{f_\mathrm{NL}}

\newcommand{\gnl}{g_\mathrm{NL}}
\newcommand{\abs}[1]{{\left \vert #1 \right \vert}}

\newcommand{\cP}{\mathcal{P}}

\newcommand{\cH}{\mathcal{H}}
\newcommand{\cS}{\mathcal{S}}
\newcommand{\cO}{\mathcal{O}}

\newcommand{\bk}{\mathbf{k}}
\newcommand{\bq}{\mathbf{q}}

\newcommand{\bx}{\mathbf{x}}
\newcommand{\bn}{\mathbf{n}}

\newcommand{\ud}{\mathrm{d}}

\newcommand{\uRD}{\mathrm{RD}}

\newcommand{\uGW}{\mathrm{gw}}
\newcommand{\ung}{\mathrm{ng}}
\newcommand{\uc}{\mathrm{c}}
\newcommand{\Beq}{\begin{align}}
\newcommand{\Eeq}{\end{align}}

\DeclareAcronym{GW}{
  short = GW,
  long = gravitational wave ,
  short-plural = s ,
}
\DeclareAcronym{LIGO}{
  short =LIGO ,
  long = Laser Interferometer Gravitational-Wave Observatory ,
  short-plural = ,
}
\DeclareAcronym{LVK}{
  short = LVK ,
  long = {LIGO, Virgo, and KAGRA},
  short-plural = ,
}

\DeclareAcronym{SGWB}{
  short = SGWB ,
  long = stochastic gravitational-wave background ,
  short-plural = s ,
}
\DeclareAcronym{GWB}{
  short = GWB ,
  long = gravitational-wave background ,
  short-plural = s ,
}
\DeclareAcronym{CBC}{
  short = CBC ,
  long = compact binary coalescence ,
  short-plural = s ,
}
\DeclareAcronym{BH}{
  short = BH ,
  long = black hole ,
  short-plural = s ,
}
\DeclareAcronym{BBH}{
  short = BBH ,
  long = binary black hole ,
  short-plural = s ,
}
\DeclareAcronym{PBH}{
  short = PBH ,
  long = primordial black hole ,
  short-plural = s ,
}
\DeclareAcronym{SNR}{
  short = SNR ,
  long = signal-to-noise ratio ,
  short-plural = s ,
}
\DeclareAcronym{IMRPPv2}{
  short = ,
  long = {\normalsize IMRP}{\footnotesize HENOM}{\normalsize P}v2 ,
  short-plural = ,
}
\DeclareAcronym{PTA}{
  short = PTA ,
  long = pulsar timing array ,
  short-plural = s ,
}
\DeclareAcronym{SFR}{
  short = SFR ,
  long = star formation rate ,
  short-plural =  ,
}
\DeclareAcronym{FRW}{
  short = FRW ,
  long = Friedmann-Robertson-Walker ,
  short-plural =  ,
}
\DeclareAcronym{IMR}{
  short = IMR ,
  long = inspiral-merger-ringdown ,
  short-plural =  ,
}
\DeclareAcronym{LISA}{
	short = LISA ,
	long  = Laser Interferometer Space Antenna,
  short-plural =  ,
}
\DeclareAcronym{ET}{
	short = ET ,
	long  = Einstein Telescope,
  short-plural =  ,
}
\DeclareAcronym{CE}{
	short = CE ,
	long  = Cosmic Explorer,
  short-plural =  ,
}
\DeclareAcronym{BBO}{
	short = BBO ,
	long  = Big Bang Observer,
  short-plural =  ,
}
\DeclareAcronym{DECIGO}{
	short = DECIGO ,
	long  = Deci-hertz Interferometer Gravitational wave Observatory,
  short-plural =  ,
}
\DeclareAcronym{ABH}{
	short = ABH ,
	long  = astrophysical black hole,
  short-plural = s ,
}

\DeclareAcronym{PNG}{
	short = PNG ,
	long  = primordial non-Gaussianity ,
  short-plural =  ,
}

\DeclareAcronym{CMB}{
	short = CMB ,
	long  = cosmic microwave background ,
  short-plural =  ,
}
\DeclareAcronym{LSS}{
	short = LSS ,
	long  = large-scale structures ,
  short-plural =  ,
}

\DeclareAcronym{PGW}{
	short = PGW ,
	long  = primordial gravitational wave ,
  short-plural = s ,
}
\DeclareAcronym{SIGW}{
	short = SIGW ,
	long  = scalar-induced gravitational wave ,
  short-plural = s ,
}

\DeclareAcronym{RD}{
	short = RD,
	long  = radiation-domination ,
  short-plural =  ,
}
\DeclareAcronym{MD}{
	short = MD,
	long  = matter-domination ,
  short-plural =  ,
}
\DeclareAcronym{eMD}{
	short = eMD,
	long  = early-matter-domination ,
  short-plural =  ,
}
\DeclareAcronym{SW}{
	short = SW,
	long  = Sachs-Wolfe ,
  short-plural =  ,
}
\DeclareAcronym{ISW}{
	short = ISW,
	long  = integrated Sachs-Wolfe ,
  short-plural =  ,
}

\DeclareAcronym{DM}{
	short = DM,
	long  = dark matter ,
  short-plural =  ,
}

\DeclareAcronym{NANOGrav}{
	short = NANOGrav ,
	long  = North American Nanohertz Observatory for Gravitational Waves ,
  short-plural =  ,
}

\DeclareAcronym{PDF}{
	short = PDF ,
	long  = probability distribution function ,
  short-plural = s ,
}

\DeclareAcronym{SMBH}{
  short = SMBH ,
  long  = supper-massive black hole ,
  short-plural = s ,
}
\DeclareAcronym{SKA}{
	short = SKA ,
	long  = Square Kilometre Array ,
  short-plural =  ,
}
\DeclareAcronym{NG15}{
  short = NG15 ,
  long  = NANOGrav 15-year ,
  short-plural =  ,
}

\title{\boldmath Complete analysis of the background and anisotropies of scalar-induced gravitational waves: primordial non-Gaussianity $f_{\mathrm{NL}}$ and $g_{\mathrm{NL}}$ considered }

\author[a,b]{Jun-Peng Li,}
\emailAdd{lijunpeng@ihep.ac.cn}
\author[a]{Sai Wang\footnote{Corresponding author},} 
\emailAdd{wangsai@ihep.ac.cn}
\author[c]{Zhi-Chao Zhao,}
\emailAdd{zhaozc@cau.edu.cn}
\author[d,e,f]{Kazunori Kohri}
\emailAdd{kazunori.kohri@gmail.com}

\affiliation[a]{Theoretical Physics Division, Institute of High Energy Physics, Chinese Academy of Sciences, 19B Yuquan Road, Shijingshan District, Beijing 100049, China}
\affiliation[b]{School of Physics, University of Chinese Academy of Sciences, 19A Yuquan Road, Shijingshan District, Beijing 100049, China}
\affiliation[c]{Department of Applied Physics, College of Science, China Agricultural University, 17 Qinghua East Road, Haidian District, Beijing 100083, China}
\affiliation[d]{Division of Science, National Astronomical Observatory of Japan (NAOJ), and SOKENDAI, 2-21-1 Osawa, Mitaka, Tokyo 181-8588, Japan}
\affiliation[e]{Theory Center, IPNS, and QUP (WPI), KEK, 1-1 Oho, Tsukuba, Ibaraki 305-0801, Japan}
\affiliation[f]{Kavli IPMU (WPI), UTIAS, The University of Tokyo, Kashiwa, Chiba 277-8583, Japan}

\abstract{

Investigation of primordial non-Gaussianity holds immense importance in testing the inflation paradigm and shedding light on the physics of the early Universe. In this study, we conduct the complete analysis of scalar-induced gravitational waves (SIGWs) by incorporating the local-type non-Gaussianity $f_{\mathrm{NL}}$ and $g_{\mathrm{NL}}$. We develop Feynman-like diagrammatic technique and derive semi-analytic formulas for both the energy-density fraction spectrum and the angular power spectrum. For the energy-density fraction spectrum, we analyze all the relevant Feynman-like diagrams, determining their contributions to the spectrum in an order-by-order fashion. As for the angular power spectrum, our focus lies on the initial inhomogeneities, giving rise to anisotropies in SIGWs, that arise from the coupling between short- and long-wavelength modes due to primordial non-Gaussianity. Our analysis reveals that this spectrum exhibits a typical multipole dependence, characterized by $\tilde{C}_{\ell}\propto[\ell(\ell+1)]^{-1}$, which plays a crucial role in distinguishing between different sources of gravitational waves. Depending on model parameters, significant anisotropies can be achieved. We also show that the degeneracies in model parameters can be broken. The findings of our study underscore the angular power spectrum as a robust probe for investigating primordial non-Gaussianity and the physics of the early Universe. Moreover, our theoretical predictions can be tested using space-borne gravitational-wave detectors and pulsar timing arrays. 

}

\pgfplotsset{compat=1.18}

\begin{document}
 
\maketitle
\flushbottom

\section{Introduction}\label{sec:intro}

Studying primordial non-Gaussianity is essential for testing the inflationary paradigm and gaining insights into the physics of the early Universe \cite{Maldacena:2002vr,Bartolo:2004if,Allen:1987vq,Bartolo:2001cw,Acquaviva:2002ud,Bernardeau:2002jy,Chen:2006nt}. Primordial non-Gaussianity refers to deviations from Gaussian statistics in the distribution of primordial perturbations that arise during inflation. In the standard inflationary scenario, these perturbations are assumed to be Gaussian, with their statistics fully described by the power spectrum \cite{Lyth:1998xn}. However, there are scenarios where these perturbations can exhibit non-Gaussian behavior, providing valuable insights into early-Universe physics \cite{Meerburg:2019qqi,Davies:2021loj}. Primordial non-Gaussianity involves higher-order correlations and statistical properties of the primordial perturbations \cite{Maldacena:2002vr}. Different inflationary models can lead to varying levels and types of non-Gaussianity, which can be characterized by non-linear parameters such as $\fnl$ and $\gnl$. Observations of the \ac{CMB} \cite{Planck:2019kim} and \ac{LSS} \cite{Castorina:2019wmr,Biagetti:2020skr} have been used for constraining these parameters on the largest observational scales. Such constraints provide valuable information about the physics of inflation and aid in distinguishing between different inflationary models.

The \acp{SIGW} \cite{Ananda:2006af,Baumann:2007zm,Espinosa:2018eve,Kohri:2018awv,Mollerach:2003nq,Assadullahi:2009jc,Domenech:2021ztg} provide a powerful tool for investigating primordial non-Gaussianity, particularly on scales that are beyond the reach of measurements from \ac{CMB} and \ac{LSS}. These \acp{GW} arise from the nonlinear interactions of linear cosmological curvature perturbations in the early Universe. Unlike photons, gravitons can freely propagate once generated, as the Universe is transparent to \acp{GW} \cite{Bartolo:2018igk,Flauger:2019cam}. As a result, the study of gravitational waves offers the potential to probe physical processes that occurred during the extremely early epochs of the Universe, corresponding to physics at incredibly small scales.

Primordial non-Gaussianity has significant effects on the energy-density fraction spectrum of \acp{SIGW}. Regarding to local-type non-Gaussianity $\fnl$, complete analysis presented in Refs.~\cite{Adshead:2021hnm,Ragavendra:2021qdu,Li:2023qua,Abe:2022xur} show that this spectrum can be substantially enhanced, potentially reaching up to two orders of magnitude larger than the Gaussian scenario, depending on the levels of $\fnl$. 
Other works related with the effects of primordial non-Gaussianity (e.g., $\fnl$ and $\gnl$) on \acp{SIGW} have been shown in Refs.~\cite{Cai:2018dig,Unal:2018yaa,Atal:2021jyo,Ragavendra:2020sop,Yuan:2020iwf,Yuan:2021qgz,Yuan:2023ofl,Garcia-Saenz:2022tzu,Zhang:2021rqs,Domenech:2017ems,Garcia-Bellido:2017aan,Nakama:2016gzw,Lin:2021vwc,Chen:2022dqr,Cai:2019amo,Cai:2019elf}.
Moreover, the presence of primordial non-Gaussianity can also impact the formation of \acp{PBH} that coexist with \acp{SIGW}. The distribution of \ac{PBH} masses is influenced by the non-Gaussian features 
\cite{Bullock:1996at,Byrnes:2012yx,Young:2013oia,Franciolini:2018vbk,Passaglia:2018ixg,Atal:2018neu,Atal:2019cdz,Taoso:2021uvl,Meng:2022ixx,Chen:2023lou,Kawaguchi:2023mgk,Fu:2020lob,Inomata:2020xad,Young:2014ana,Choudhury:2023kdb,Garcia-Bellido:2017aan,Nakama:2016gzw,Ferrante:2022mui,Green:2020jor,Carr:2020gox,Escriva:2022duf,Escriva:2022pnz,Ezquiaga:2019ftu,Kehagias:2019eil,Cai:2021zsp,Cai:2022erk,Yi:2020cut,Zhang:2021vak}. 
Experimental tests of these theoretical predictions can be conducted through multi-band \ac{GW} observations. Notably, \ac{PTA} experiments \cite{Xu:2023wog,EPTA:2023fyk,NANOGrav:2023gor,Reardon:2023gzh} have provided compelling evidence for the existence of a \ac{GWB}, which is speculated to originate from \acp{SIGW} \cite{Antoniadis:2023xlr,NANOGrav:2023hvm,Franciolini:2023pbf,Inomata:2023zup,Cai:2023dls,Wang:2023ost,Liu:2023ymk,Abe:2023yrw,Ebadi:2023xhq,Figueroa:2023zhu,Yi:2023mbm,Madge:2023cak,Firouzjahi:2023lzg,Zhu:2023faa,You:2023rmn,Ye:2023xyr,HosseiniMansoori:2023mqh,Balaji:2023ehk,Das:2023nmm,Bian:2023dnv,Jin:2023wri,Zhao:2023joc,Liu:2023pau,Yi:2023tdk,Frosina:2023nxu,Choudhury:2023hfm,Ellis:2023oxs,Kawasaki:2023rfx,Yi:2023npi,Harigaya:2023pmw,An:2023jxf,Gangopadhyay:2023qjr,Chang:2023ist}. They provide an avenue for testing and validating the presence of \acp{SIGW} and their associated non-Gaussian features.

Studies have demonstrated that the non-Gaussianity $\fnl$ can induce significant inhomogeneities and anisotropies in \acp{SIGW} \cite{Bartolo:2019zvb,Li:2023qua}. 
Unlike \ac{CMB}, where the angular power spectrum is determined by the two-point correlators of linear curvature perturbations, the angular power spectrum of \acp{SIGW} is determined by the four-point correlators of these perturbations, inherently incorporating the effects of primordial non-Gaussianity $\fnl$. 
This implies that the energy density of \acp{SIGW} produced by short-wavelength modes is redistributed by long-wavelength modes due to the coupling between these modes. The parameter $\fnl$ was initially investigated as a means to explore this possibility in Refs.~\cite{Bartolo:2019zvb}, and subsequently, a comprehensive analysis on this topic was conducted for the first time in Ref.~\cite{Li:2023qua}. Other relevant works can be found in Refs.~\cite{ValbusaDallArmi:2020ifo,Dimastrogiovanni:2021mfs,LISACosmologyWorkingGroup:2022kbp,LISACosmologyWorkingGroup:2022jok,Unal:2020mts,Malhotra:2020ket,Carr:2020gox,Cui:2023dlo,Malhotra:2022ply,ValbusaDallArmi:2023nqn}. In particular, Ref.~\cite{Wang:2023ost} showed that if the \ac{PTA} signal originates from scalar-induced gravitational waves, it can be tested using the \ac{SKA} \cite{2009IEEEP..97.1482D,Weltman:2018zrl,Moore:2014lga,Janssen:2014dka} by measuring the corresponding angular power spectrum. Furthermore, the angular power spectrum was found to break degeneracies in cosmological parameters for the energy-density fraction spectrum \cite{Li:2023qua}, making it a powerful tool for probing primordial non-Gaussianity and exploring physics in the early Universe. 
However, to our knowledge, there are not works extending to consider the non-Gaussianity $\gnl$.

In this study, we present the first complete analysis of \acp{SIGW}, specifically examining both the energy-density fraction spectrum and the angular power spectrum, by incorporating the local-type primordial non-Gaussianity, specifically the parameters $\fnl$ and $\gnl$. This work builds upon our previous research \cite{Li:2023qua} and extends it to include a broader scope. To begin, we investigate the energy-density fraction spectrum of \acp{SIGW} at a homogeneous and isotropic level using a Feynman-like diagrammatic approach. We derive semi-analytic formulas for this spectrum and analyze the impact of non-Gaussianity on its behavior. As part of a comprehensive analysis, we establish a systematic classification of Feynman-like diagrams and develop a dictionary that relates these diagrams to their corresponding integrals. Subsequently, we focus on studying the angular power spectrum of \acp{SIGW}, which is influenced by both the initial inhomogeneities and the propagation effects. While the propagation effects are similar to those observed in \ac{CMB}, the analysis of initial inhomogeneities requires an extension of the Feynman-like diagrammatic technique. This is because practical calculations involve a large number of Feynman-like diagrams, making it challenging to straightforwardly expand the multi-point correlators using Wick's theorem. To address this, we develop a systematic method for evaluating the integrals associated with these Feynman-like diagrams, enabling a comprehensive analysis of the angular power spectrum. We derive semi-analytic formulas for the angular power spectrum, which can be expressed as a combination of the Feynman-like diagrams in our established dictionary. Furthermore, we perform numerical studies to explore the implications of the angular power spectrum for primordial non-Gaussianity. We also demonstrate how our theoretical results can be tested through future multi-band \ac{GW} observations. Overall, this work presents a thorough analysis of both the energy-density fraction spectrum and the angular power spectrum of \acp{SIGW}, incorporating the effects of primordial non-Gaussianity $\fnl$ and $\gnl$. It provides semi-analytic formulas and numerical results that contribute to our understanding of the implications of non-Gaussianity in the early Universe and suggests potential avenues for experimental validation through \ac{GW} observations.

The paper is structured as follows. In \cref{sec:ogw}, we investigate the energy-density fraction spectra of homogeneous \acp{SIGW} at the background level. In \cref{sec:aps}, we analyze the angular power spectrum of anisotropic \acp{SIGW}, considering the contributions from both the initial inhomogeneities and the propagation effects at the perturbation level. In \cref{sec:conc}, we provide a discussion and conclusion of our findings. The supplemental materials, including additional details on the theoretical basics, numerical integration techniques, and the Boltzmann equation, are summarized in \cref{sec:basic,sec:num-int,sec:Boltz}, respectively.

\section{Energy-density fraction spectrum}
\label{sec:ogw}

In this section, we initially disregard the presence of large-scale inhomogeneities in \acp{SIGW}, but we will address them in the next section. Therefore, in this section, we treat \acp{SIGW} as a homogeneous and isotropic \ac{GWB} on large scales. However, small-scale inhomogeneities can still exist within \acp{SIGW}. Nevertheless, due to the angular resolution limitations of gravitational wave detectors, the observed signal along a line-of-sight is a combination of \acp{SIGW} originating from numerous small-scale regions. Consequently, the observed signal represents an ensemble average over these regions, resulting in identical signals along different line-of-sights. This perspective simplifies the evaluation of the energy-density fraction spectrum. By assuming the small-scale regions to be identical (either isotropic or anisotropic, although we adopt isotropy for simplicity), we can effectively obtain the spectrum by evaluating it in any one region. These regions can be viewed as the Hubble horizons at the time of \ac{SIGW} production. We carry out this evaluation in the subsequent analysis.

The energy-density fraction spectrum $\bar{\Omega}_{\mathrm{GW}} (\eta,q)$ for \acp{SIGW} on subhorizon scales is defined as follows \cite{Maggiore:1999vm}  
\begin{equation}\label{eq:Omega-def}
    \bar{\rho}_\uGW (\eta) 
    = \rho_\uc(\eta) \int \ud \ln q \, \bar{\Omega}_{\uGW} (\eta,q)\ ,
\end{equation}
where $\rho_\uc=3\cH^2\mpl^2$ denotes the critical energy density of the Universe, $\mpl=1/\sqrt{8\pi G}$ is the Planck mass, $\cH(\eta)$ stands for the conformal Hubble parameter at conformal time $\eta$, and the overbar signifies physical quantities at the background level. Here and hereinafter, the wavenumber $q$ is denoted as the magnitude of the 3-momentum $\mathbf{q}$, namely, $q=|\mathbf{q}|$. The energy density of \acp{SIGW} is defined as follows \cite{Maggiore:1999vm}
\begin{equation}\label{eq:rho-def}
    \bar{\rho}_\uGW (\eta) 
    = \frac{\mpl^2}{16 a^2(\eta)} \langle\overbar{\partial_l h_{ij}(\eta,\bx) \partial_l h_{ij}(\eta,\bx)}\rangle\ ,
\end{equation}
where $a(\eta)$ is the scale factor of the Universe, the long overbar labels a temporal average over oscillations, and the angle brackets stand for the ensemble average.
It is useful to expand \acp{SIGW} in Fourier mode as follows 
\begin{equation}\label{eq:h-Fourier}
h_{ij}(\eta,\bx) = \sum_{\lambda=+,\times} 
\int \frac{\ud^3 \bq}{(2\pi)^{3/2}} e^{i\bq\cdot\bx} \epsilon_{ij}^{\lambda}(\bq) h_\lambda(\eta, \bq)\ ,
\end{equation}
where the polarization tensors are defined as $\epsilon^+_{ij}(\bq) = \left[\epsilon_i(\bq) \epsilon_j(\bq) - \bar{\epsilon}_i(\bq) \bar{\epsilon}_j(\bq)\right] /\sqrt{2}$ and $\epsilon^\times_{ij}(\bq) =  \left[\epsilon_i(\bq) \bar{\epsilon}_j(\bq) + \bar{\epsilon}_i(\bq) \epsilon_j(\bq) \right]/\sqrt{2}$, with  $\epsilon_{i}(\bq)$ and $\bar{\epsilon}_{i}(\bq)$ being a set of orthonormal basis vectors that are perpendicular to the wavevector $\bq$. 
This definition implies that we have $\epsilon_{ij}^\lambda \epsilon_{ij}^{\lambda'} = \delta^{\lambda\lambda'}$ and $\epsilon_{ij}^\lambda (\bq)$ are transverse to $\bq$. 
The power spectrum of $h_\lambda$ is defined as 
\begin{equation}\label{eq:h-cor} 
    \langle h_\lambda (\eta,\bq) h_{\lambda'} (\eta,\bq')\rangle 
    = \delta_{\lambda\lambda'} \delta^{(3)} (\bq+\bq') P_{h_\lambda} (\eta,q) \ . 
\end{equation}  
Combining the above formulae, the energy-density fraction spectrum is expressed in terms of the power spectrum of \acp{SIGW}, namely, \cite{Inomata:2016rbd} 
\begin{equation}\label{eq:Omegabar-h} 
    \bar{\Omega}_\uGW (\eta, q)
    = \frac{q^5}{96\pi^2 \cH^2} 
        \overbar{P_{h} (\eta,q)}\ ,
\end{equation}
where we introduce $P_{h} = \sum_{\lambda=+,\times} P_{h_\lambda}$ for convenience.

The equation of motion for \acp{SIGW} was originally derived in the literature \cite{Ananda:2006af,Baumann:2007zm}. Semi-analytic derivations can be found in Refs.~\cite{Espinosa:2018eve,Kohri:2018awv}. We provide a brief overview of the derivation results in Appendix \ref{sec:basic}, following the conventions of Ref.~\cite{Li:2023qua}. We have a solution to the equation of motion of \acp{SIGW} on subhorizon scales, given by 
\begin{equation}\label{eq:h} 
    h_\lambda(\eta, \bq) 
    = 4 \int \frac{\ud^3 \bq_a}{(2\pi)^{3/2}} 
        \zeta(\bq_a) \zeta(\bq-\bq_a) Q_{\lambda}(\bq,\bq_a) 
        \hat{I} (\abs{\bq - \bq_a},q,\eta)\ , 
\end{equation}
where $\zeta(\mathbf{q})$ denotes the linear primordial curvature perturbations, $Q_{\lambda}(\mathbf{q}, \mathbf{q}_a)$ denotes the projection factor as shown in Eq.~(\ref{eq:Qsai}), and $\hat{I} (|\mathbf{q} - \mathbf{q}_a|,q,\eta)$ represents the kernel function as shown in Eqs.~(\ref{eq:Ihat-Iuv}) and (\ref{eq:I-RD}).

In a schematic manner, the equations above, Eqs.~(\ref{eq:h-Fourier}), (\ref{eq:h-cor}), (\ref{eq:Omegabar-h}), and (\ref{eq:h}), suggest that $\bar{\Omega}_\uGW\sim\langle\zeta^4\rangle$. To be more specific, we find
\begin{eqnarray}\label{eq:Ph-zeta}
    \langle
        h_{ij}(\eta,\bq)
        h_{ij}(\eta,\bq')
    \rangle
    & = & 16
        \int \frac{\ud^3 \bq_1}{(2\pi)^{3/2}} \frac{\ud^3 \bq_2}{(2\pi)^{3/2}} 
        \langle
            \zeta(\bq_1) \zeta(\bq - \bq_1) \zeta(\bq_2) \zeta(\bq' - \bq_2)
        \rangle \\
        & &\ \times \sum_{\lambda,\lambda'} \epsilon_{ij}^{\lambda} \epsilon_{ij}^{\lambda'}
        Q_{\lambda}(\bq, \bq_1)
        \hat{I} (\abs{\bq - \bq_1}, q_1, \eta) 
        Q_{\lambda'}(\bq', \bq_2)
        \hat{I} (\abs{\bq' - \bq_2}, q_2, \eta) \ .\nonumber
\end{eqnarray}
This result indicates the necessity to handle the four-point correlators of $\zeta$. The evaluation of this correlator will be discussed in detail in the remaining part of this section.

The nature of primordial curvature perturbations, such as non-Gaussianity, introduces complexity in calculating these correlators. By adopting the Feynman-like diagrammatic technique, the four-point correlator of Gaussian curvature perturbations can be described by a single diagram \cite{Espinosa:2018eve,Kohri:2018awv}. However, in the case of non-Gaussianity with a parameter $\fnl$, the derivation process becomes significantly more involved, requiring the evaluation of a total of seven diagrams to calculate the four-point correlator of linear curvature perturbations \cite{Adshead:2021hnm,Ragavendra:2021qdu,Li:2023qua,Abe:2022xur}. The first complete analysis of this calculation can be found in Ref.~\cite{Adshead:2021hnm}, and other relevant works on this topic can be found in Refs.~\cite{Ragavendra:2021qdu,Li:2023qua,Cai:2018dig,Unal:2018yaa,Atal:2021jyo,Zhang:2021rqs,Ragavendra:2020sop,Nakama:2016gzw,Abe:2022xur}. Furthermore, relevant works on the primordial non-Gaussian parameter $\gnl$ can be found in Refs.~\cite{Yuan:2020iwf,Yuan:2021qgz,Abe:2022xur,Yuan:2023ofl,Nakama:2016gzw}.

In this study, we employ the diagrammatic approach, which has been extensively utilized in the literature \cite{Adshead:2021hnm,Ragavendra:2021qdu,Li:2023qua,Abe:2022xur}, to investigate the local-type primordial non-Gaussianity, parameterized by $\fnl$ and $\gnl$, of curvature perturbations within the theoretical framework of \acp{SIGW}. This work serves as a natural extension of our previous research \cite{Li:2023qua}, which focused solely on $\fnl$. The primary objective of our study is to derive the formulas for the angular power spectrum of anisotropic \acp{SIGW}. Additionally, it is crucial to independently derive the formulas for $\bar{\Omega}_{\uGW}$ using the diagrammatic approach. Guided by these objectives, we conduct a comprehensive analysis of \acp{SIGW}, incorporating both $\fnl$ and $\gnl$. We provide formulae for the energy-density fraction spectrum and the angular power spectrum, and systematically compare our results with those presented in numerous pioneering papers. Our work can serve as a valuable resource for readers interested in the same topic, akin to a dictionary of sorts.

By incorporating the local-type primordial non-Gaussianity parameterized by $\fnl$ and $\gnl$, we express the linear curvature perturbations $\zeta$ in terms of their Gaussian components, denoted as $\zeta_g$, as follows \cite{Komatsu:2001rj} 
\begin{equation}\label{eq:Fnl-Gnl-def}
     \zeta (\bx) = \zeta_g (\bx) + \frac{3}{5}\fnl \left[ \zeta_g^2(\bx) - \langle \zeta_g^{2}(\bx) \rangle \right] + \frac{9}{25}\gnl \zeta_g^3(\bx) \ .
\end{equation}
Furthermore, it is customary to decompose $\zeta_g$ into short-wavelength modes $\zeta_{gS}$ and long-wavelength modes $\zeta_{gL}$, i.e., \cite{Tada:2015noa}
\begin{equation}\label{eq:shortlong}
\zeta_g=\zeta_{gS}+\zeta_{gL}\ .
\end{equation} 
As demonstrated in Refs.~\cite{Bartolo:2019zvb,Li:2023qua,Wang:2023ost}, the short-wavelength modes induce a \ac{GWB}, contributing to the homogeneous and isotropic energy density. On the other hand, the long-wavelength modes redistribute the energy density spatially, resulting in inhomogeneities and anisotropies within this \ac{GWB}. In this section, we will focus on studying the former, while the latter will be examined in the subsequent section.

Since $\zeta_{gS}$ (as well as $\zeta_{gL}$) is Gaussian, it is sufficient to determine its power spectrum, defined as 
\begin{equation}\label{eq:PgS-def}
\langle \zeta_{gS} (\bq) \zeta_{gS} (\bq') \rangle = \delta^{(3)} (\bq+\bq') P_{gS} (q)\ ,
\end{equation}
where the dimensionless power spectrum is defined as 
\begin{equation}
    \Delta^2_{S} (q)=\frac{q^3}{2\pi^2} P_{gS} (q)\ .
\end{equation}
Similarly, we define $\Delta_{L}^{2}(q)$ as the power spectrum of $\zeta_{gL}$. In this study, we assume $\Delta_{S}^{2}$ to follow a normal distribution with respect to $\ln q$, given by 
\begin{equation}\label{eq:Lognormal}
    \Delta^2_{S} (q) = \frac{A_S}{\sqrt{2\pi\sigma^2}}\exp\left[-\frac{\ln^2 (q/q_\ast)}{2 \sigma^2}\right]\ ,
\end{equation}
where $\sigma$ represents the spectral width, and $A_S$ is the spectral amplitude at the peak wavenumber $q_\ast$. The wavenumber $q$ can be directly converted into the frequency $\nu$, i.e., $q=2\pi\nu$. 
In this study, we consider a range of spectral amplitudes $A_S$ on the order of $10^{-4}$ to $10^{-1}$, which is particularly relevant in the context of \ac{PBH} formation scenarios (see reviews in Refs.~\cite{Green:2020jor,Carr:2020gox,Escriva:2022duf,Saikawa:2018rcs} and references therein). 
For simplicity, we set $\sigma$ to unity, but it can be easily generalized to other values. 
It is important to note that perturbativity imposes the conditions $1 > 3|\fnl| \sqrt{A_S}/5 + 9|\gnl| A_S/25$ and $3|\fnl| \sqrt{A_S}/5 > 9|\gnl| A_S/25$. 
One should note that some of the most extreme value, such as $(3|\fnl| \sqrt{A_S}/5 + 9|\gnl| A_S/25) \rightarrow 1$, may already exceed the reliable range of perturbativity.
In the next section, we will also consider $\Delta_{L}^{2}\simeq2.1\times10^{-9}$ on the largest observable scales \cite{Planck:2018vyg}.

\subsection{Feynman-like Rules and Diagrams}\label{subsec:frd}

By substituting Eq.~(\ref{eq:Fnl-Gnl-def}) into the four-point correlator $\langle\zeta^4\rangle$ in Eq.~(\ref{eq:Ph-zeta}), we can expand this correlator as a series of two-point correlators involving $\zeta_{gS}$. When studying Gaussian curvature perturbations, Wick's theorem is a valuable tool for such expansions \cite{Ananda:2006af,Baumann:2007zm,Espinosa:2018eve,Kohri:2018awv}. However, when considering non-Gaussianity, we encounter challenges in conducting such expansions due to the involvement of a large number of two-point correlators. This difficulty is particularly pronounced when studying the angular power spectrum, as it requires consideration of multi-point correlators.

\begin{figure}
    \centering
    \includegraphics[width =1 \columnwidth]{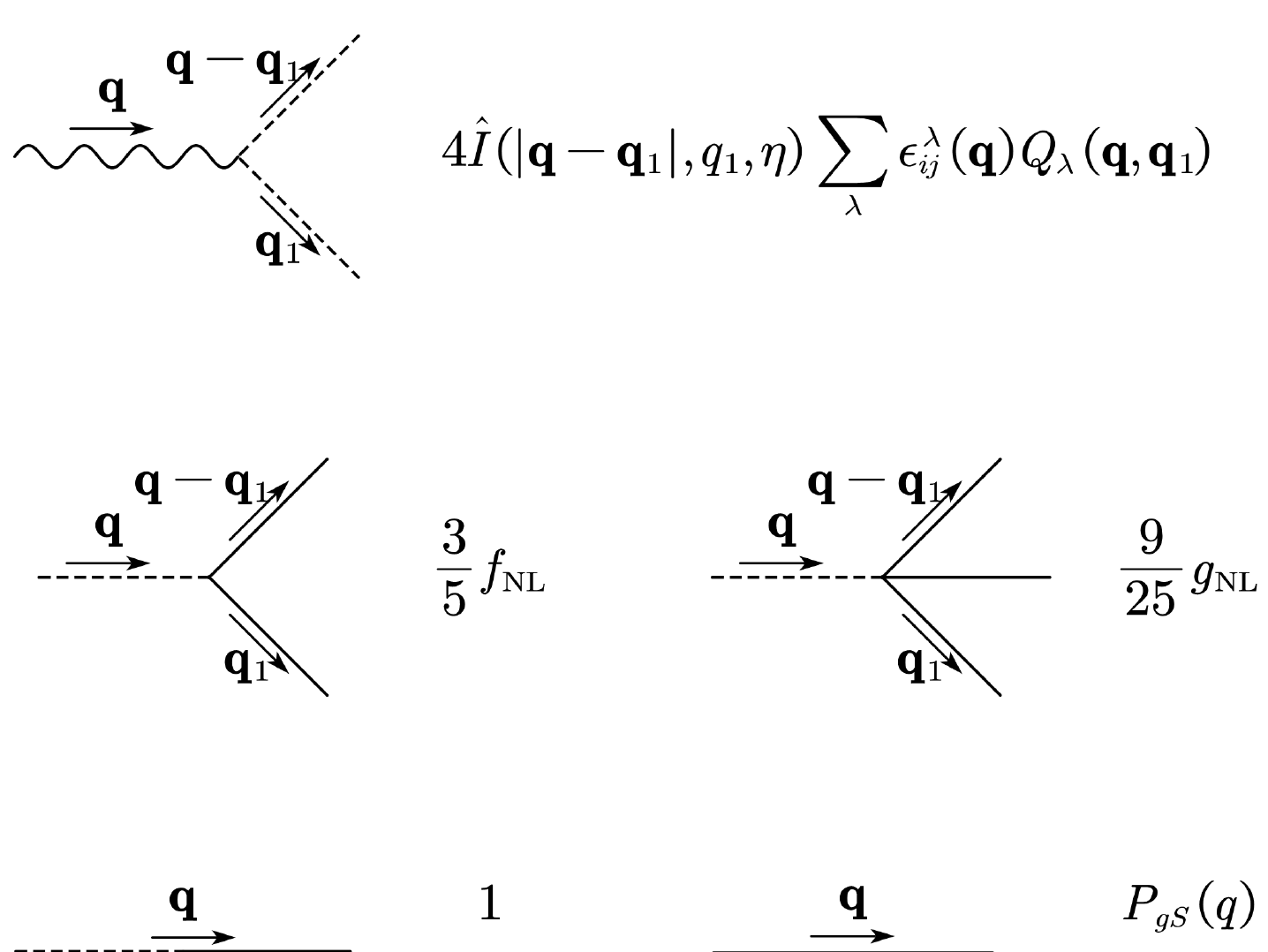}
    \caption{The Feynman-like rules for evaluation of Eq.~(\ref{eq:Ph-zeta}) by incorporating Eq.~(\ref{eq:Fnl-Gnl-def}). Wavy lines represent GWs, dashed lines denote the transfer functions, and solid lines stand for the primordial curvature power spectra. Arrows indicate the flow of comoving 3-momenta, which are conserved at each vertex. The total 3-momentum vanishes for each individual diagram, and all the loop 3-momenta should be integrated over. 
}\label{fig:FD_Rules}
\end{figure}

In this work, we employ the Feynman-like diagrammatic technique to efficiently and accurately evaluate the multi-point correlators. The diagrammatic technique has been widely used in quantum field theory and cosmology. Recently, it has been applied to calculate the energy-density fraction spectrum of \acp{SIGW} \cite{Adshead:2021hnm,Ragavendra:2021qdu,Abe:2022xur}, as well as the angular power spectrum \cite{Bartolo:2019zvb,Li:2023qua}. This technique has the potential to overcome the challenges encountered when expanding multi-point correlators using Wick's theorem. By utilizing this method, we can more easily evaluate the energy-density fraction spectrum and, in particular, the angular power spectrum.

To evaluate Eq.~(\ref{eq:Ph-zeta}) by incorporating Eq.~(\ref{eq:Fnl-Gnl-def}), we establish a set of Feynman-like rules, which are summarized in Figure~\ref{fig:FD_Rules}. The vertices in Figure~\ref{fig:FD_Rules} are labeled as the $h$-vertex\footnote{ Notably, we have included the  sum of polarization modes in the $h$-vertex in this study. This convention differs slightly from that in our existing paper \cite{Li:2023qua}. Nevertheless, this difference would not affect our results, but would simplify subsequent calculations, since the power spectra of the two polarization modes are the same, i.e., $P_{h_+} (\eta,q) = P_{h_\times} (\eta,q)$, in this work.} (upper panel), $\fnl$-vertex (middle left panel), $\gnl$-vertex (middle right panel), and Gaussian-vertex (bottom left panel), respectively. The propagator is shown in the bottom right panel. 

\begin{figure}
    \centering
    \includegraphics[width =1 \columnwidth]{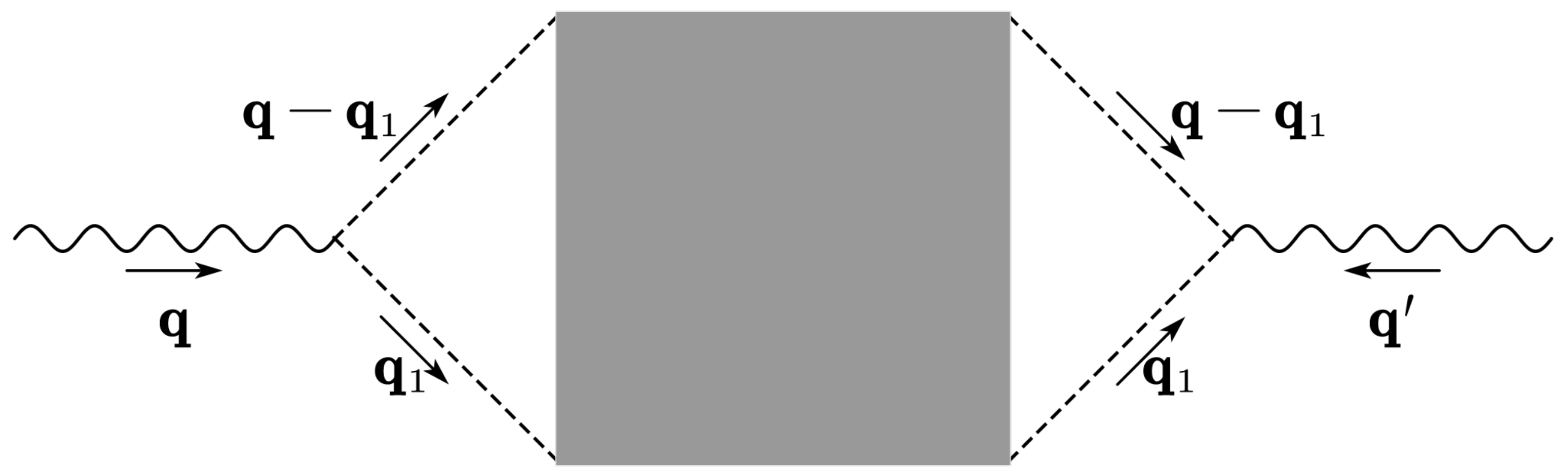}
    \caption{The Feynman-like diagrams for evaluating Eq.~(\ref{eq:Ph-zeta}) by incorporating Eq.~(\ref{eq:Fnl-Gnl-def}). 
    The shadow square should be replaced with the panels of Figure~\ref{fig:Feynman_Diagrams}. }\label{fig:FD_Omega_Frame}
\end{figure}

\begin{figure}[htbp]
\centering
    \subcaptionbox*{$G$}{\includegraphics[width = 0.103\columnwidth]{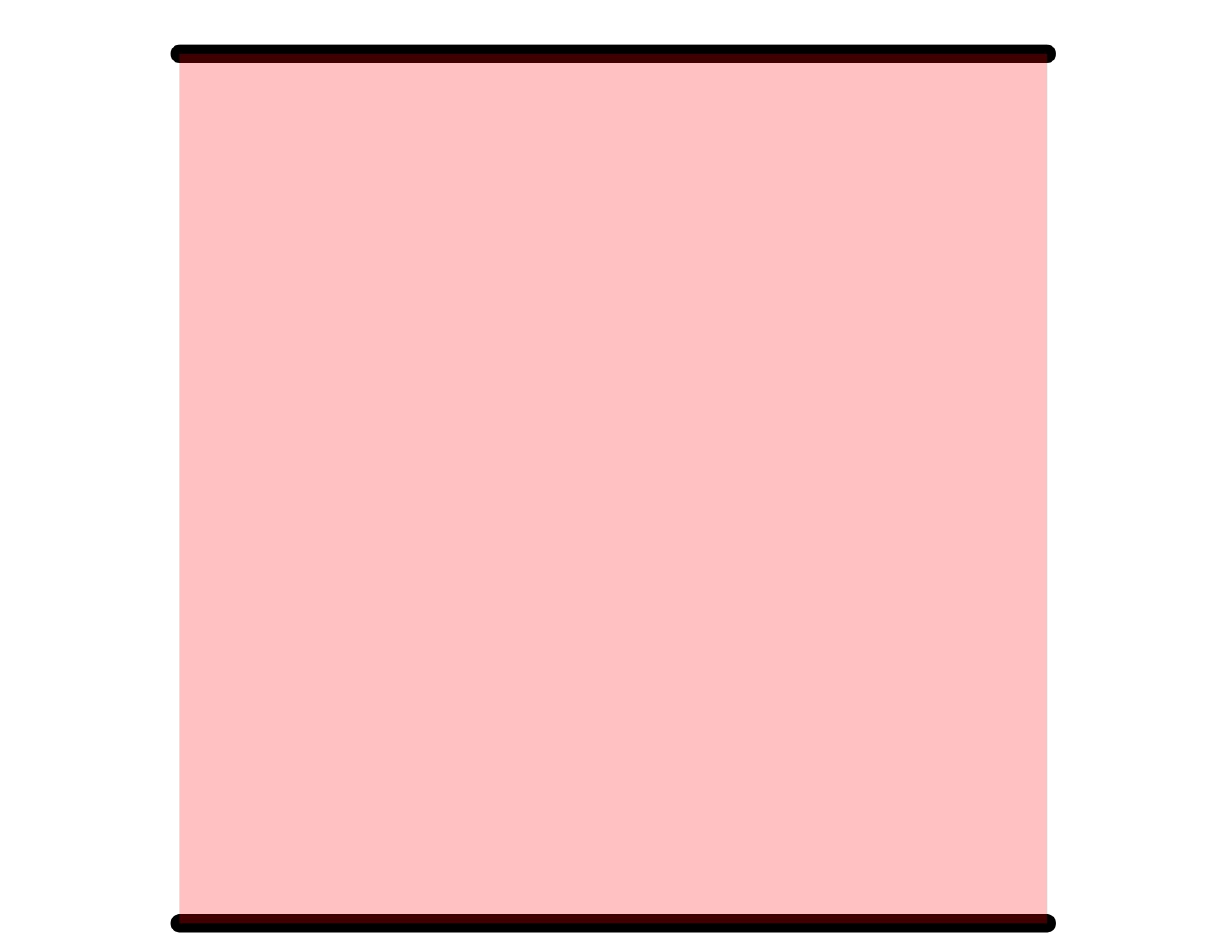}}
    \subcaptionbox*{$Gl$}{\includegraphics[width = 0.103\columnwidth]{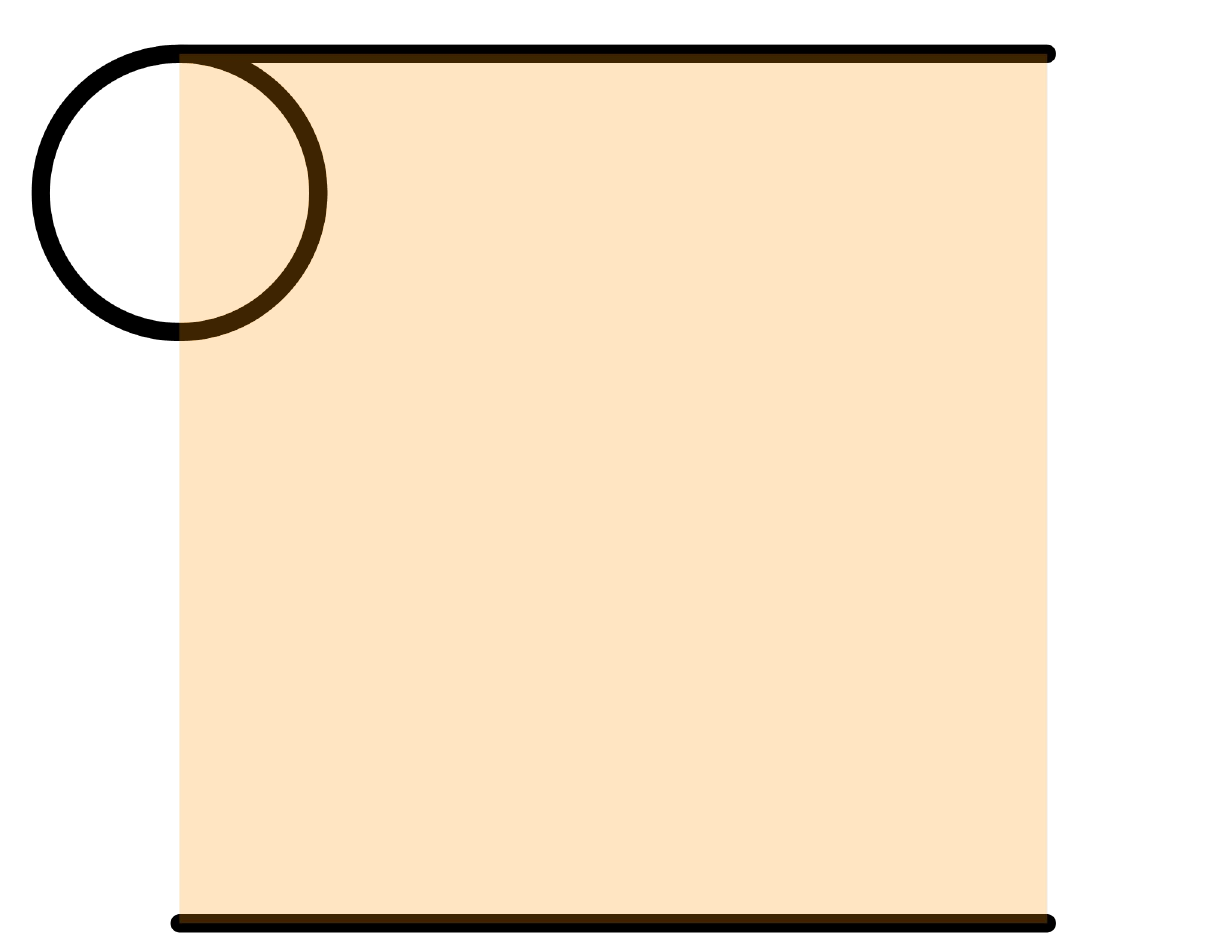}}
    \subcaptionbox*{$Gl^H$}{\includegraphics[width = 0.103\columnwidth]{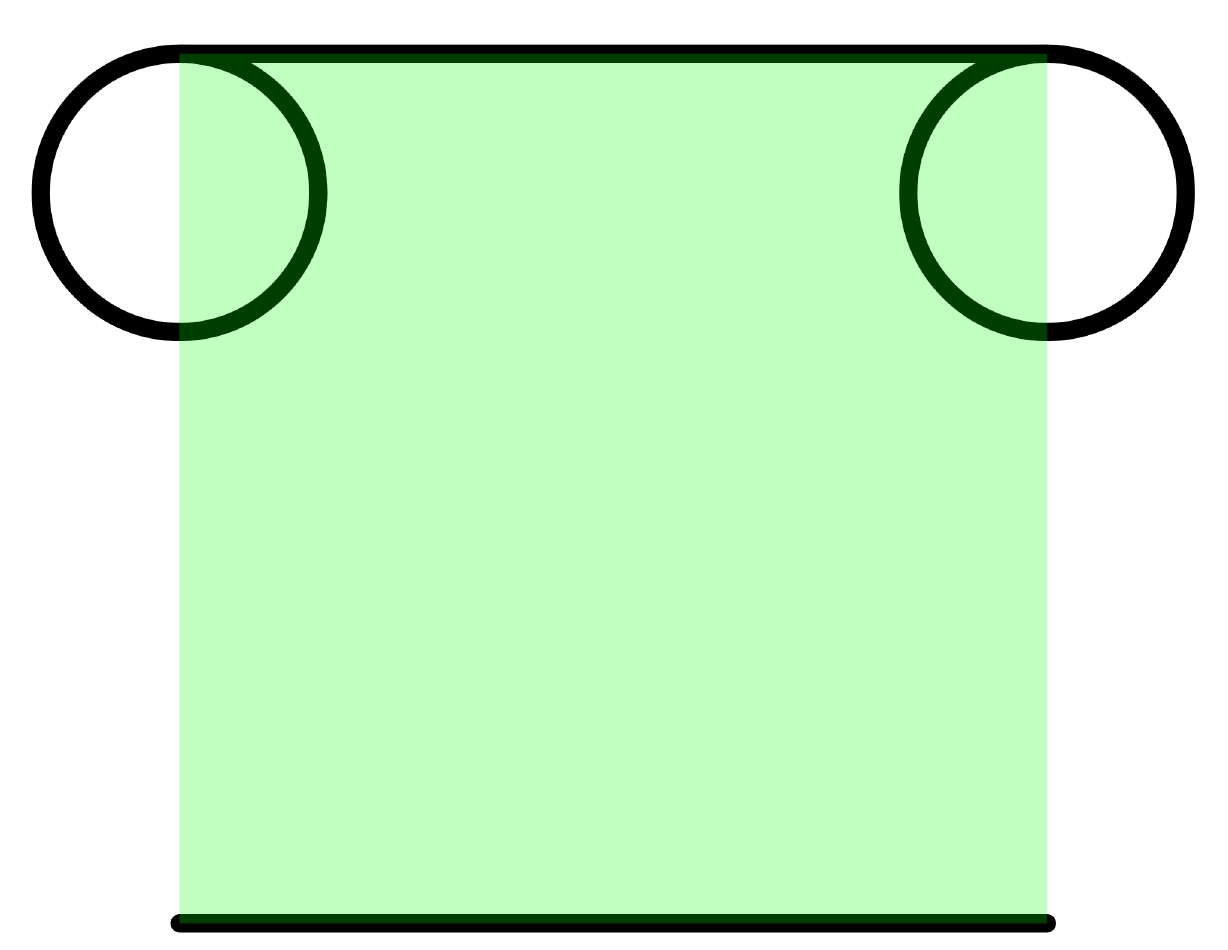}}
    \subcaptionbox*{$Gl^C$}{\includegraphics[width = 0.103\columnwidth]{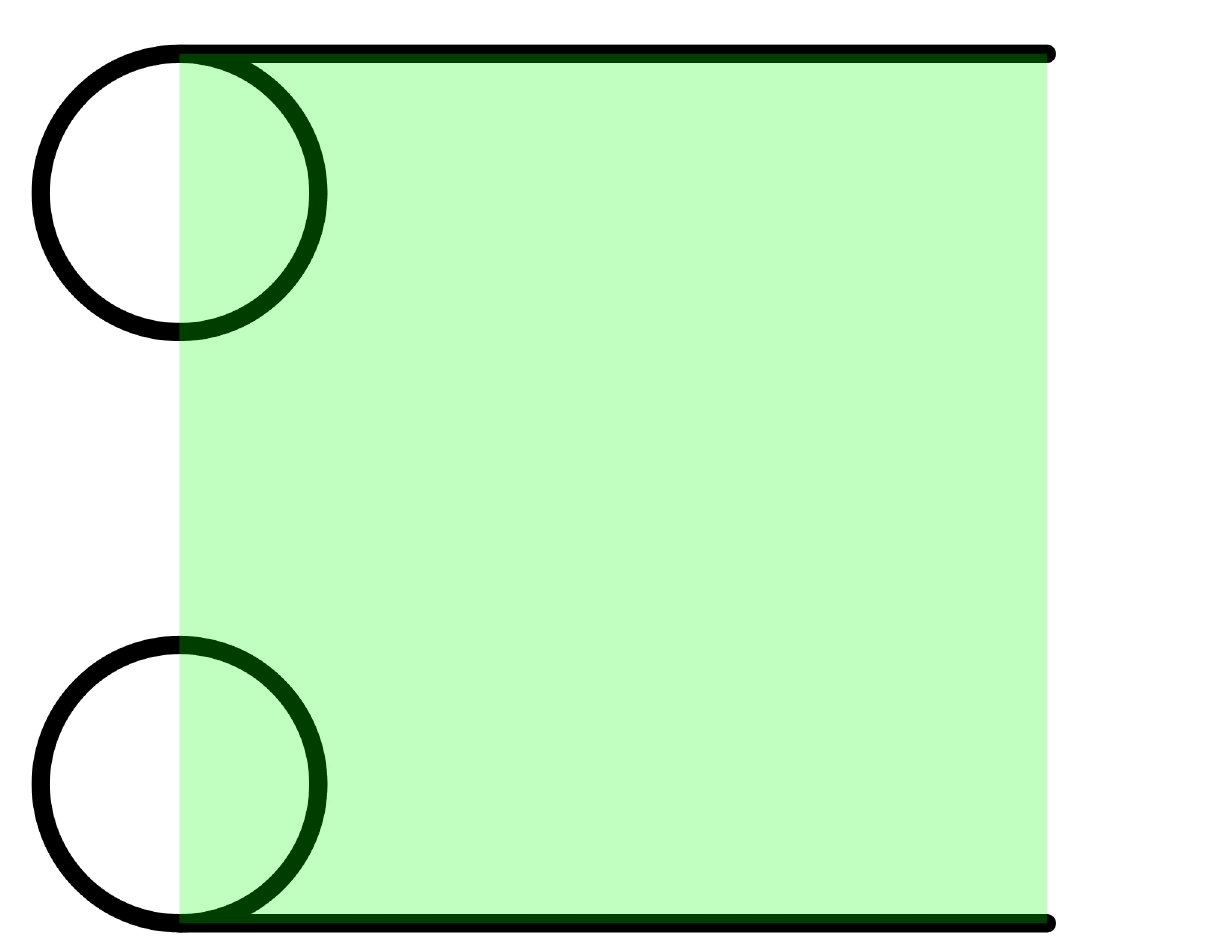}}
    \subcaptionbox*{$Gl^Z$}{\includegraphics[width = 0.103\columnwidth]{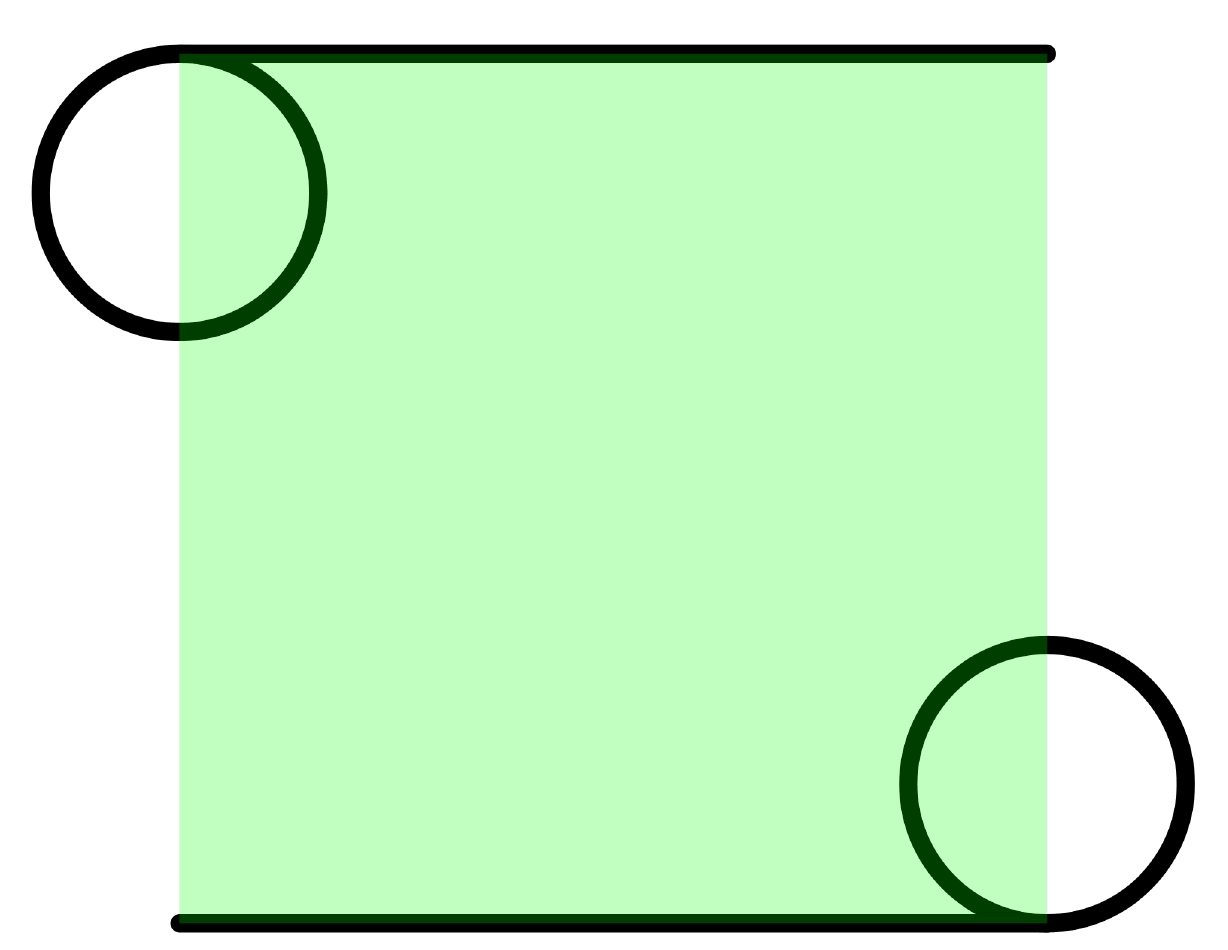}}
    \subcaptionbox*{$Gl^3$}{\includegraphics[width = 0.103\columnwidth]{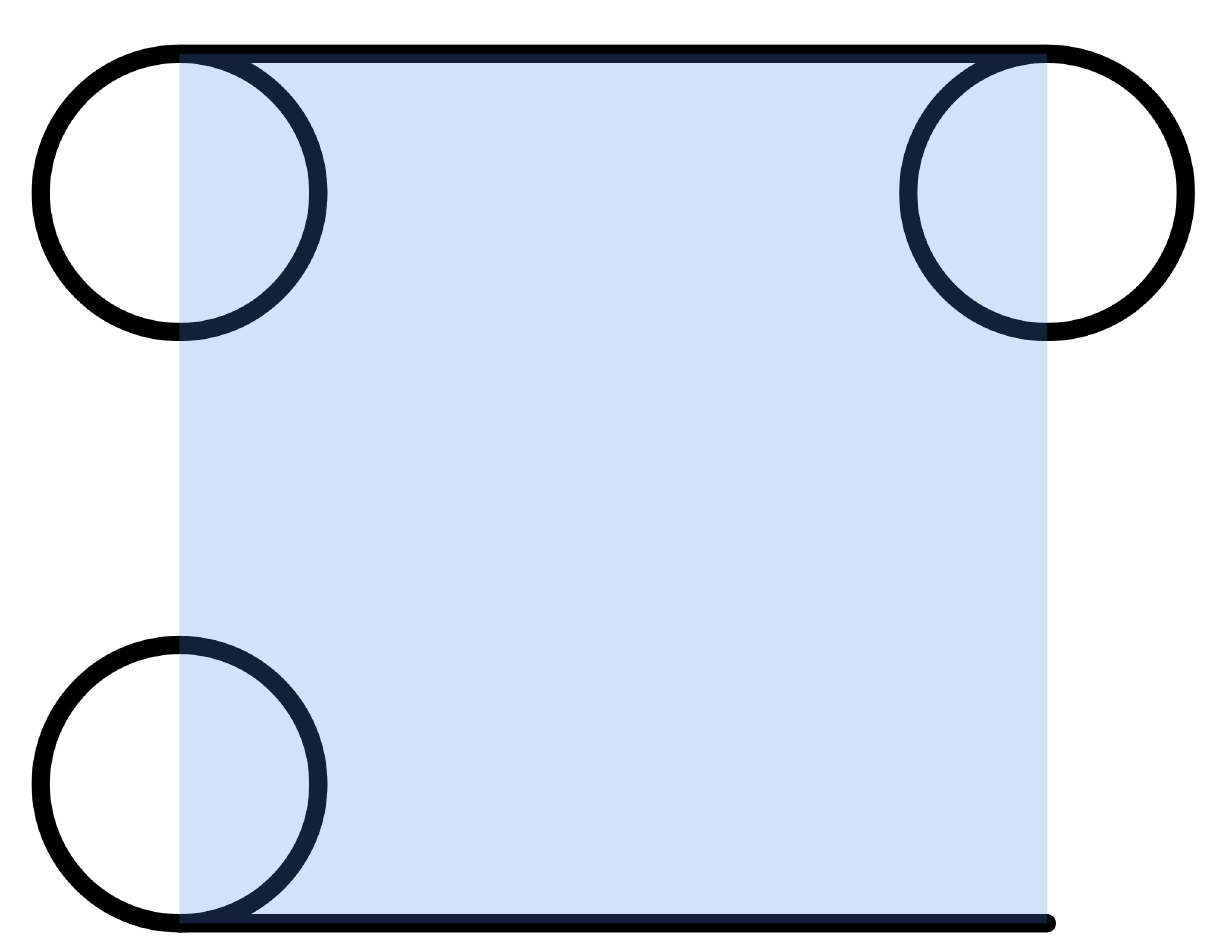}}
    \subcaptionbox*{$Gl^4$}{\includegraphics[width = 0.103\columnwidth]{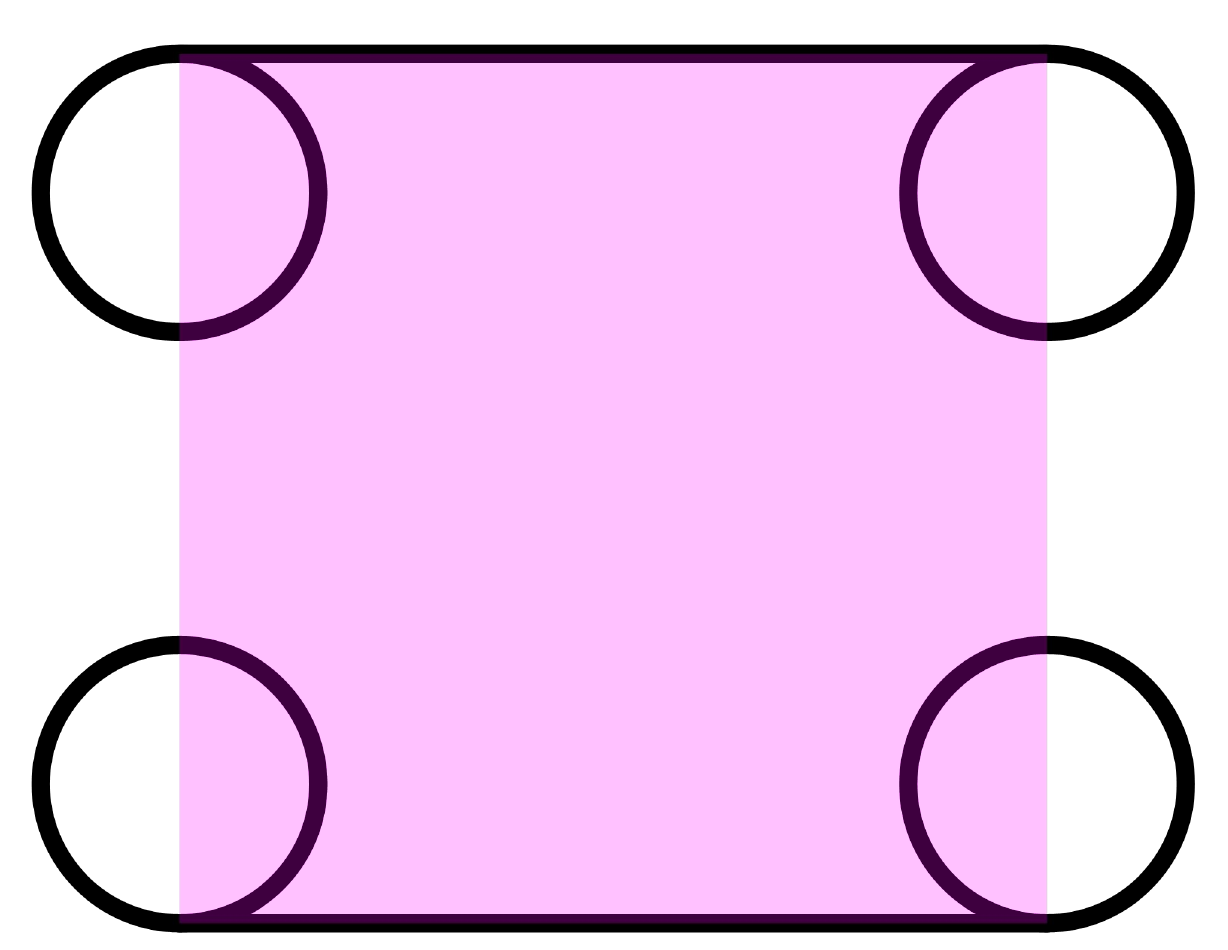}}\\
    \subcaptionbox*{$H$}{\includegraphics[width = 0.103\columnwidth]{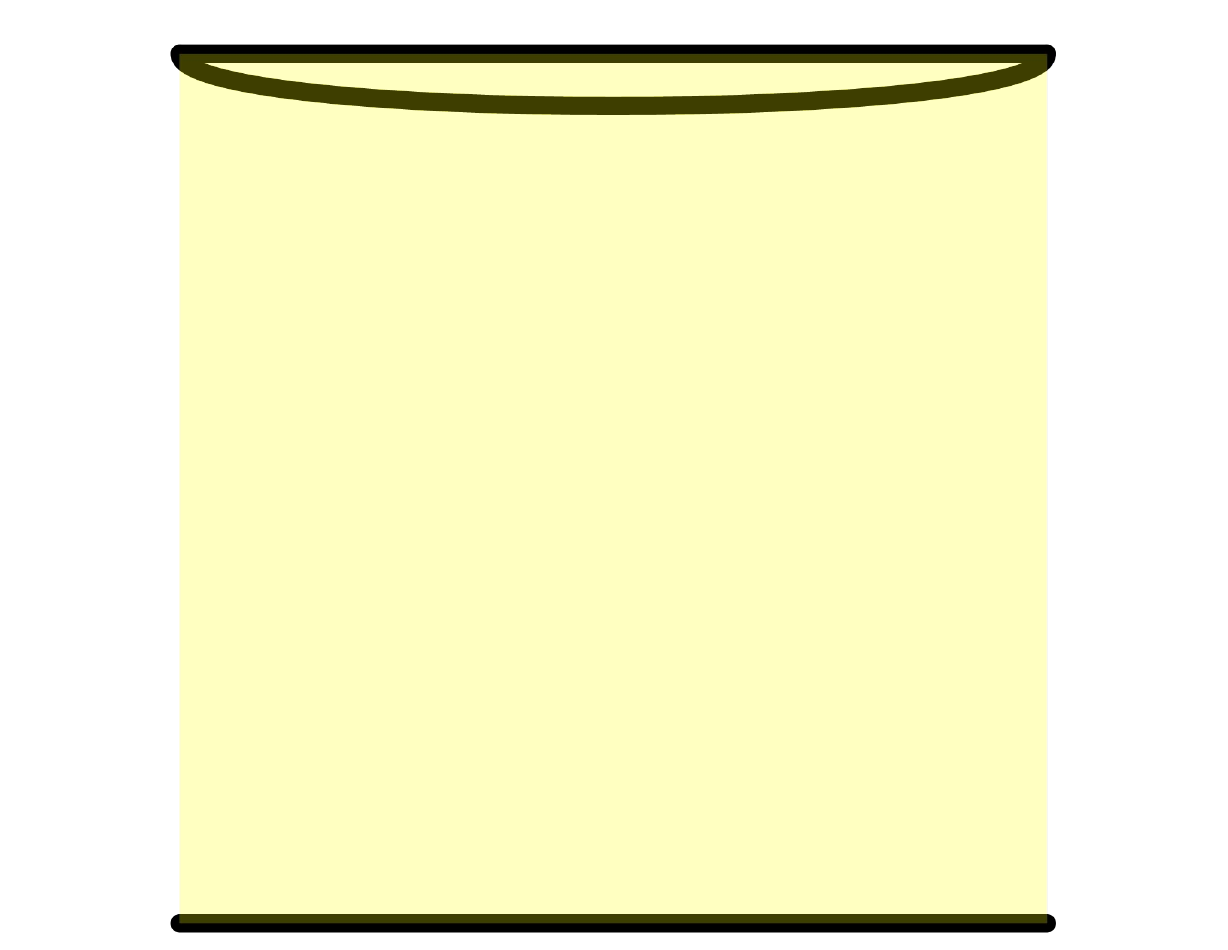}}
    \subcaptionbox*{$Hl$}{\includegraphics[width = 0.103\columnwidth]{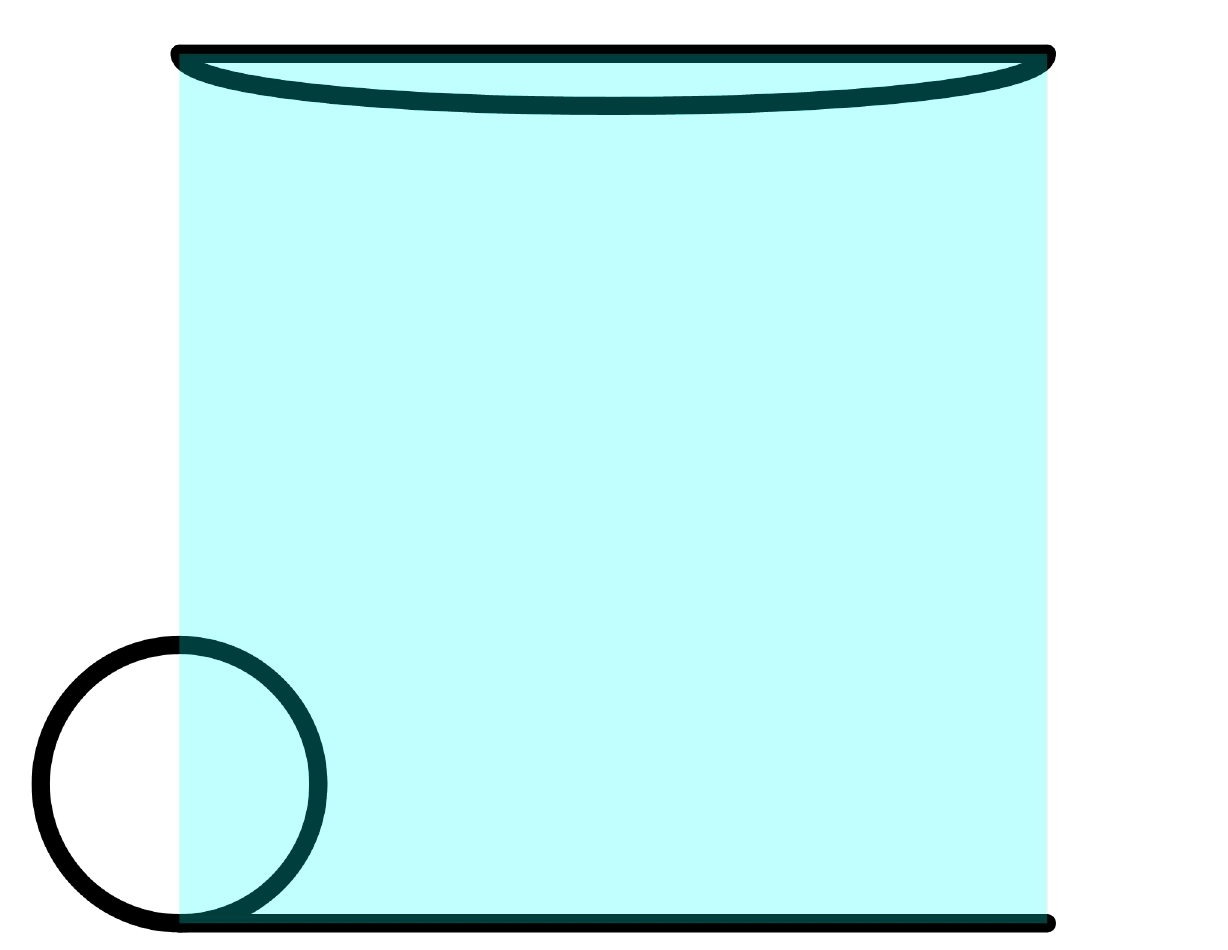}}
    \subcaptionbox*{$Hl^2$}{\includegraphics[width = 0.103\columnwidth]{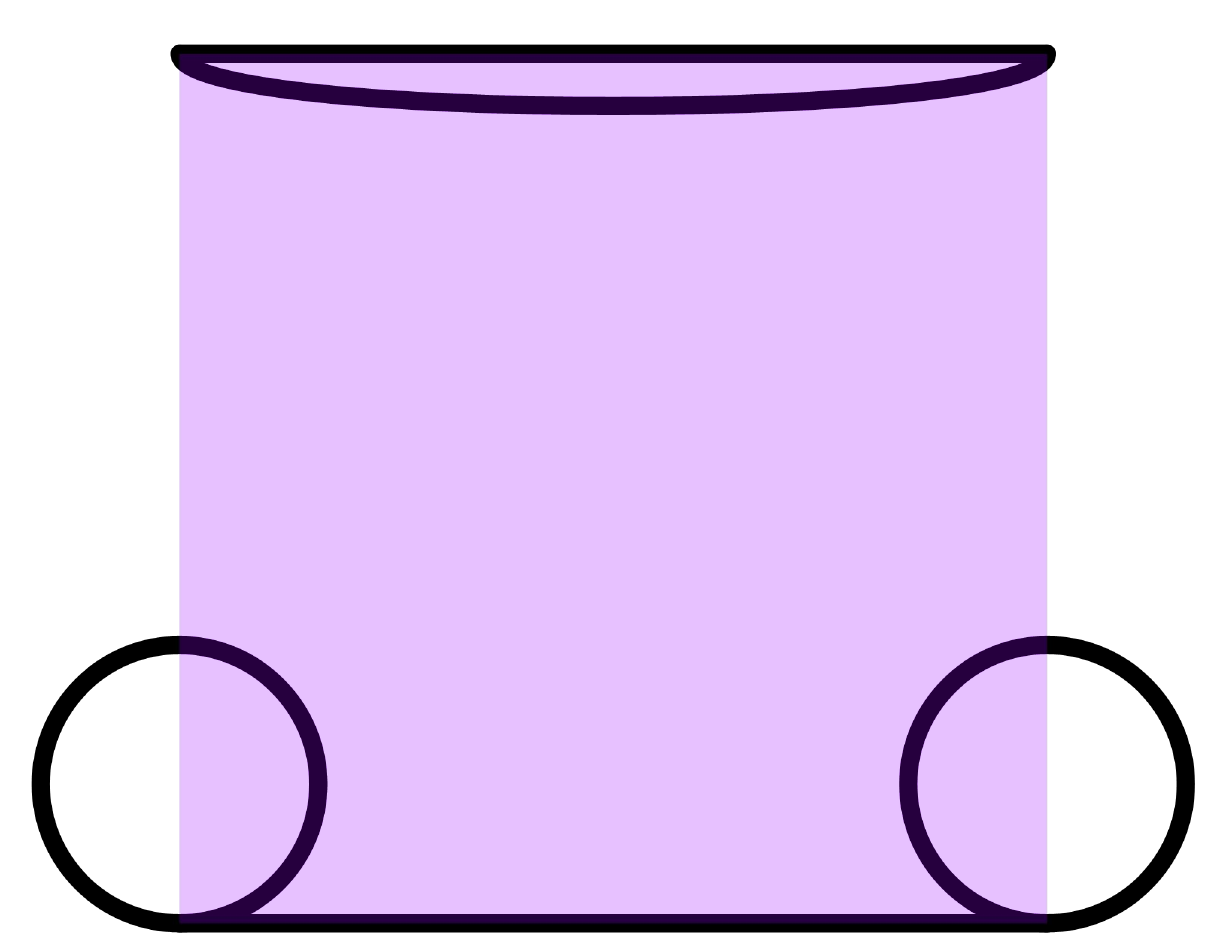}}
    \subcaptionbox*{$H^2$}{\includegraphics[width = 0.103\columnwidth]{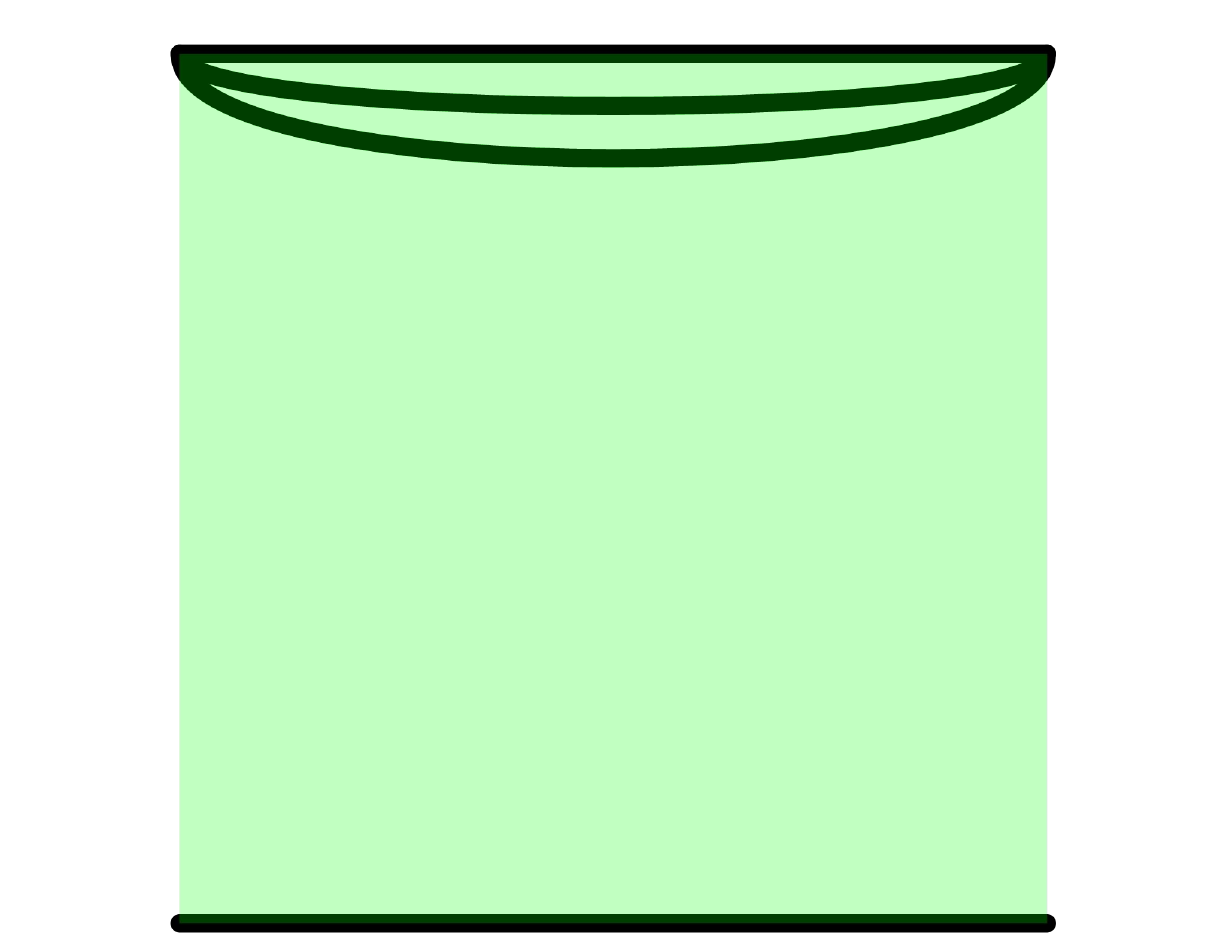}}
    \subcaptionbox*{$H^2 l$}{\includegraphics[width = 0.103\columnwidth]{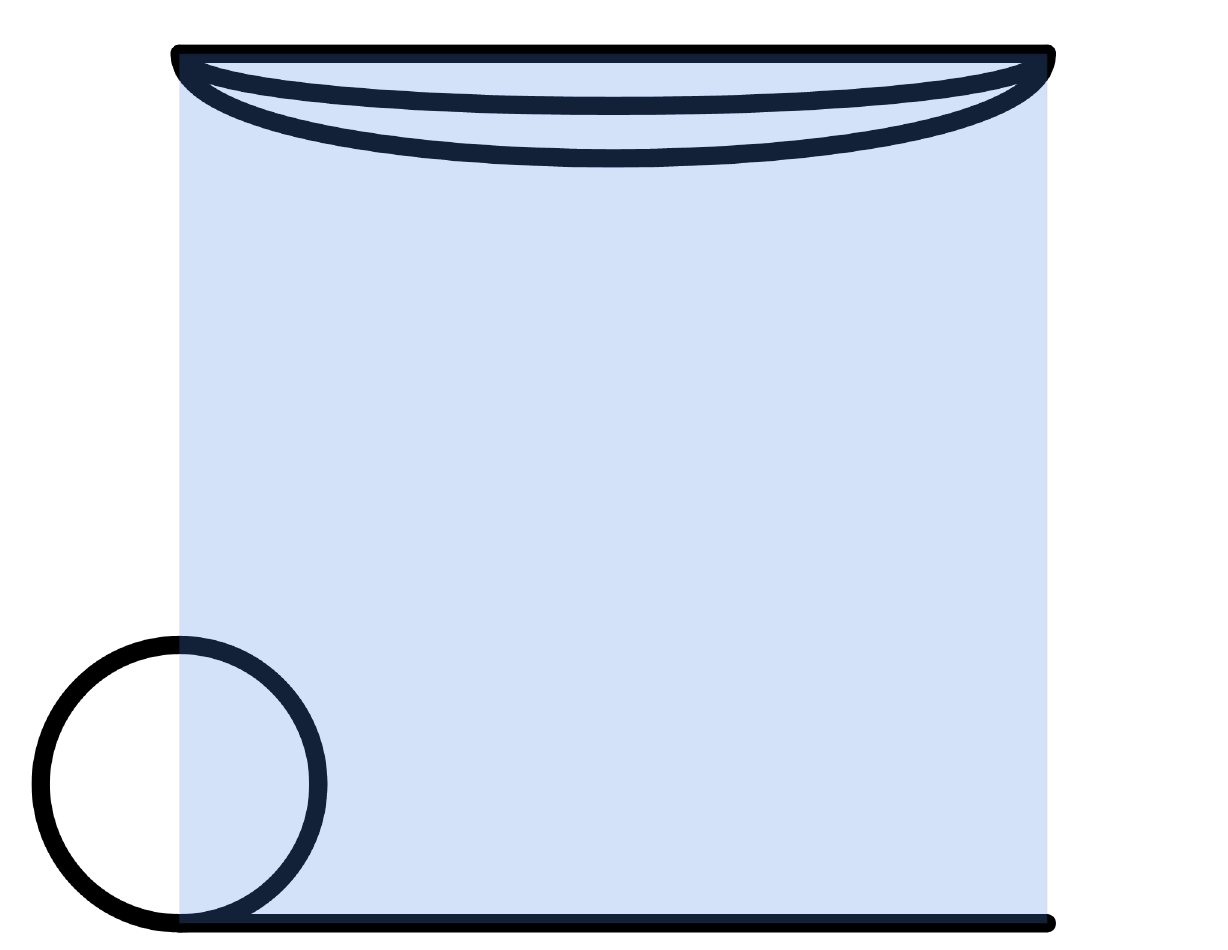}}
    \subcaptionbox*{$H^2 l^2$}{\includegraphics[width = 0.103\columnwidth]{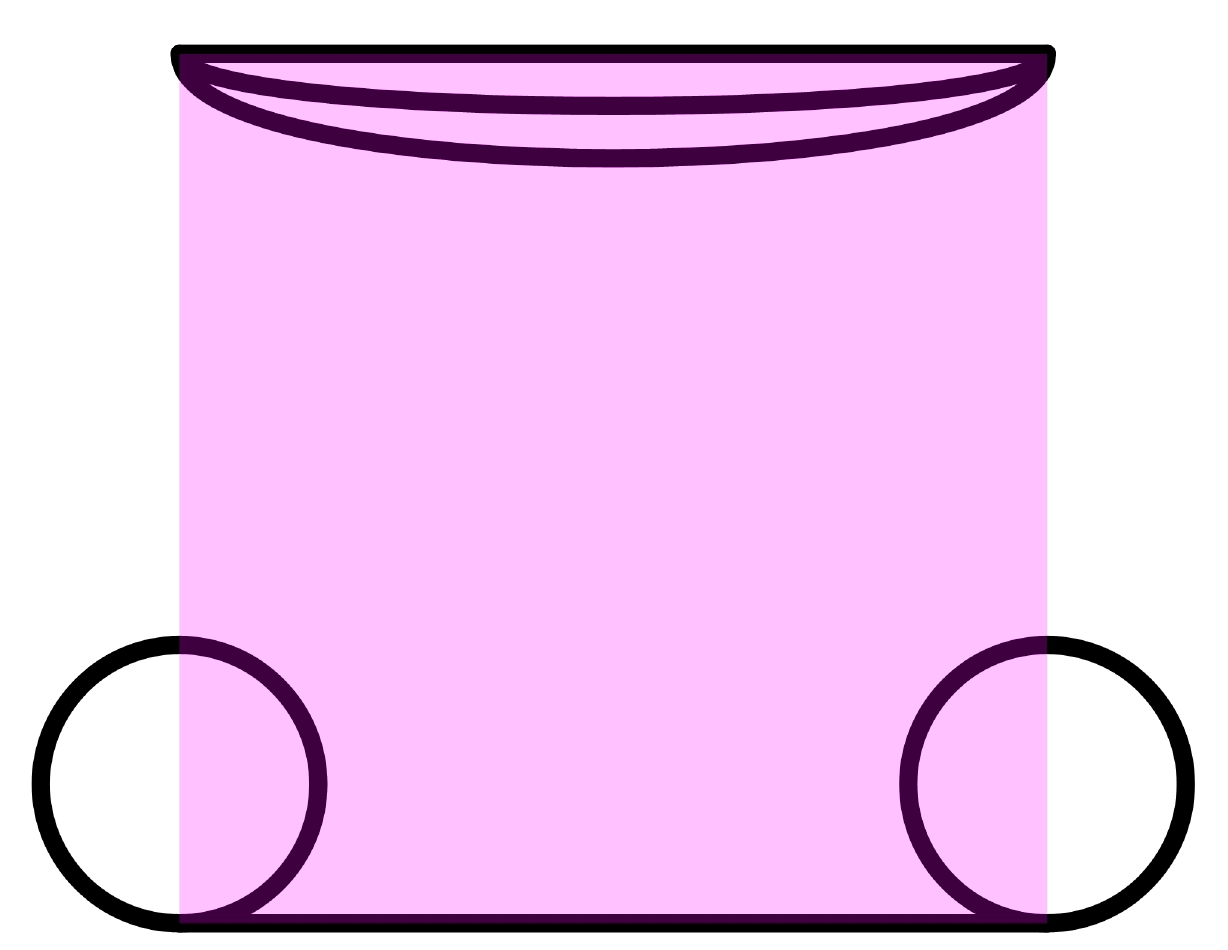}}
    \subcaptionbox*{$R$}{\includegraphics[width = 0.103\columnwidth]{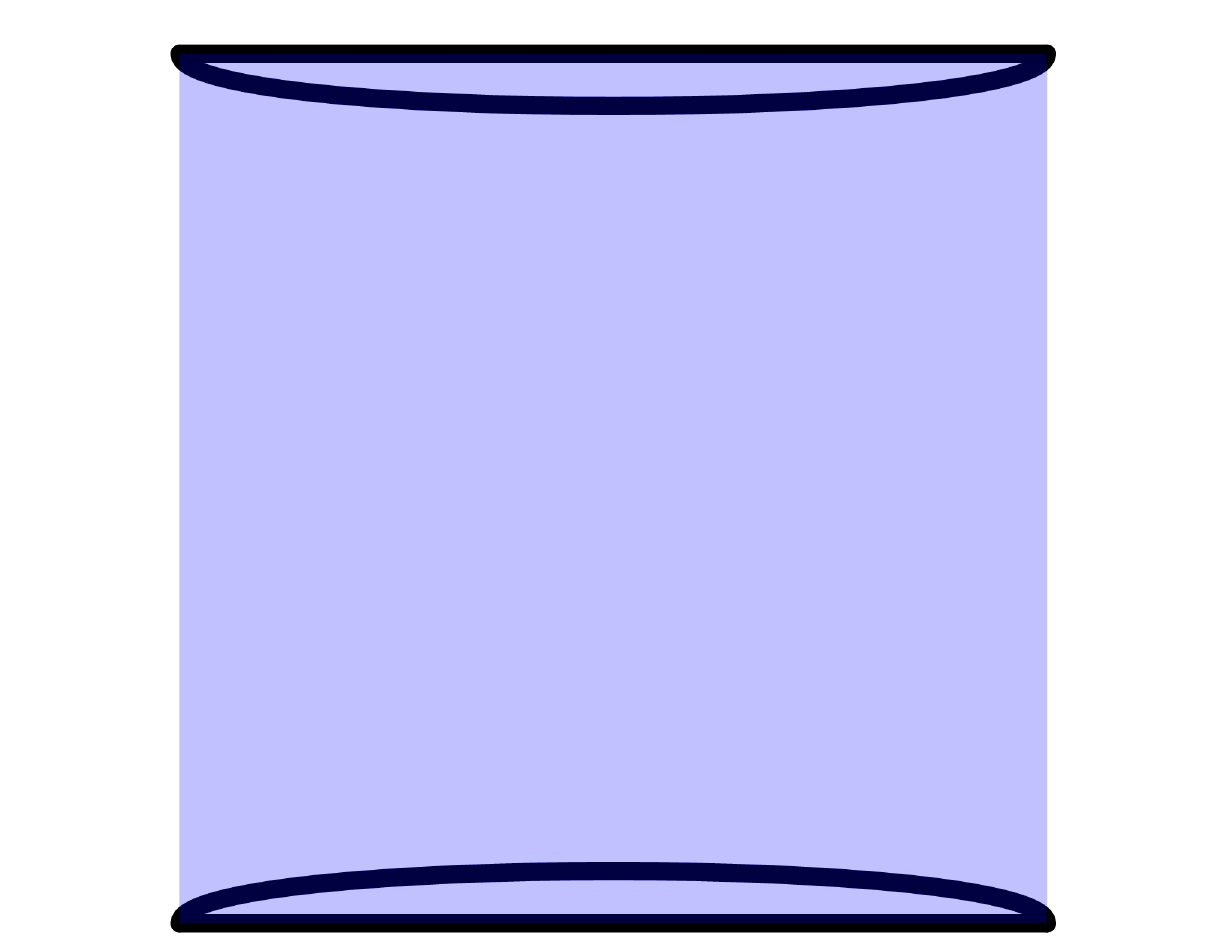}}
    \subcaptionbox*{$RH$}{\includegraphics[width = 0.103\columnwidth]{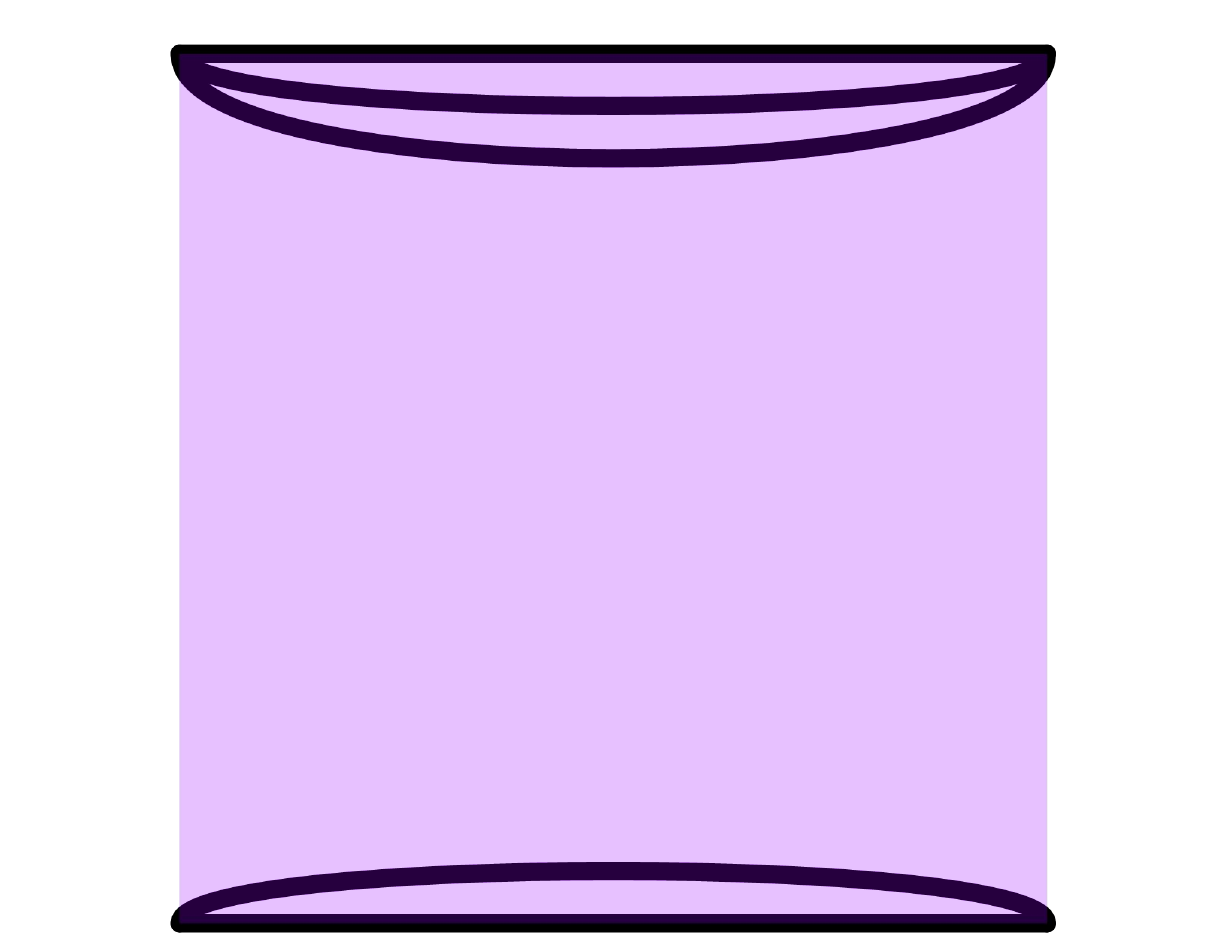}}
    \subcaptionbox*{$R^2$}{\includegraphics[width = 0.103\columnwidth]{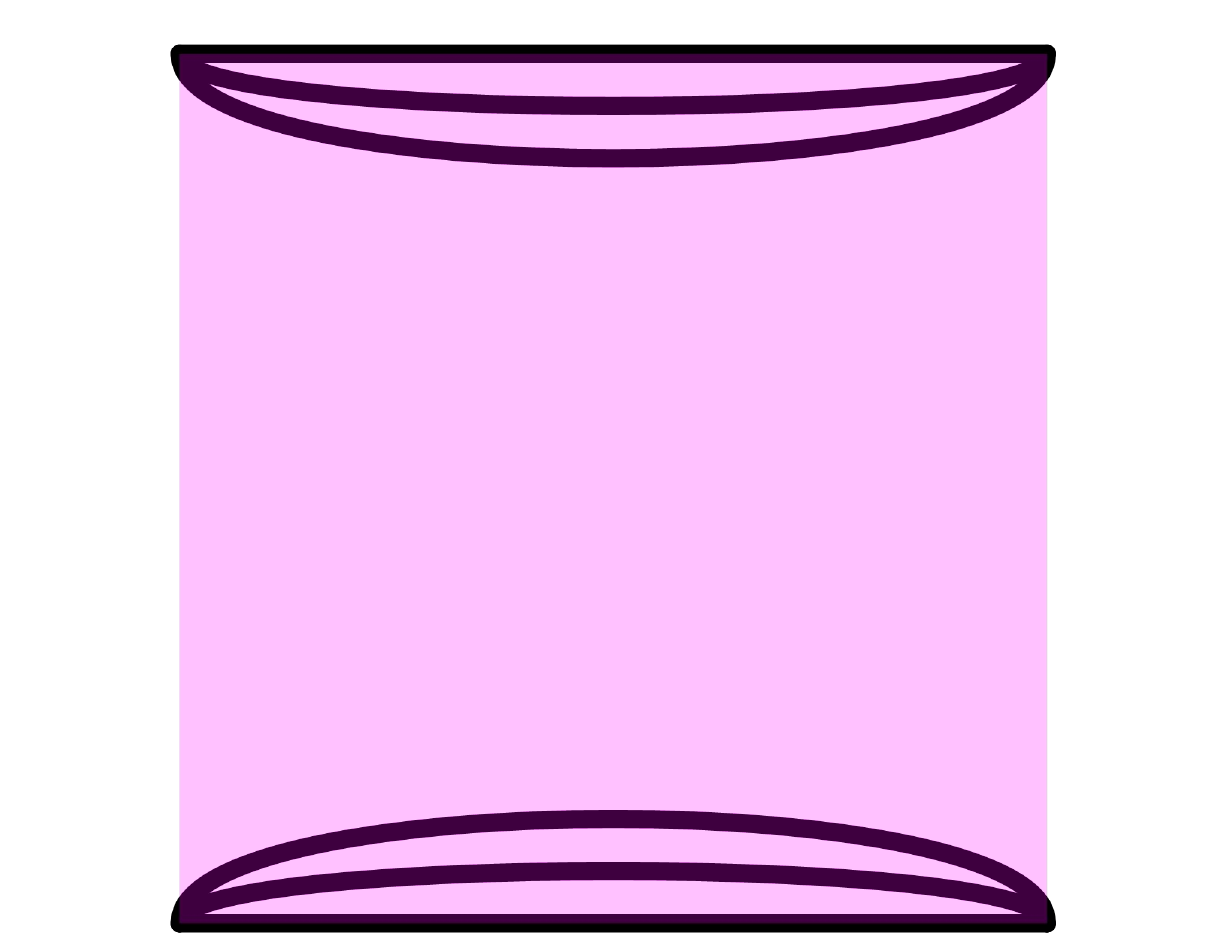}}\\
    \subcaptionbox*{$C$}{\includegraphics[width = 0.103\columnwidth]{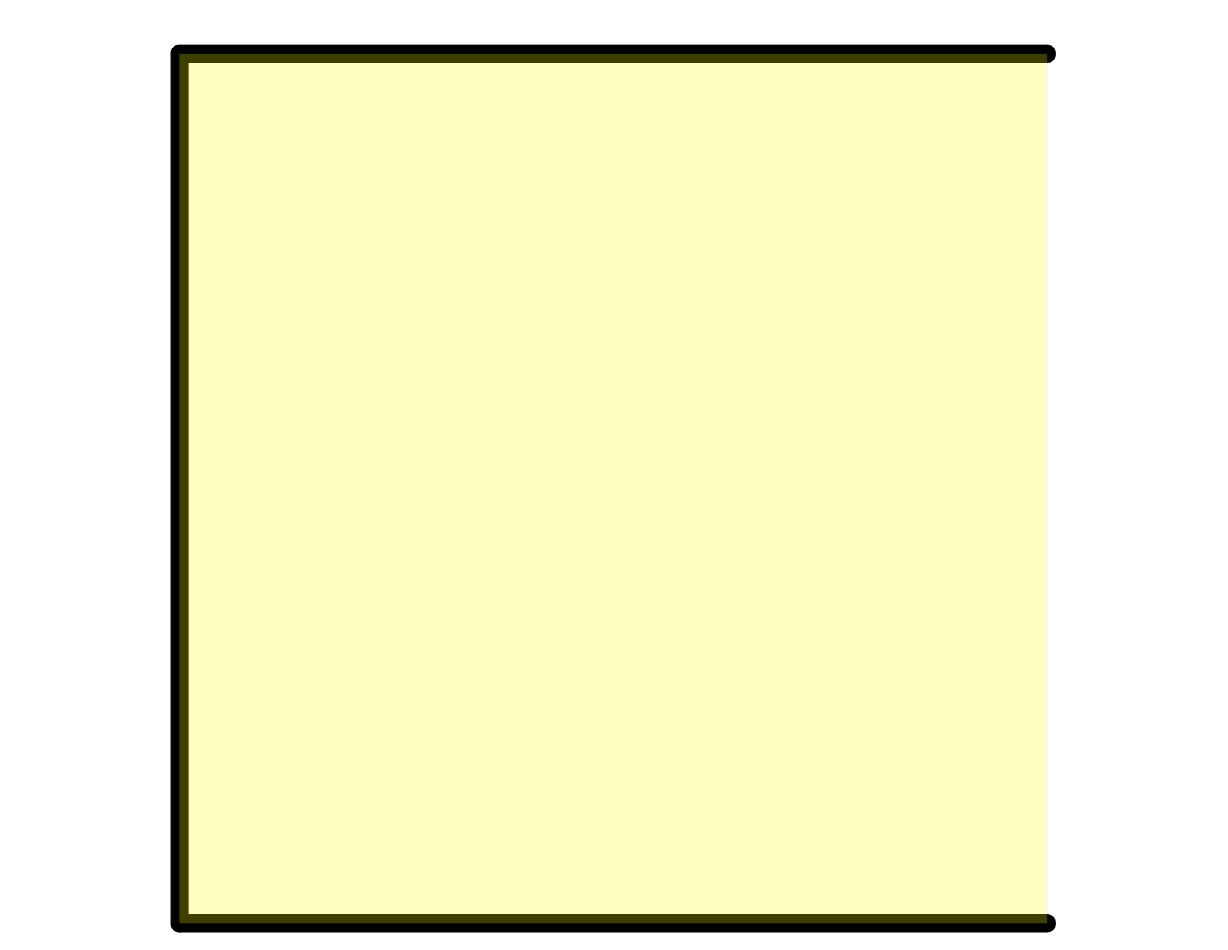}}
    \subcaptionbox*{$Cl$}{\includegraphics[width = 0.103\columnwidth]{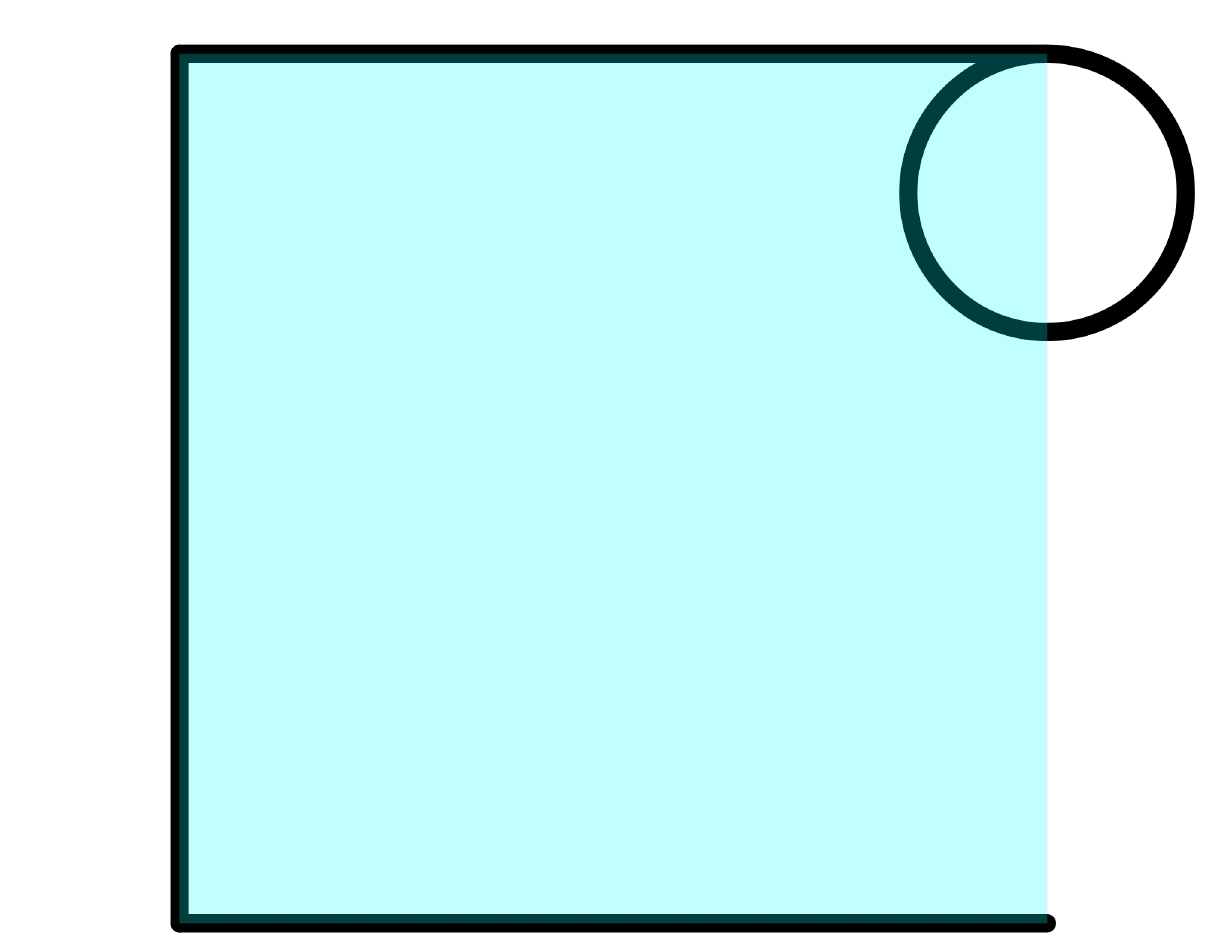}}
    \subcaptionbox*{$Cl^2$}{\includegraphics[width = 0.103\columnwidth]{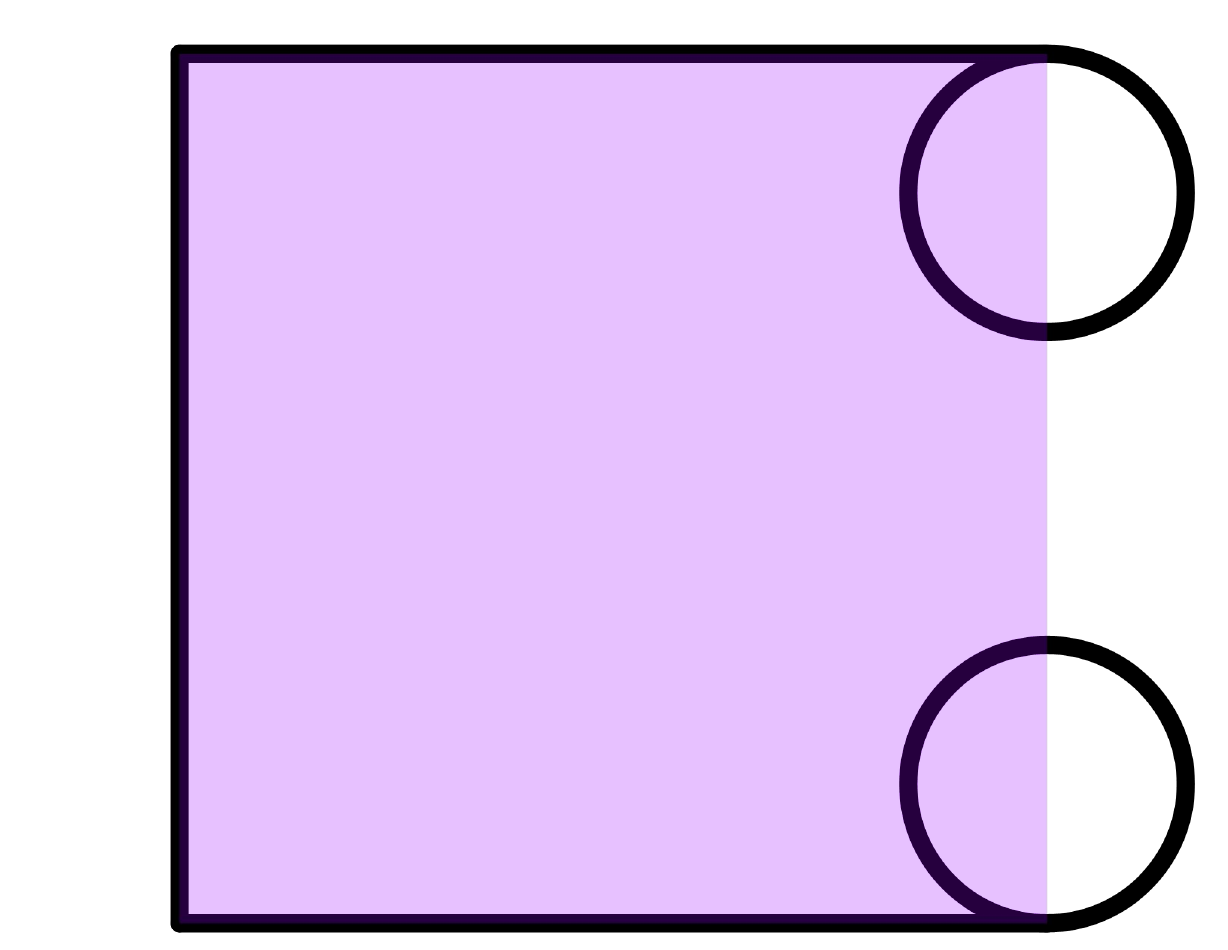}}
    \subcaptionbox*{$CH$}{\includegraphics[width = 0.103\columnwidth]{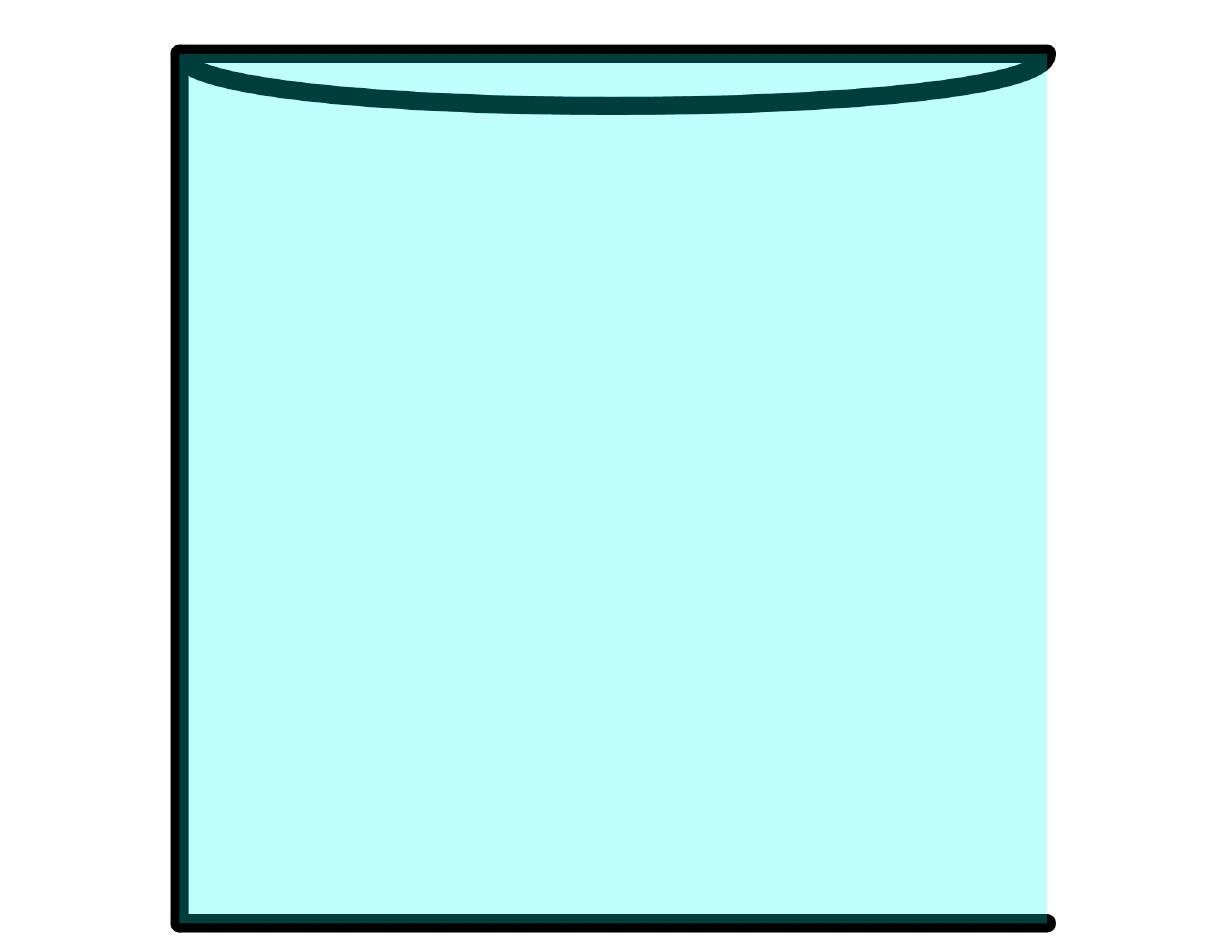}}
    \subcaptionbox*{$CHl$}{\includegraphics[width = 0.103\columnwidth]{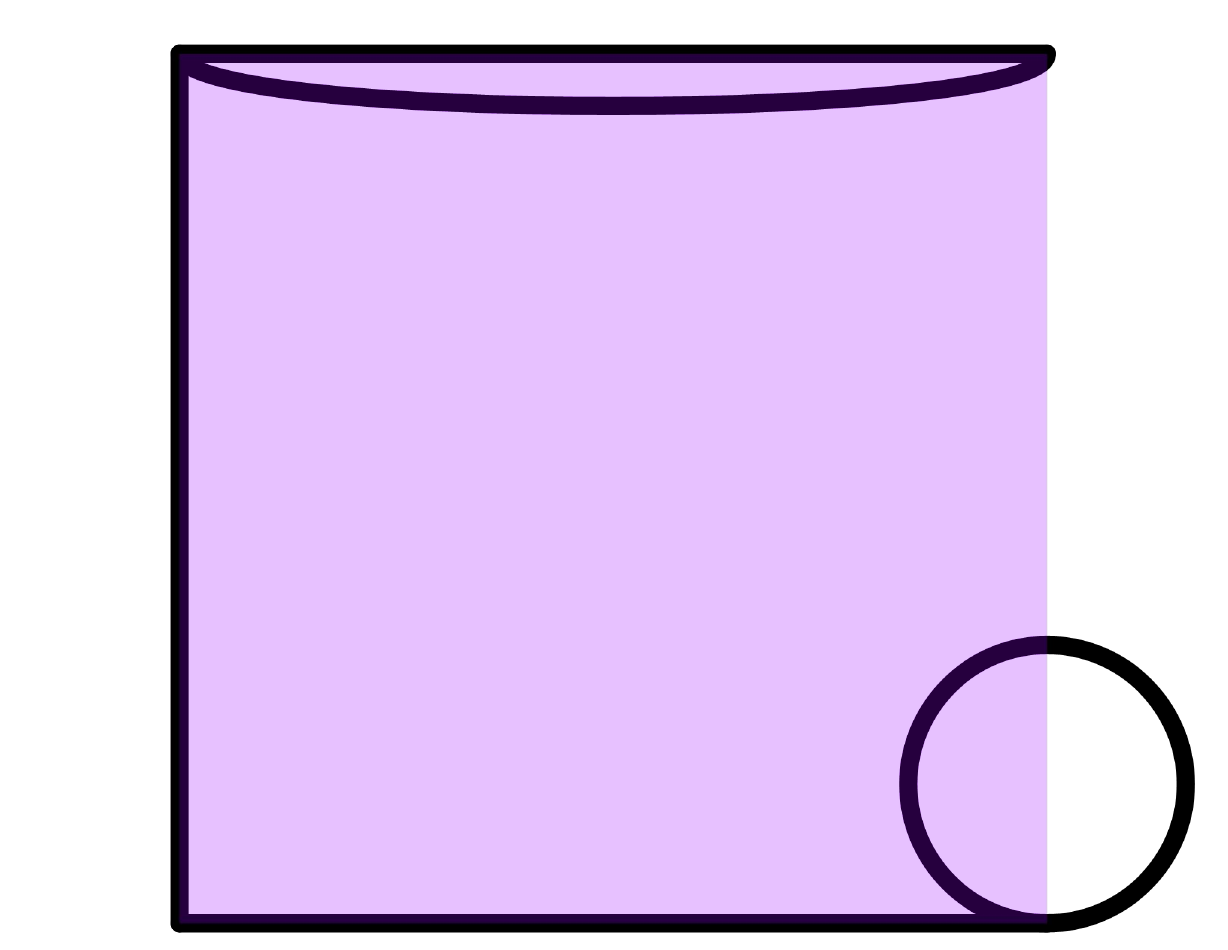}}
    \subcaptionbox*{$CR$}{\includegraphics[width = 0.103\columnwidth]{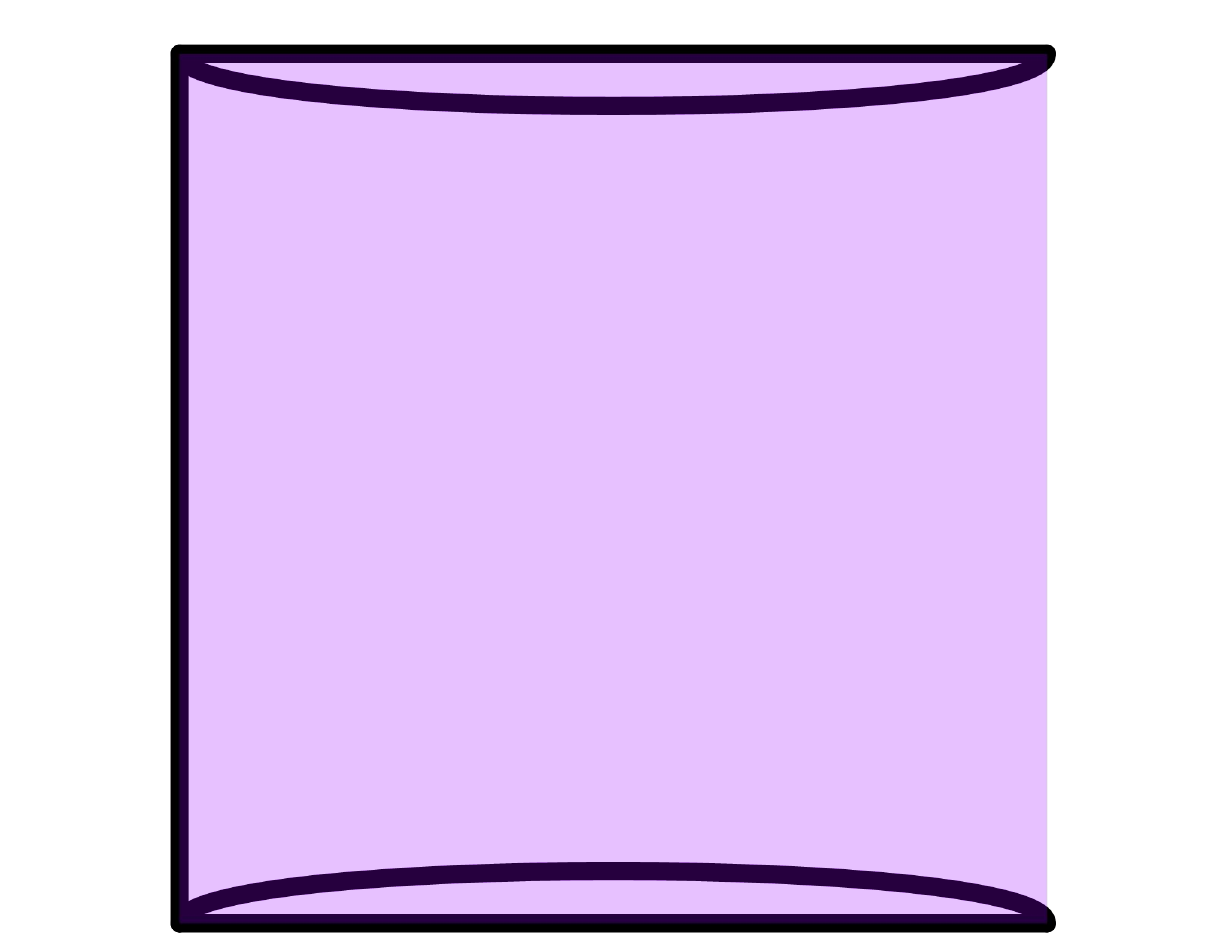}}
    \subcaptionbox*{$C^2$}{\includegraphics[width = 0.103\columnwidth]{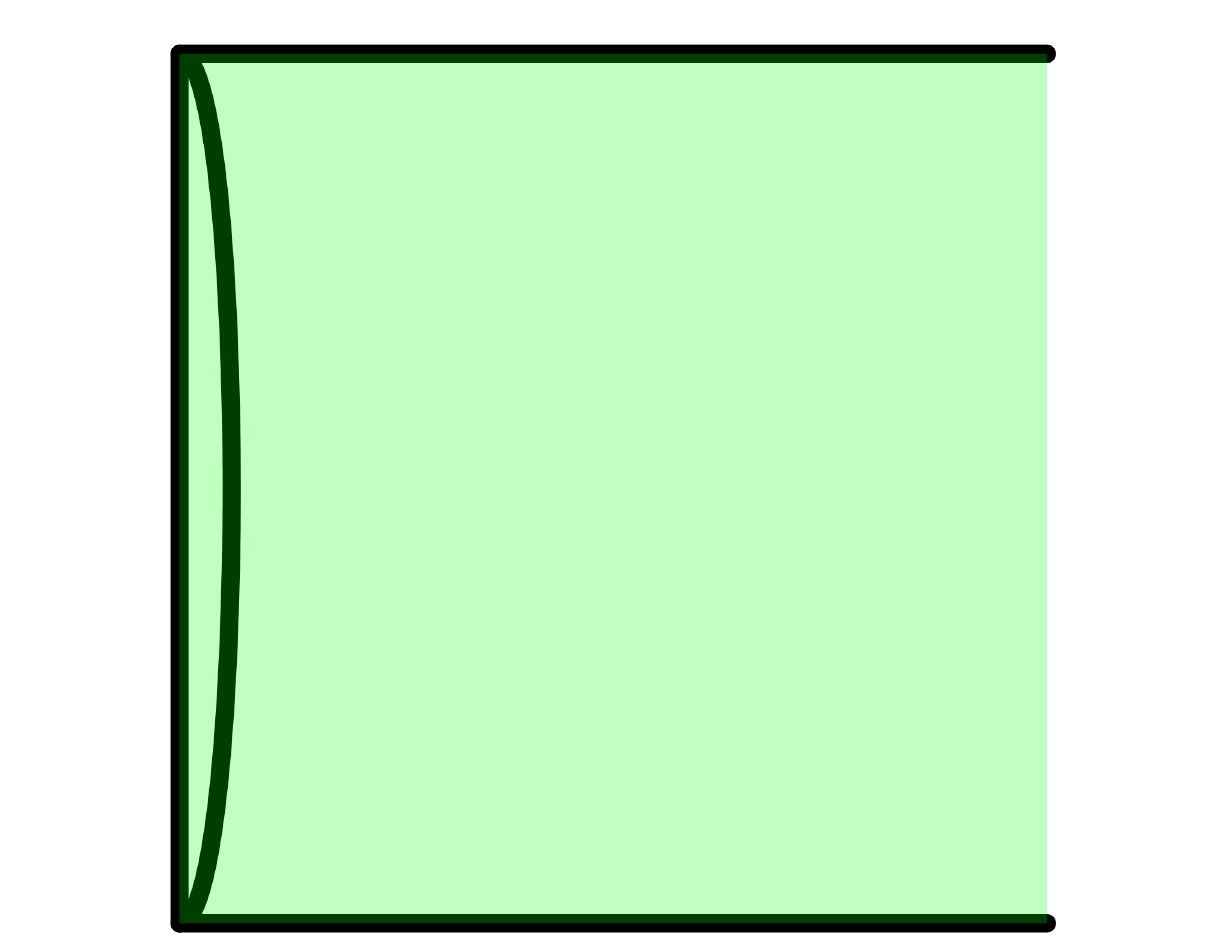}}
    \subcaptionbox*{$C^2 l$}{\includegraphics[width = 0.103\columnwidth]{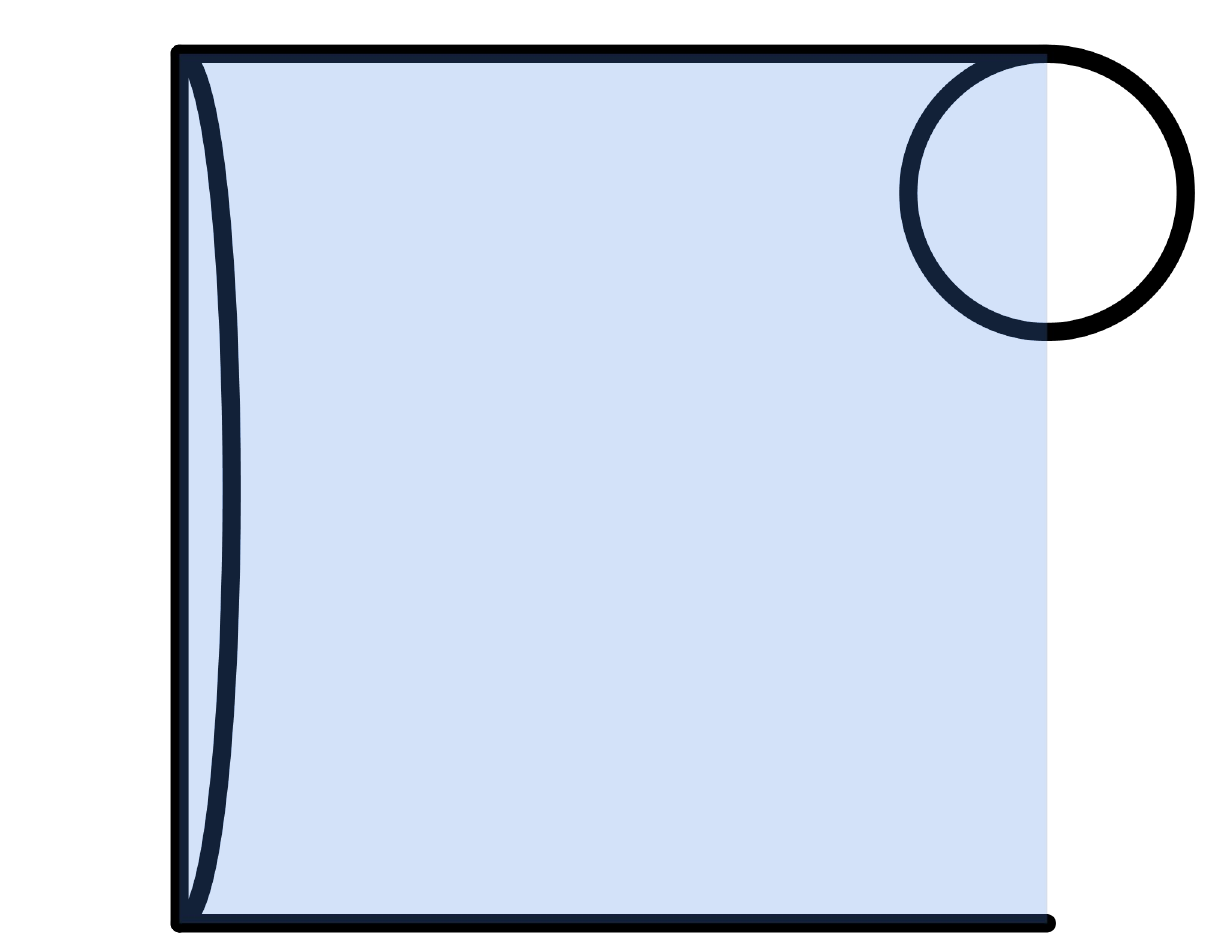}}
    \subcaptionbox*{$C^2 l^2$}{\includegraphics[width = 0.103\columnwidth]{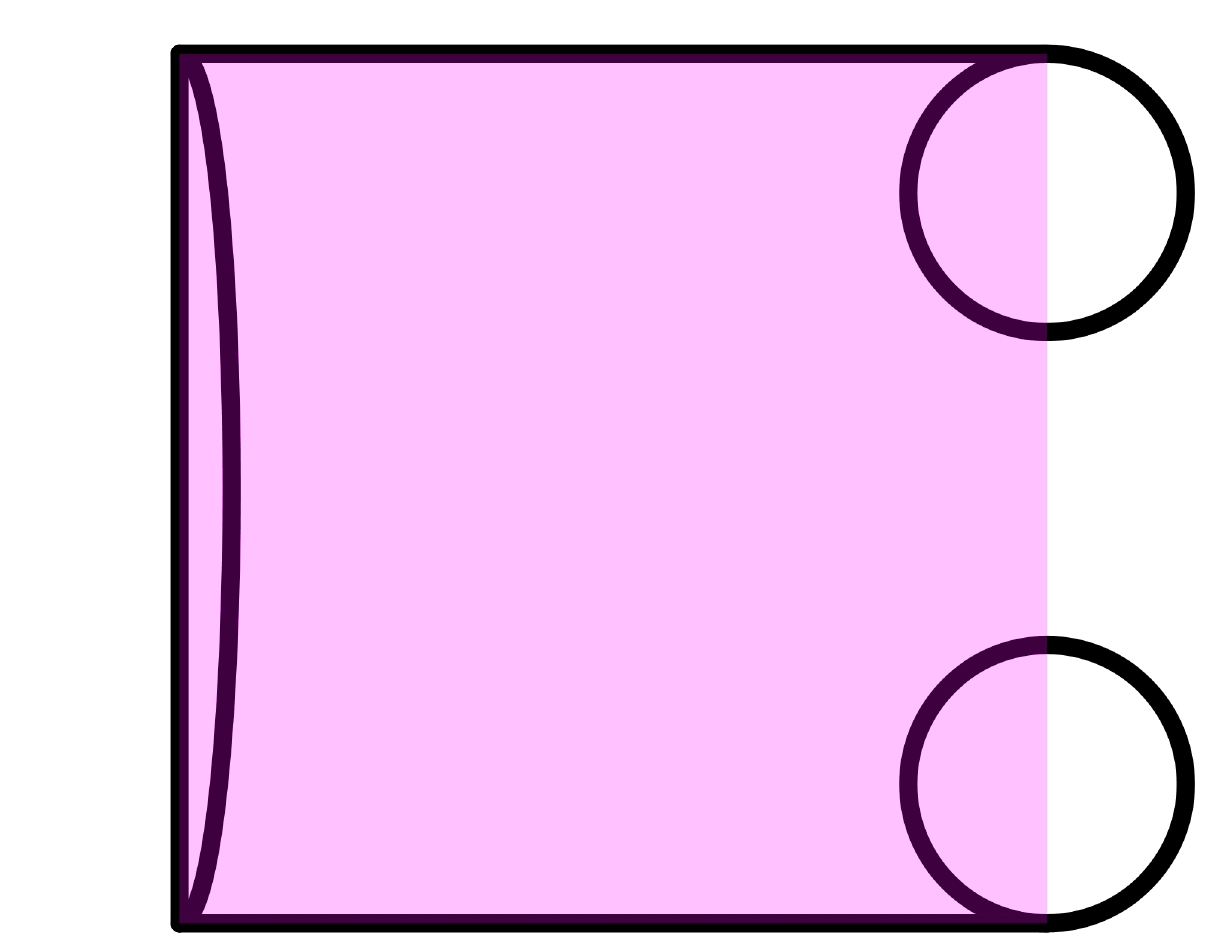}}\\
    \subcaptionbox*{$Z$}{\includegraphics[width = 0.103\columnwidth]{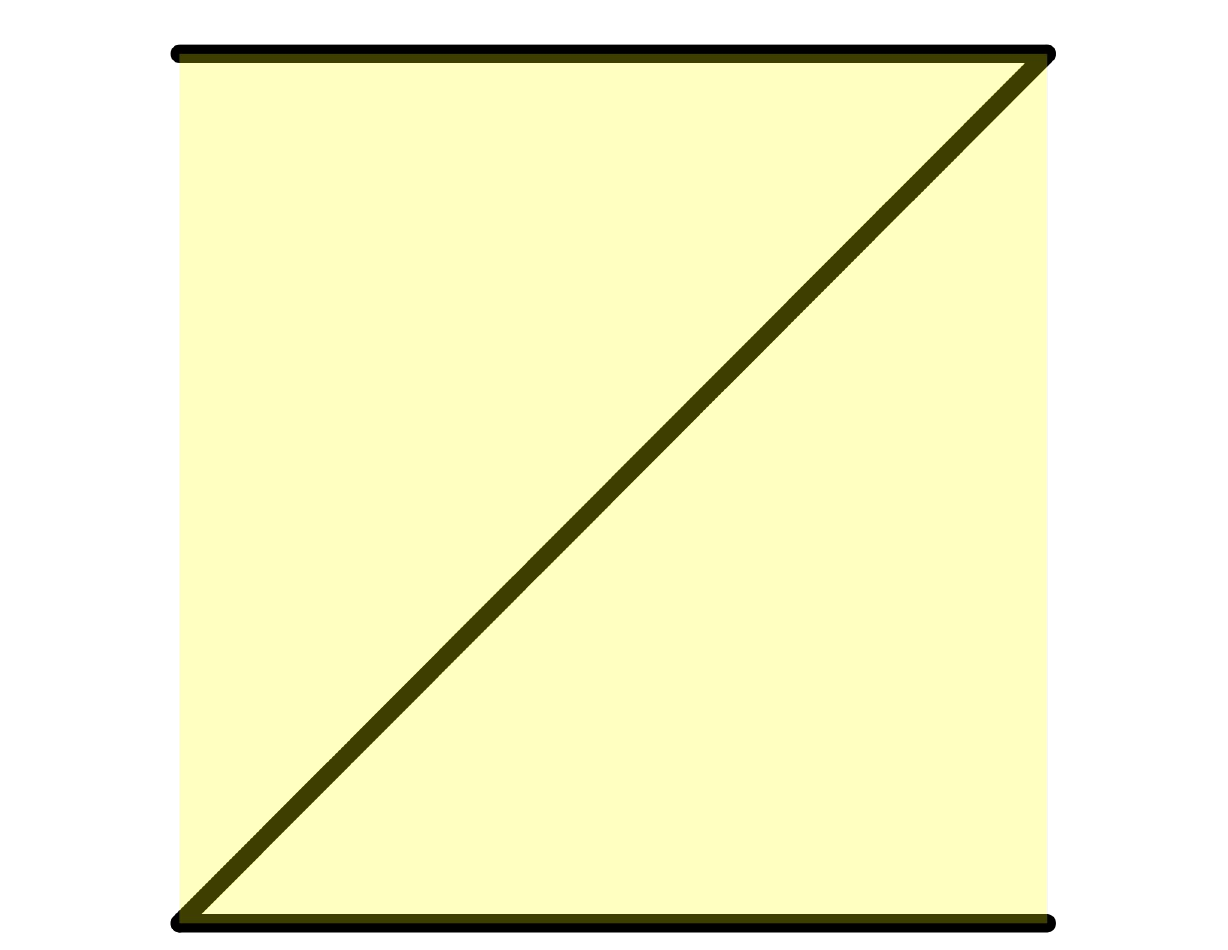}}
    \subcaptionbox*{$Zl$}{\includegraphics[width = 0.103\columnwidth]{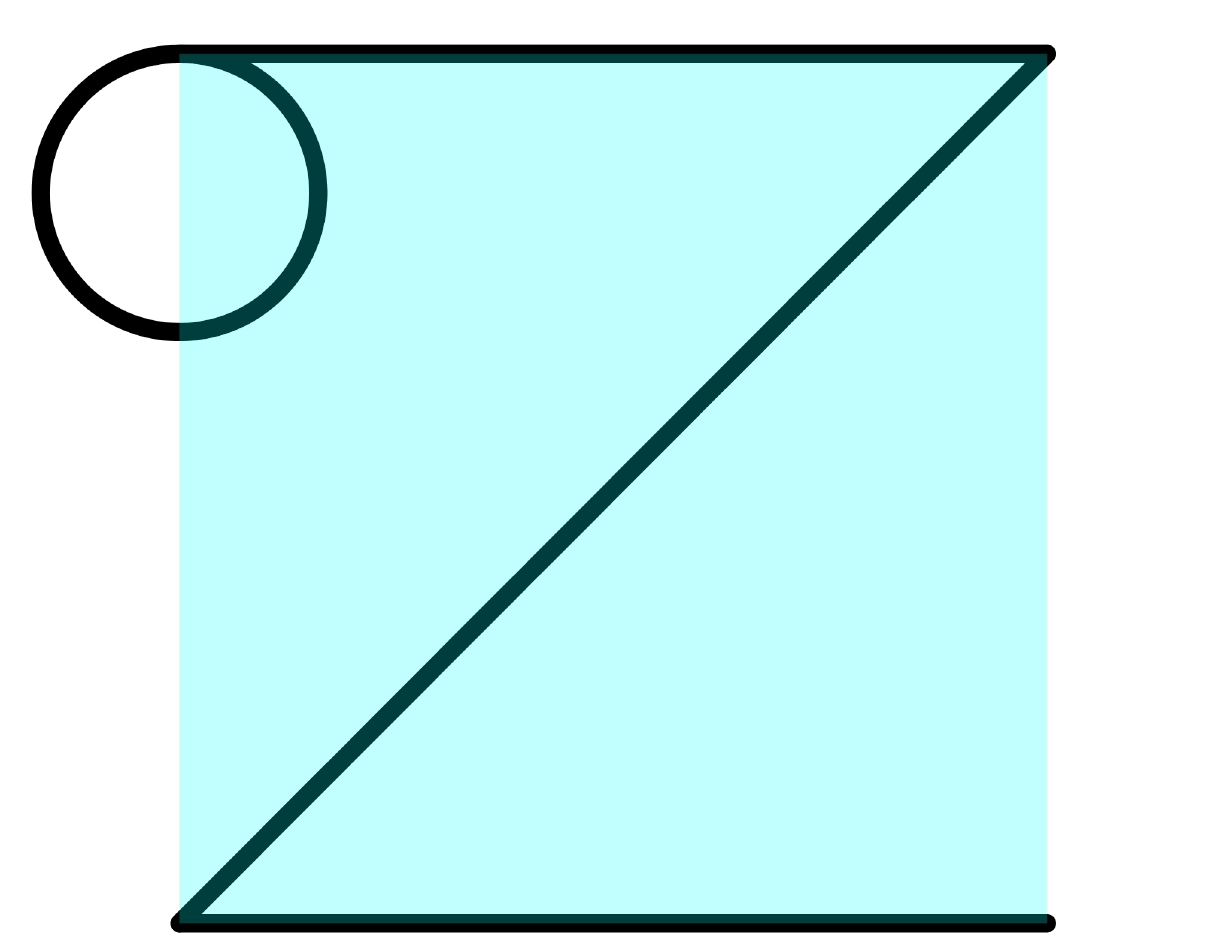}}
    \subcaptionbox*{$Zl^2$}{\includegraphics[width = 0.103\columnwidth]{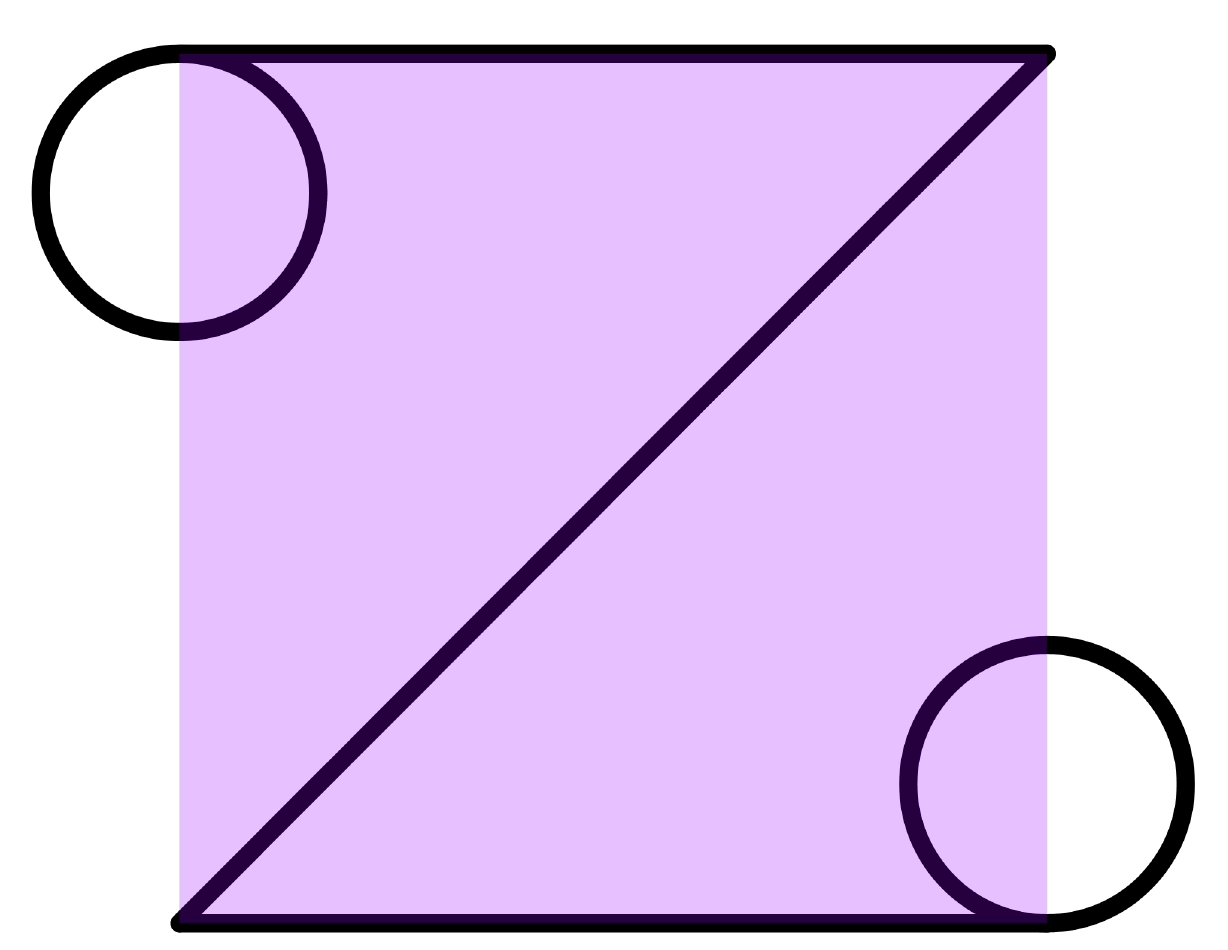}}
    \subcaptionbox*{$ZH$}{\includegraphics[width = 0.103\columnwidth]{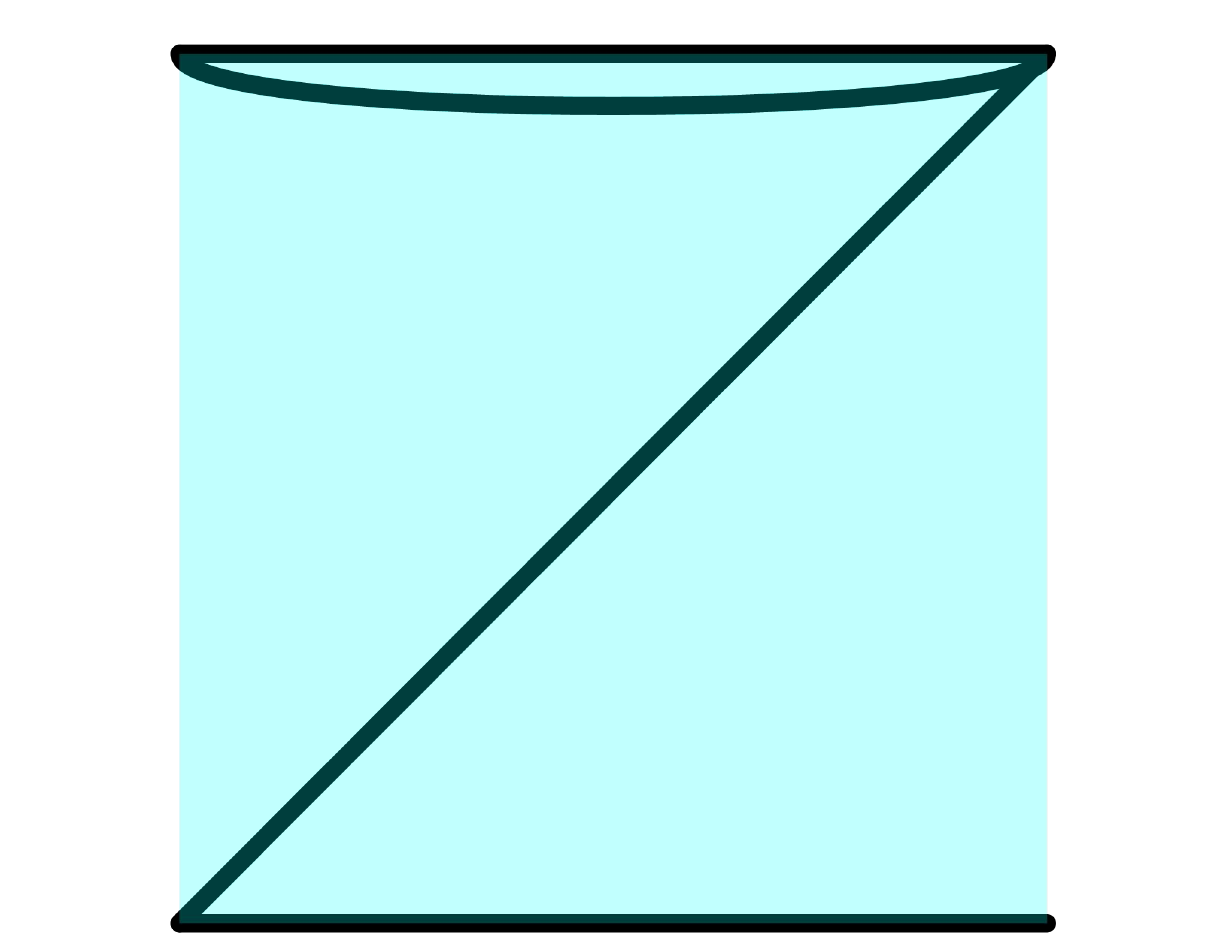}}
    \subcaptionbox*{$ZHl$}{\includegraphics[width = 0.103\columnwidth]{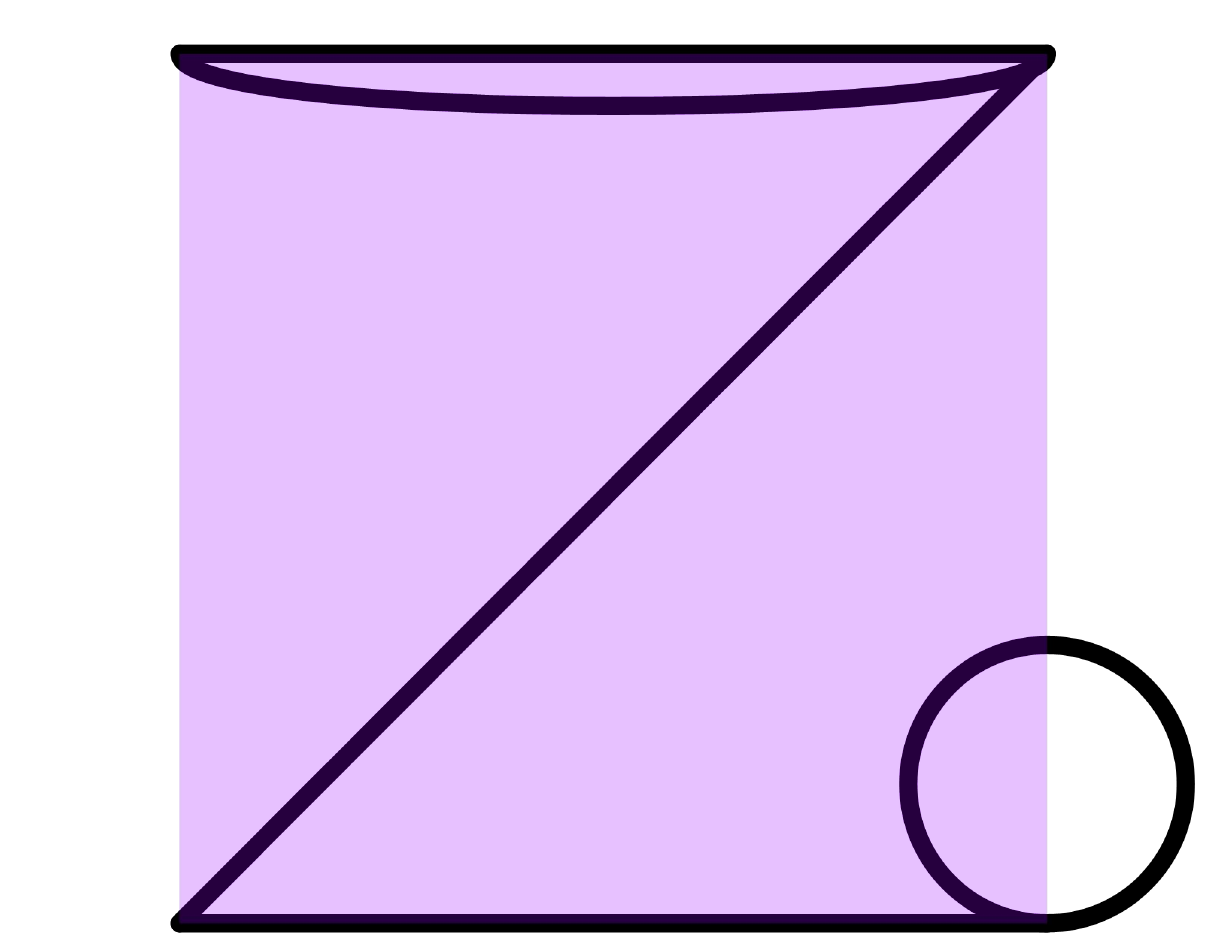}}
    \subcaptionbox*{$ZR$}{\includegraphics[width = 0.103\columnwidth]{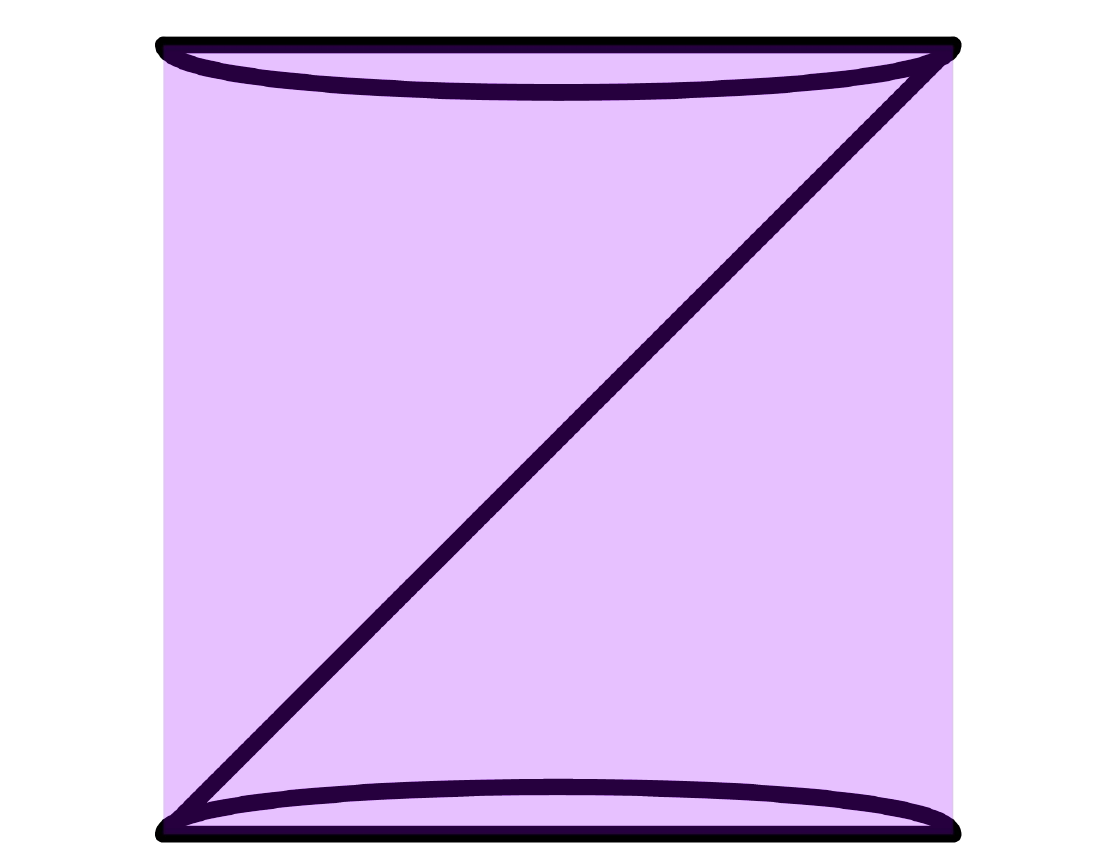}}
    \subcaptionbox*{$Z^2$}{\includegraphics[width = 0.103\columnwidth]{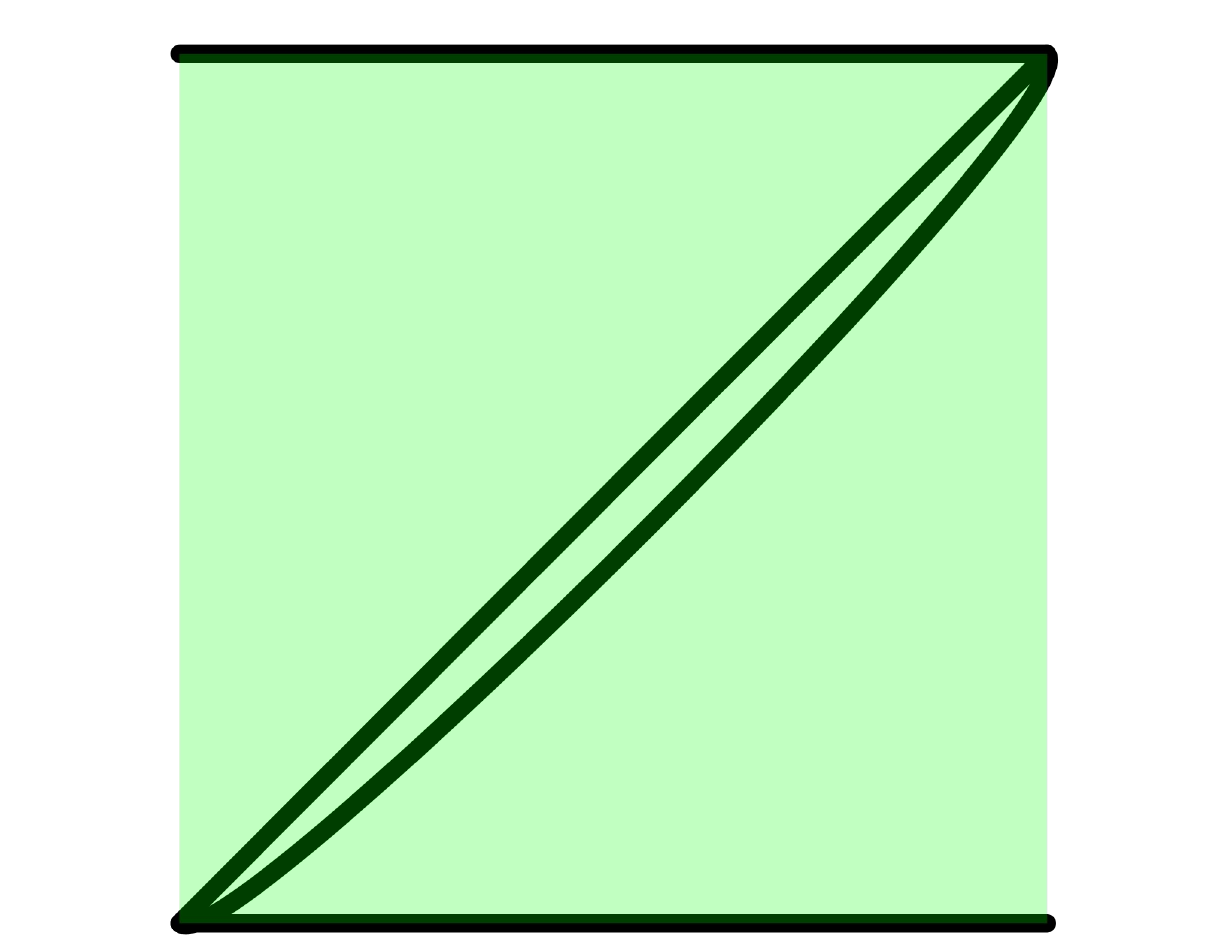}}
    \subcaptionbox*{$Z^2 l$}{\includegraphics[width = 0.103\columnwidth]{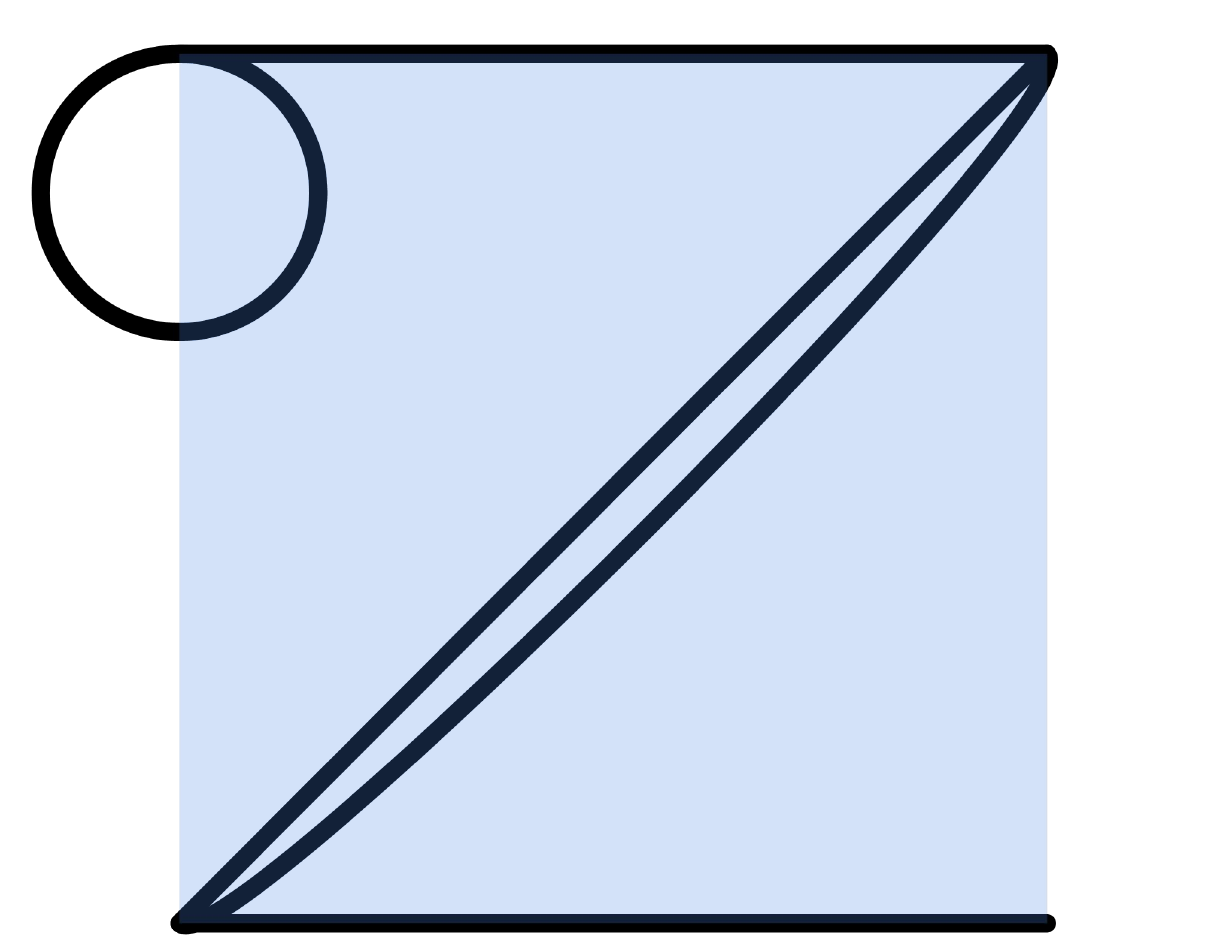}}
    \subcaptionbox*{$Z^2 l^2$}{\includegraphics[width = 0.103\columnwidth]{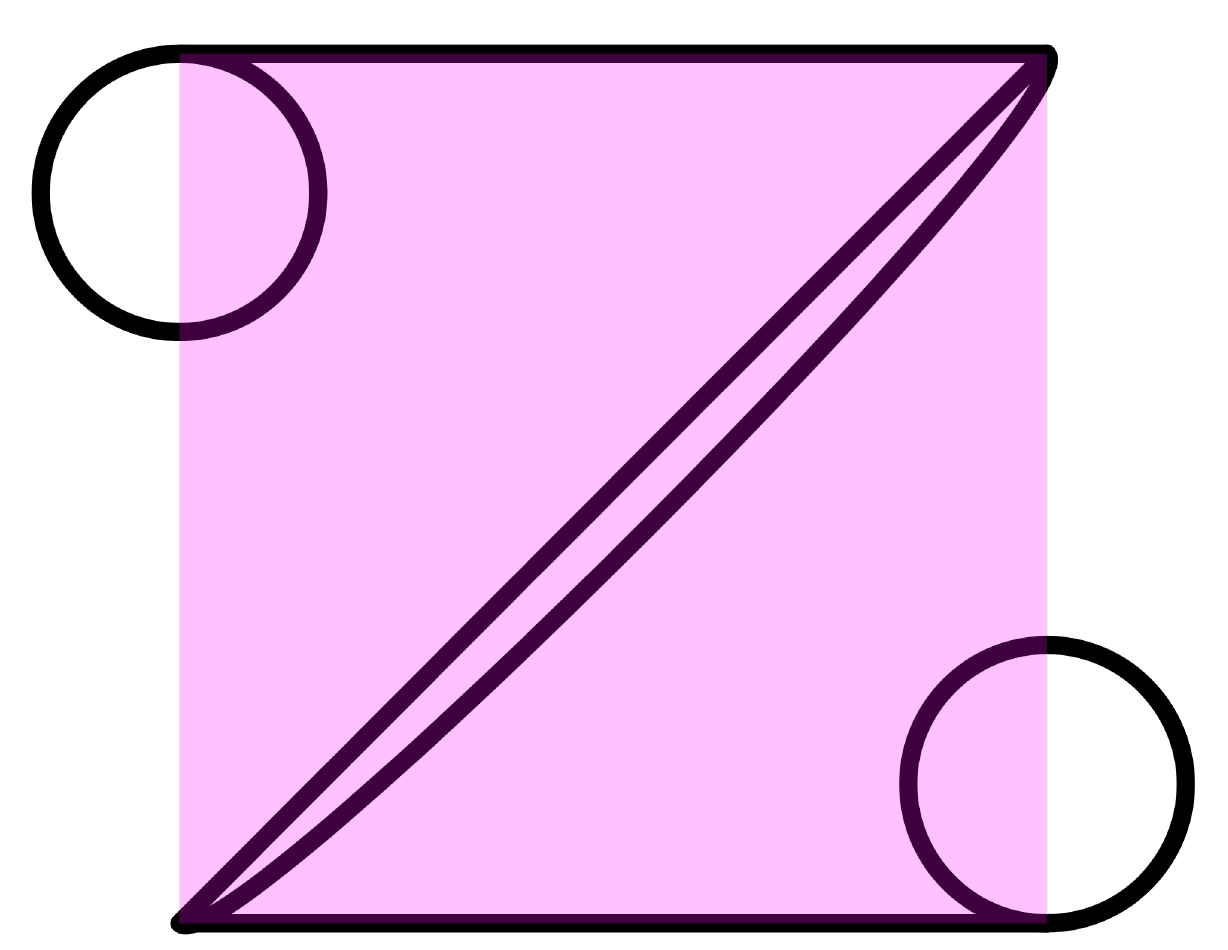}}\\
    \subcaptionbox*{$P$}{\includegraphics[width = 0.103\columnwidth]{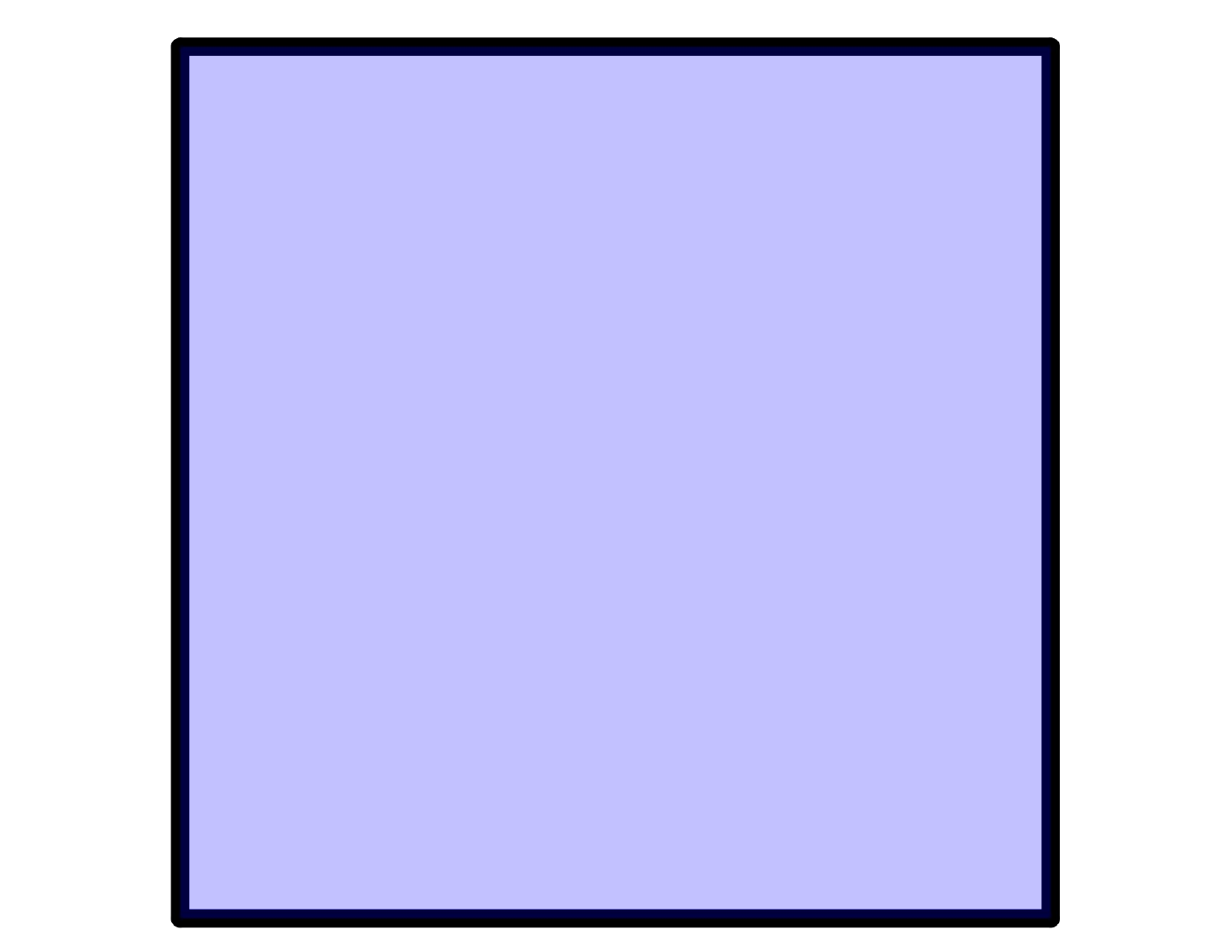}}
    \subcaptionbox*{$PH$}{\includegraphics[width = 0.103\columnwidth]{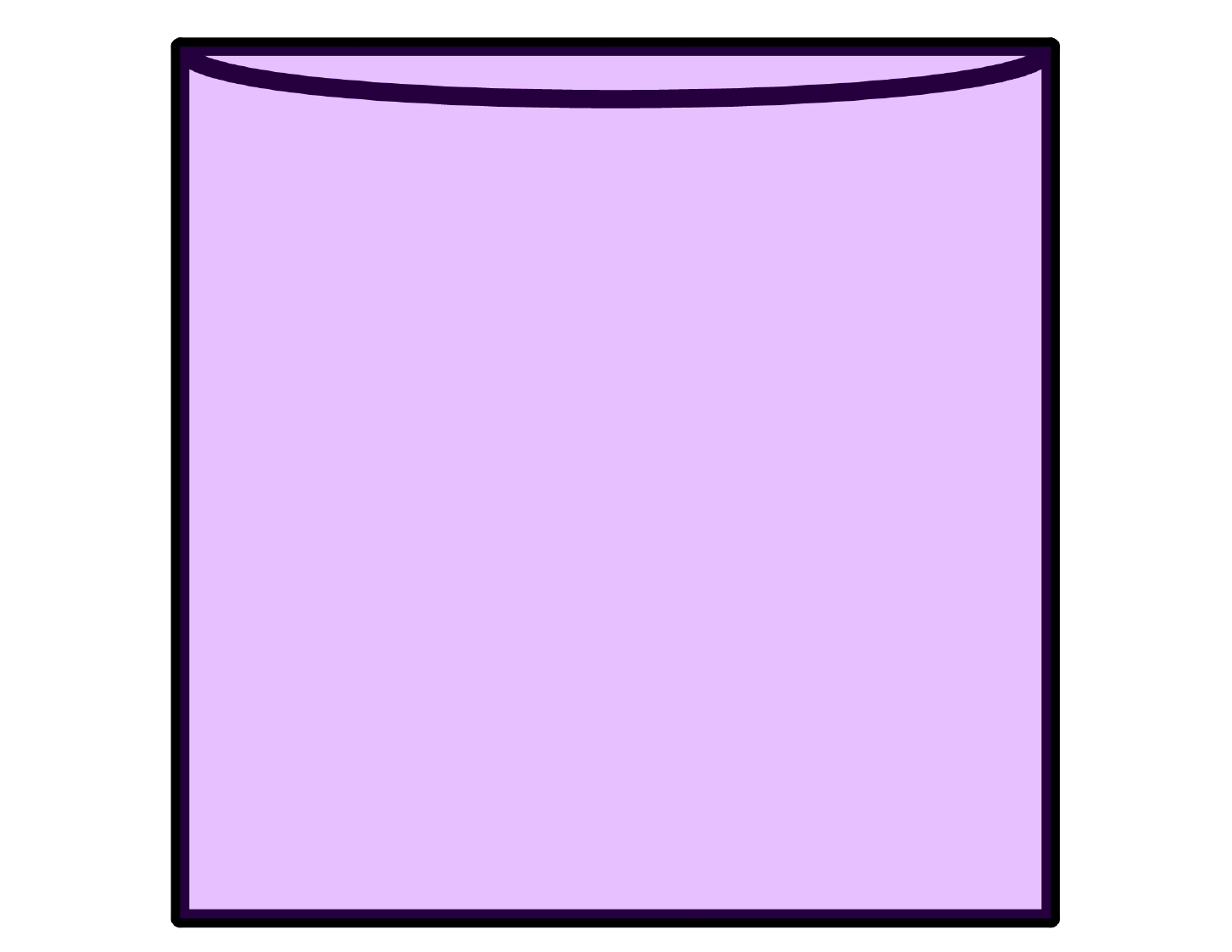}}
    \subcaptionbox*{$PR$}{\includegraphics[width = 0.103\columnwidth]{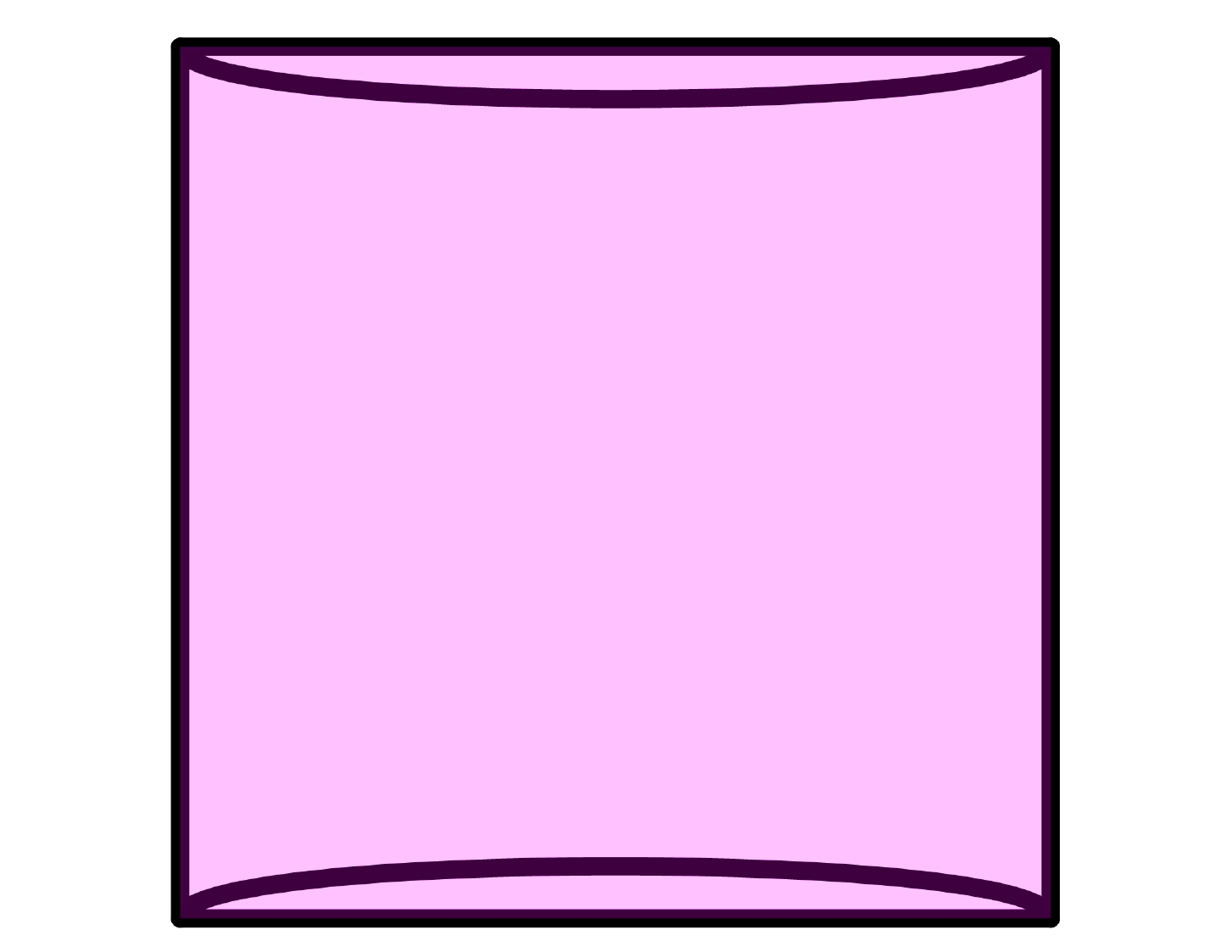}}
    \subcaptionbox*{$PC$}{\includegraphics[width = 0.103\columnwidth]{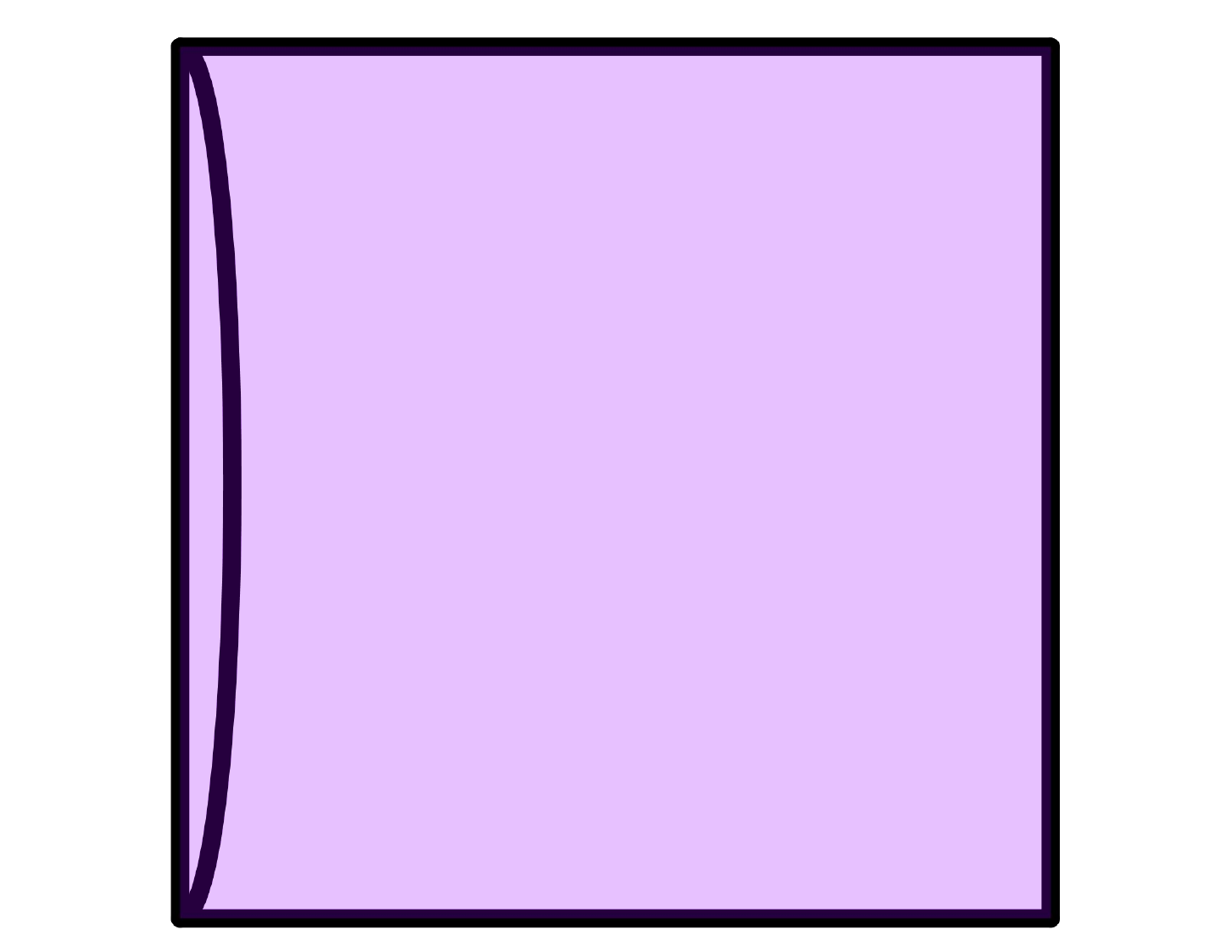}}
    \subcaptionbox*{$PP$}{\includegraphics[width = 0.103\columnwidth]{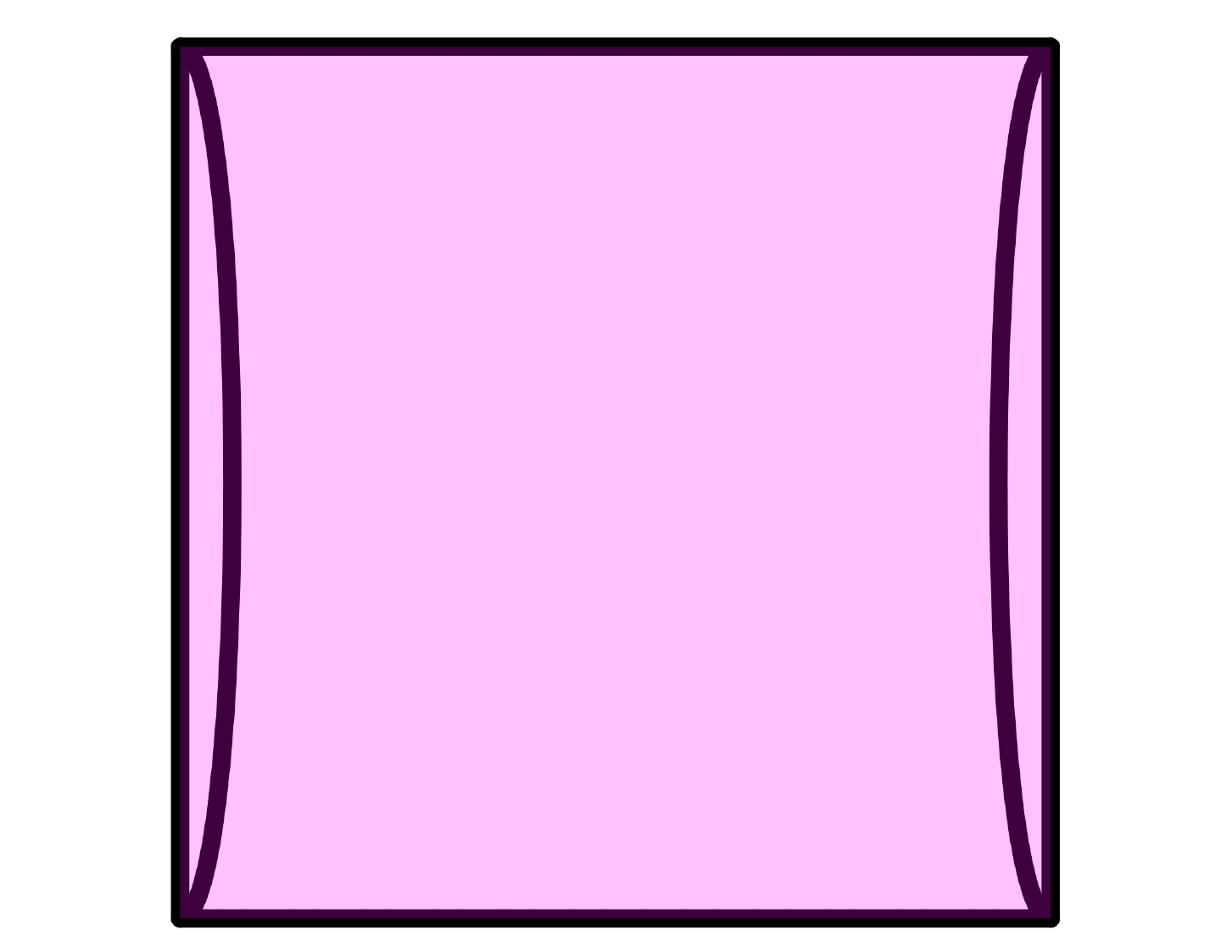}}\\
    \subcaptionbox*{$N$}{\includegraphics[width = 0.103\columnwidth]{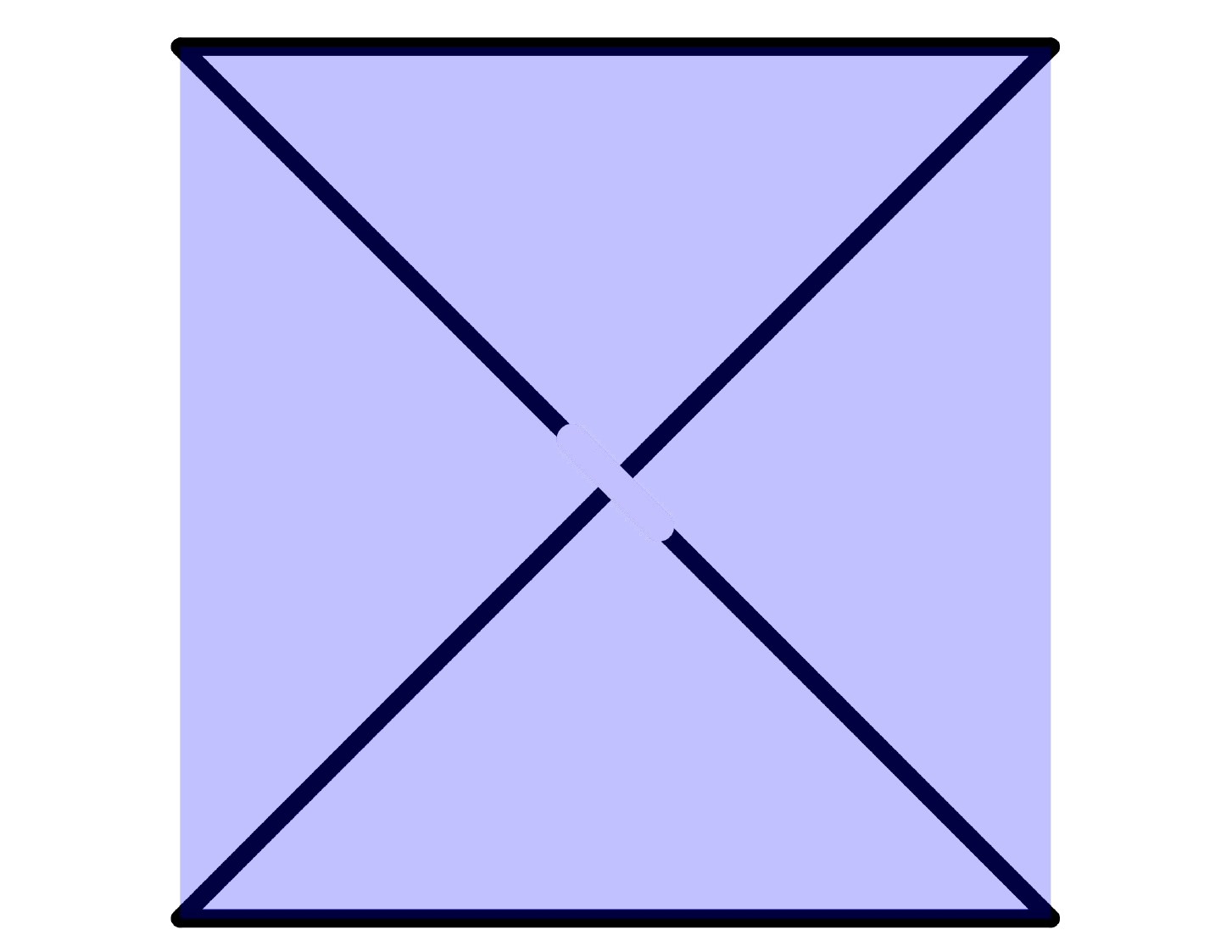}}
    \subcaptionbox*{$NH$}{\includegraphics[width = 0.103\columnwidth]{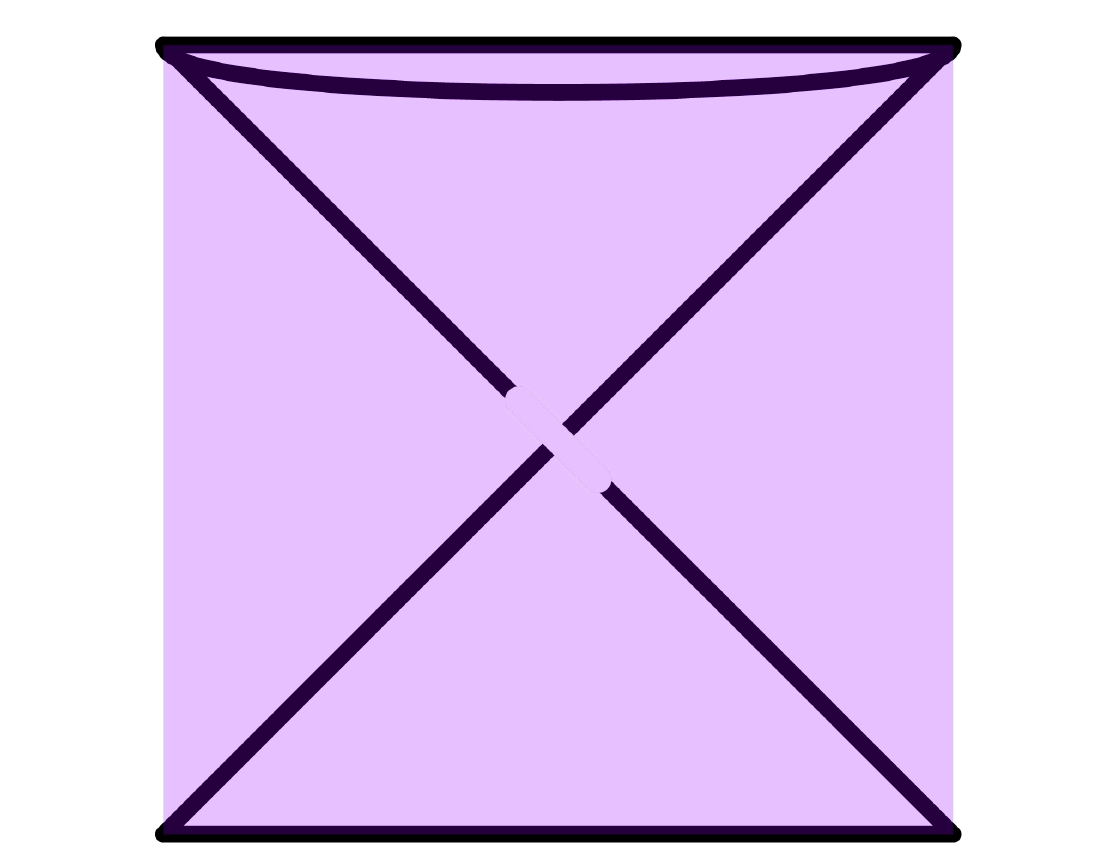}}
    \subcaptionbox*{$NR$}{\includegraphics[width = 0.103\columnwidth]{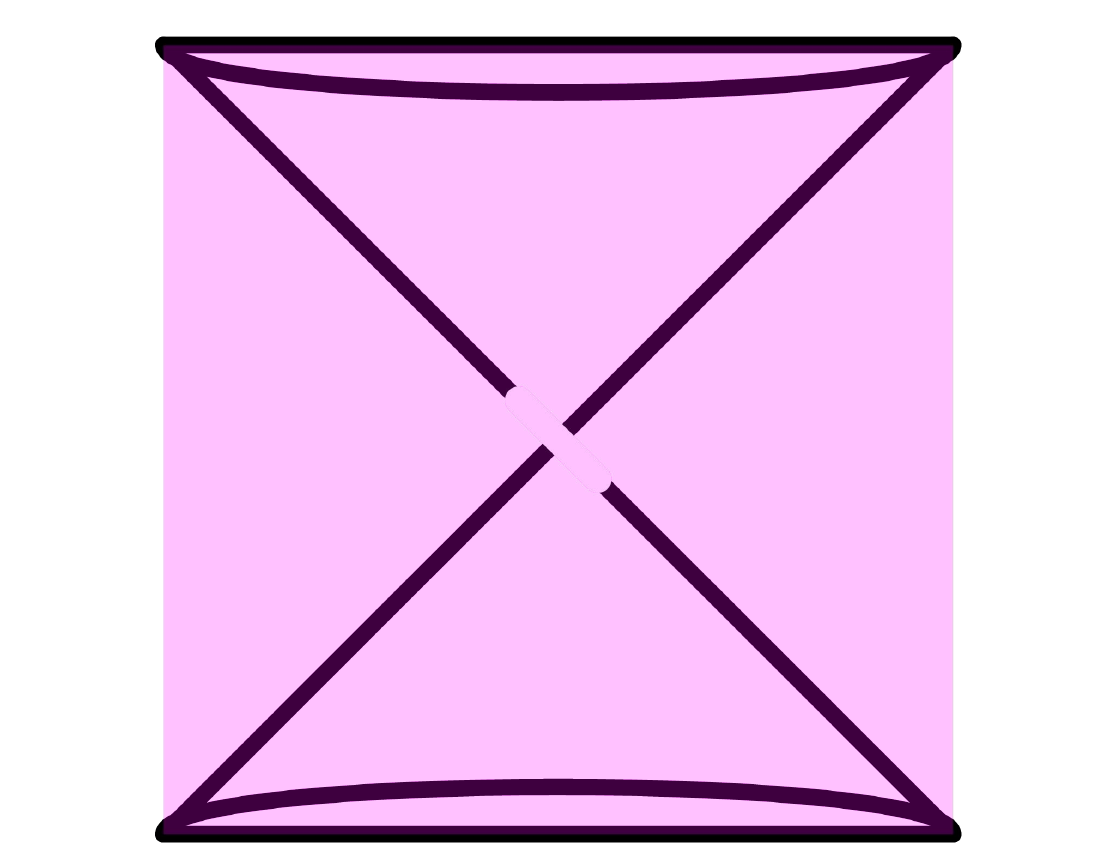}}\\
    \subcaptionbox*{$CZ$}{\includegraphics[width = 0.103\columnwidth]{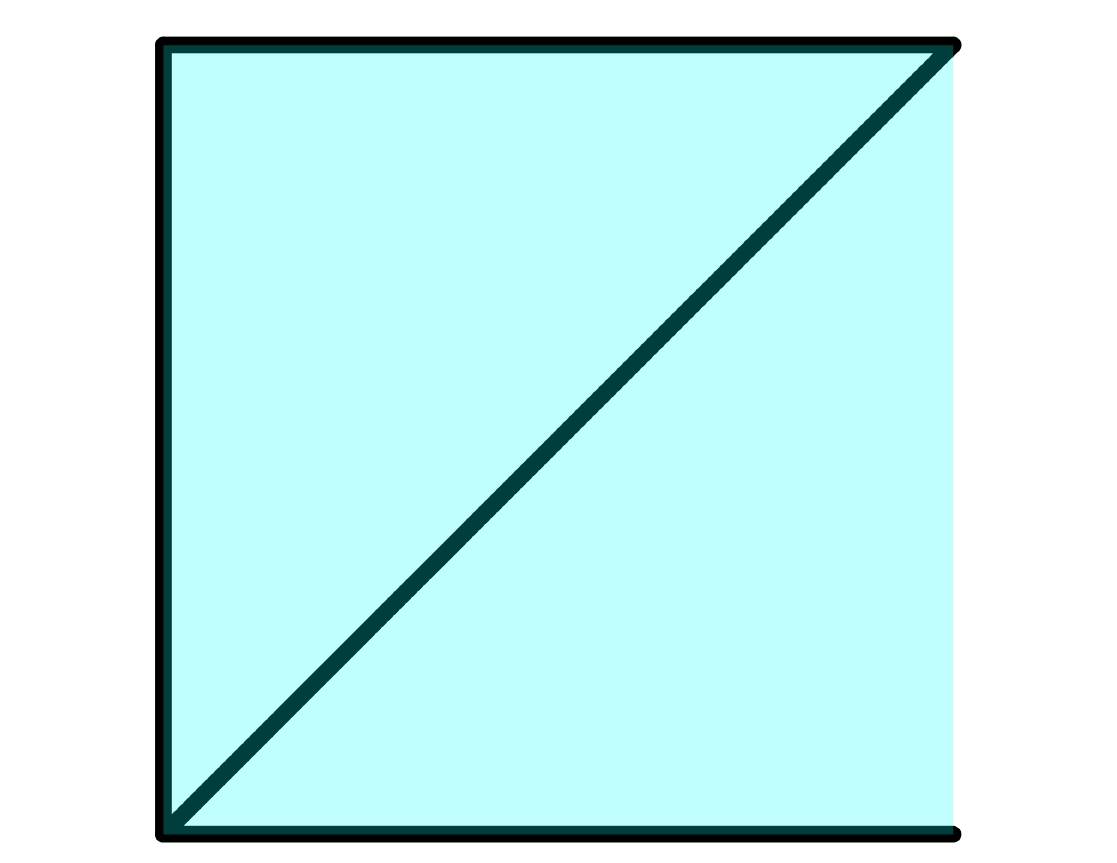}}
    \subcaptionbox*{$CZl$}{\includegraphics[width = 0.103\columnwidth]{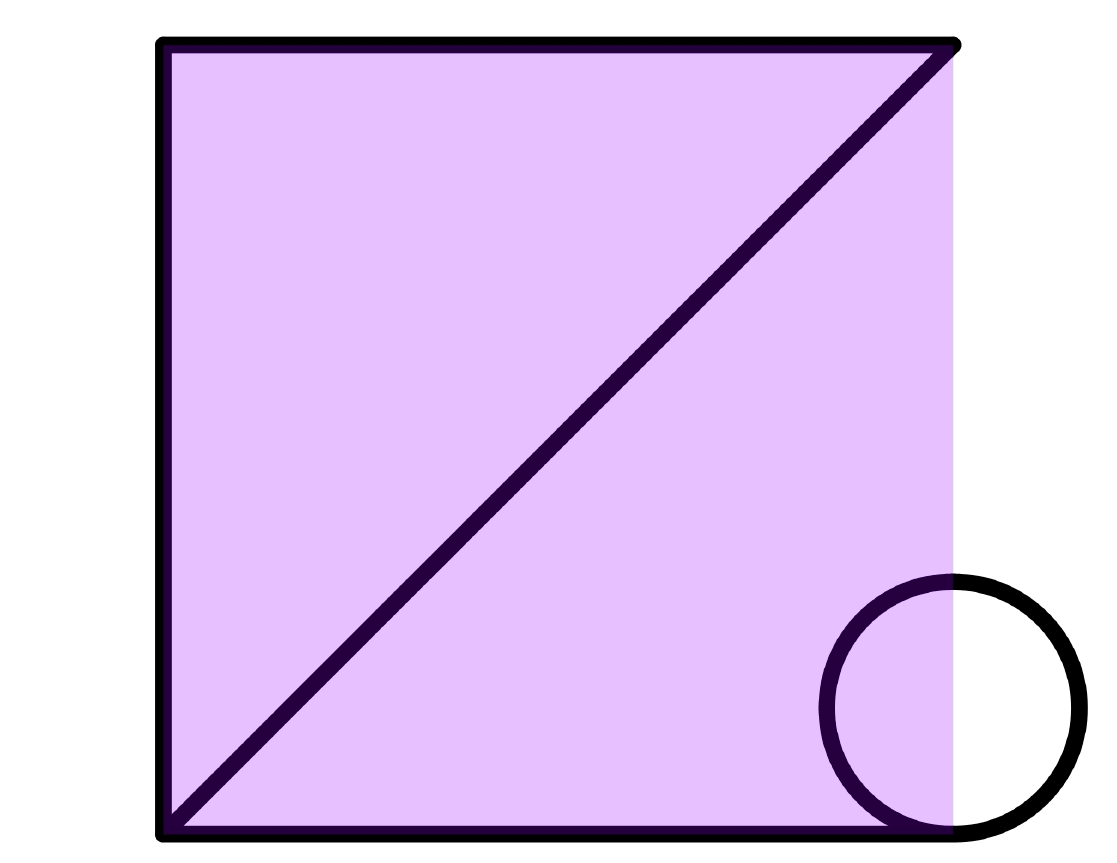}}
    \subcaptionbox*{$CZH$}{\includegraphics[width = 0.103\columnwidth]{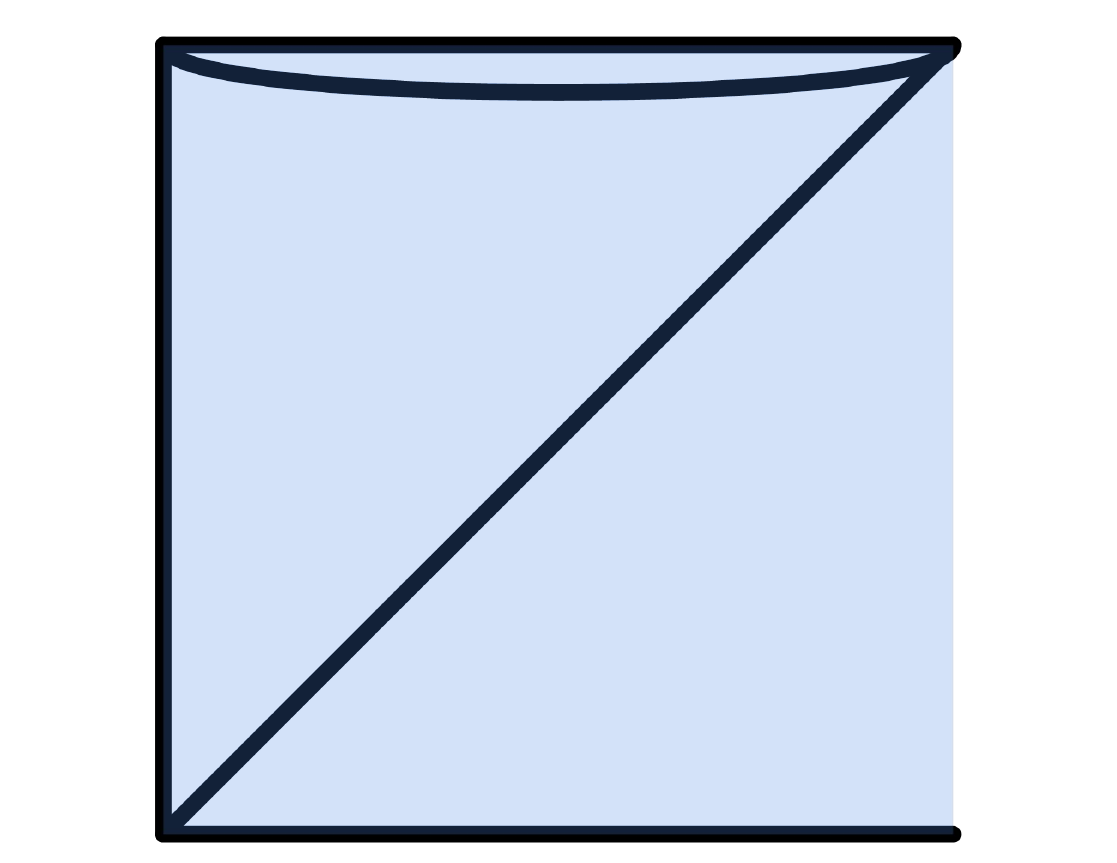}}
    \subcaptionbox*{$CZHl$}{\includegraphics[width = 0.103\columnwidth]{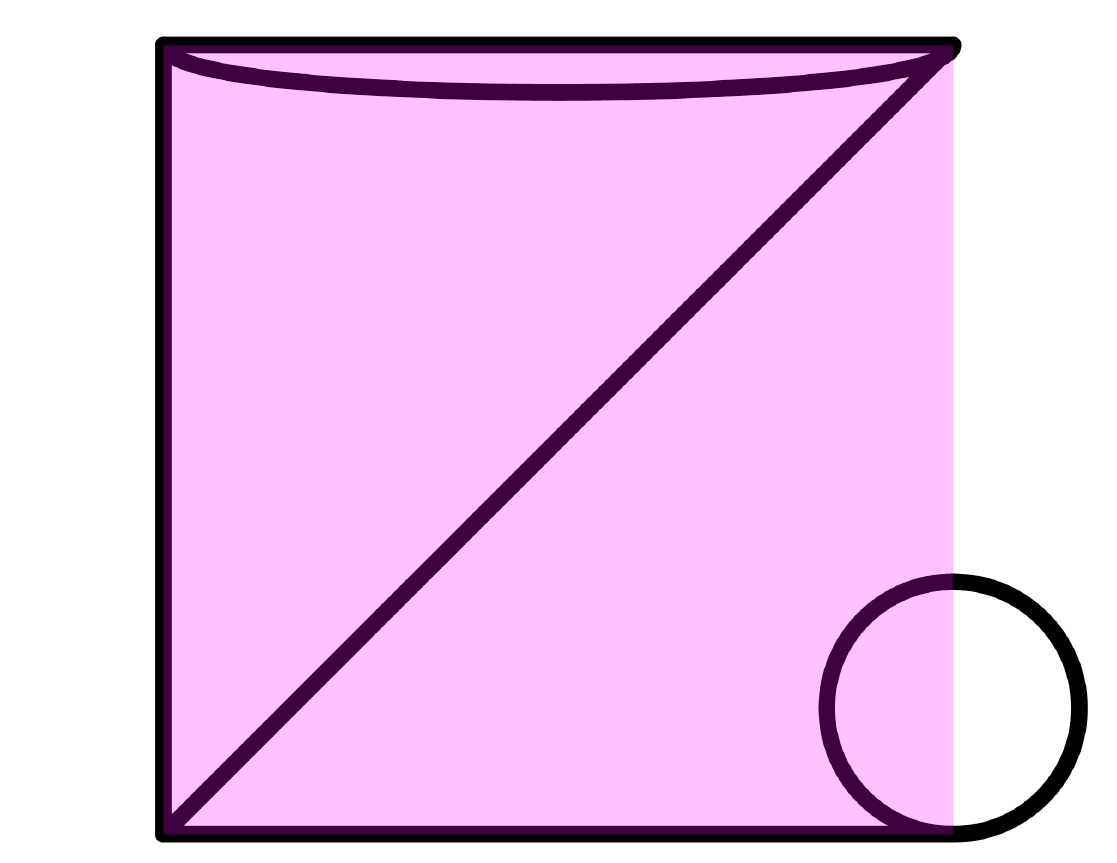}}\\
    \subcaptionbox*{$PZ$}{\includegraphics[width = 0.103\columnwidth]{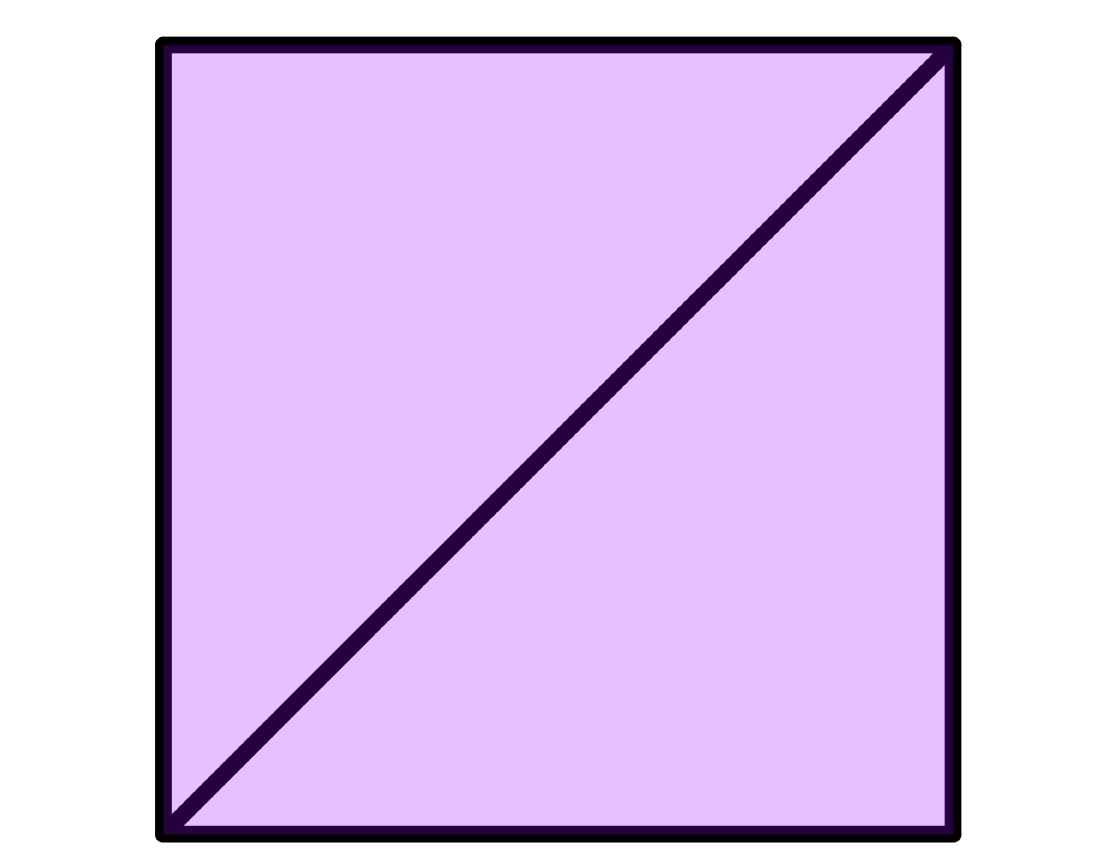}}\\
    \subcaptionbox*{$NC$}{\includegraphics[width = 0.103\columnwidth]{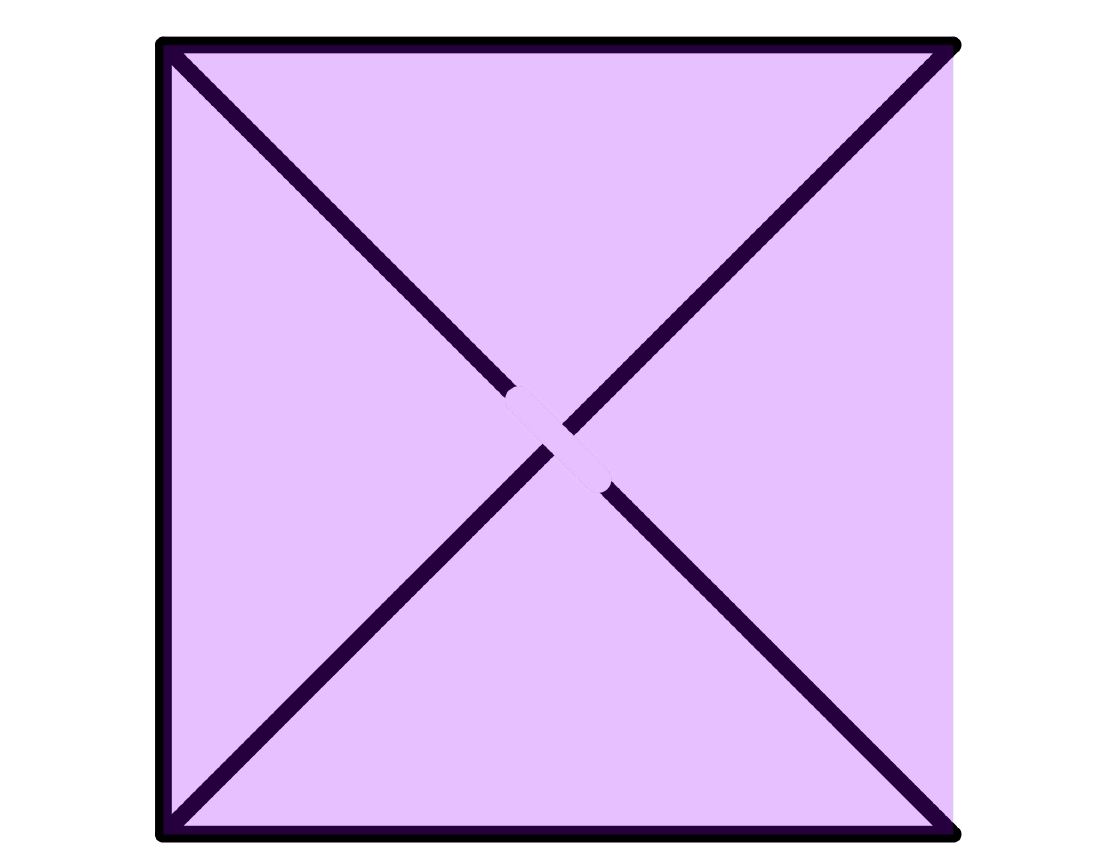}}\\
    \subcaptionbox*{$PN$}{\includegraphics[width = 0.103\columnwidth]{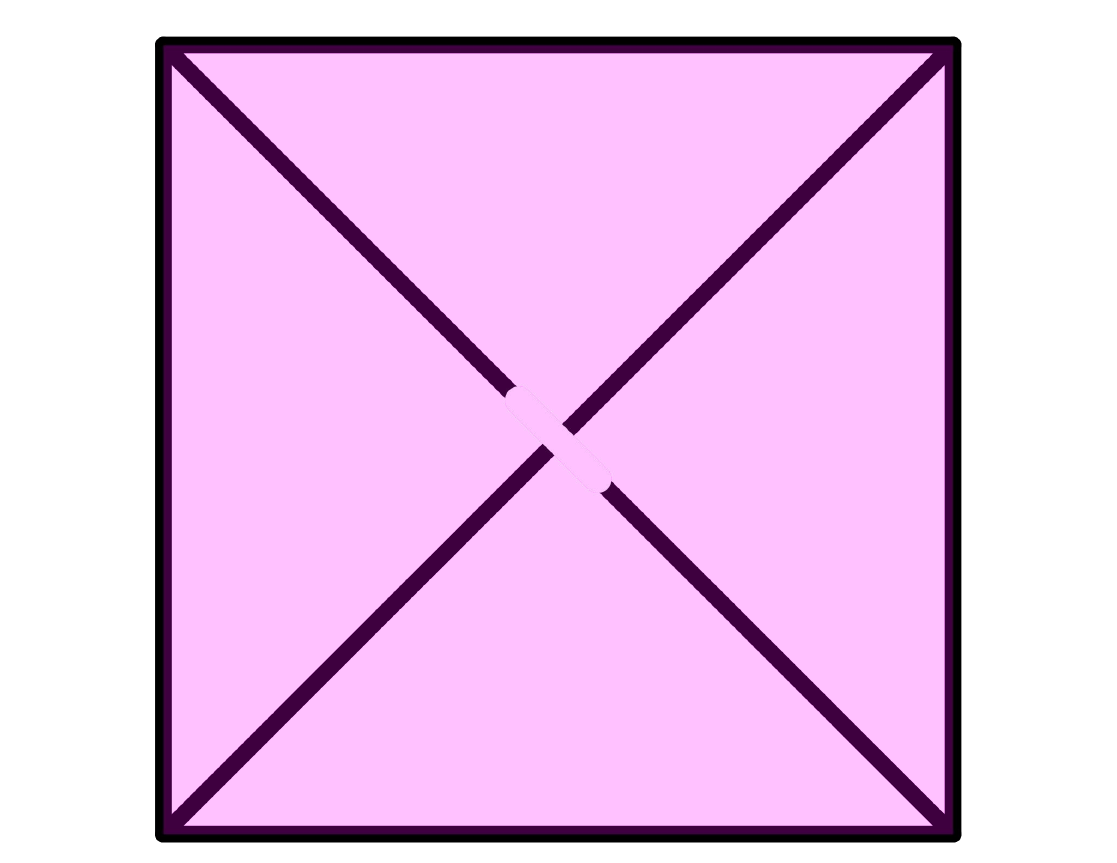}}
\caption{Diagrams to replace the shaded square(s) in Figure~\ref{fig:FD_Omega_Frame} (\ref{fig:FD_Cl_Frame}). 
Panels labeled with the same color are classified into the same category (see Table~\ref{tab:order-FD} for the category classification). 
Panels in the same row are classified into the same family (see Table~\ref{tab:FD_class} for the family classification). 
Here, momenta flows along propagators are omitted for the sake of brevity.  
}\label{fig:Feynman_Diagrams}
\end{figure}

By incorporating the primordial non-Gaussianity defined in Eq.~(\ref{eq:Fnl-Gnl-def}), we can express the power spectrum of Eq.~(\ref{eq:Ph-zeta}) as a series of Feynman-like diagrams, once the Feynman-like rules are known. These diagrams are thoroughly displayed in Figure~\ref{fig:FD_Omega_Frame} and Figure~\ref{fig:Feynman_Diagrams}. To obtain an individual diagram, we replace the shaded square in Figure~\ref{fig:FD_Omega_Frame} with a panel from Figure~\ref{fig:Feynman_Diagrams}. For the sake of brevity, we have omitted the momenta flows in each panel of Figure~\ref{fig:Feynman_Diagrams}, as they can be straightforwardly deduced from momentum conservation at each vertex. In total, there are 49 Feynman-like diagrams that need to be considered in this work.

\begin{table}[htbp]
\centering
\renewcommand\arraystretch{1.6}
\begin{tabular}[width=.3\columnwidth]{l|c|c|c}
\hline
$P_h^X$     & Integral & Vertices & S.F. \\
\hline
$P_h^G$      & $\cP_{G}^{[1,1]}$ & 1 & 2\\
$P_h^{Gl}$  & $A_S\cP_{G}^{[1,1]}$ & $\left(\frac{3}{5}\right)^{2}\gnl$ & 24 \\
$P_h^{Gl^H}$ & $A_S^2 \cP_{G}^{[1,1]}$ & $\left(\frac{3}{5}\right)^{4}\gnl^2$ & 36 \\
$P_h^{Gl^C}$ & $A_S^2 \cP_{G}^{[1,1]}$ & $\left(\frac{3}{5}\right)^{4}\gnl^2$ & 36 \\
$P_h^{Gl^Z}$ & $A_S^2 \cP_{G}^{[1,1]}$ & $\left(\frac{3}{5}\right)^{4}\gnl^2$ & 36\\
$P_h^{Gl^3}$ & $A_S^3 \cP_{G}^{[1,1]}$ & $\left(\frac{3}{5}\right)^{6}\gnl^3$ & 216\\
$P_h^{Gl^4}$ & $A_S^4 \cP_{G}^{[1,1]}$ & $\left(\frac{3}{5}\right)^{8}\gnl^4$ & 162\\
$P_h^{H}$ & $\cP_{G}^{[1,2]}$ & $\left(\frac{3}{5}\right)^{2}\fnl^2$ & 8\\
$P_h^{Hl}$ & $A_S \cP_{G}^{[1,2]}$ & $\left(\frac{3}{5}\right)^{4}\fnl^2 \gnl$ & 48\\
$P_h^{Hl^2}$ & $A_S^2 \cP_{G}^{[1,2]}$ & $\left(\frac{3}{5}\right)^{6}\fnl^2 \gnl^2$ & 72\\
$P_h^{H^2}$ & $\cP_{G}^{[1,3]}$ & $\left(\frac{3}{5}\right)^{4}\gnl^2$ & 24\\
$P_h^{H^2 l}$ & $A_S \cP_{G}^{[1,3]}$ & $\left(\frac{3}{5}\right)^{6}\gnl^3$ & 144\\
$P_h^{H^2 l^2}$ & $A_S^2 \cP_{G}^{[1,3]}$ & $\left(\frac{3}{5}\right)^{8}\gnl^4$ & 216\\
$P_h^{R}$ & $\cP_{G}^{[2,2]}$ & $\left(\frac{3}{5}\right)^{4}\fnl^4$ & 8\\
$P_h^{RH}$ & $\cP_{G}^{[2,3]}$ & $\left(\frac{3}{5}\right)^{6}\fnl^2 \gnl^2$ & 48\\
$P_h^{R^2}$ & $\cP_{G}^{[3,3]}$ & $\left(\frac{3}{5}\right)^{8}\gnl^4$ & 72\\
\hline
$P_h^{C}$ & $\cP_{C}^{[1,1;1]}$ & $\left(\frac{3}{5}\right)^{2}\fnl^2$ & 16\\
$P_h^{Cl}$ & $A_S \cP_{C}^{[1,1;1]}$ & $\left(\frac{3}{5}\right)^{4}\fnl^2 \gnl$ & 96\\
$P_h^{Cl^2}$ & $A_S^2 \cP_{C}^{[1,1;1]}$ & $\left(\frac{3}{5}\right)^{6}\fnl^2 \gnl^2$ & 144 \\
$P_h^{CH}$ & $\cP_{C}^{[1,2;1]}$ & $\left(\frac{3}{5}\right)^{4}\fnl^2 \gnl$ & 96\\
$P_h^{CHl}$ & $A_S \cP_{C}^{[1,2;1]}$ & $\left(\frac{3}{5}\right)^{6}\fnl^2 \gnl^2$ & 288\\
$P_h^{CR}$ & $\cP_{C}^{[2,2;1]}$ & $\left(\frac{3}{5}\right)^{6}\fnl^2 \gnl^2$ & 144\\
$P_h^{C^2}$ & $\cP_{C}^{[1,1;2]}$ & $\left(\frac{3}{5}\right)^{4}\gnl^2$ & 72\\
$P_h^{C^2 l}$ & $A_S \cP_{C}^{[1,1;2]}$ & $\left(\frac{3}{5}\right)^{6}\gnl^3 $ & 432\\
$P_h^{C^2 l^2}$ & $A_S^2 \cP_{C}^{[1,1;2]}$ & $\left(\frac{3}{5}\right)^{8}\gnl^4$ & 648\\
\hline
\end{tabular}
\quad\quad
\begin{tabular}[width=.3\columnwidth]{l|c|c|c}
\hline
\multicolumn{4}{c}{continued}\\
\hline
$P_h^X$     & Integral & Vertices & S.F. \\
\hline
$P_h^{Z}$ & $\cP_{Z}^{[1,1;1]}$ & $\left(\frac{3}{5}\right)^{2}\fnl^2$ & 16\\
$P_h^{Zl}$ & $A_S \cP_{Z}^{[1,1;1]}$ & $\left(\frac{3}{5}\right)^{4}\fnl^2 \gnl$ & 96\\
$P_h^{Zl^2}$ & $A_S^2 \cP_{Z}^{[1,1;1]}$ & $\left(\frac{3}{5}\right)^{6}\fnl^2 \gnl^2$ & 144 \\
$P_h^{ZH}$ & $\cP_{Z}^{[1,2;1]}$ & $\left(\frac{3}{5}\right)^{4}\fnl^2 \gnl$ & 96\\
$P_h^{ZHl}$ & $A_S \cP_{Z}^{[1,2;1]}$ & $\left(\frac{3}{5}\right)^{6}\fnl^2 \gnl^2$ & 288\\
$P_h^{ZR}$ & $\cP_{Z}^{[2,2;1]}$ & $\left(\frac{3}{5}\right)^{6}\fnl^2 \gnl^2$ & 144\\
$P_h^{Z^2}$ & $\cP_{Z}^{[1,1;2]}$ & $\left(\frac{3}{5}\right)^{4}\gnl^2$ & 72\\
$P_h^{Z^2 l}$ & $A_S \cP_{Z}^{[1,1;2]}$ & $\left(\frac{3}{5}\right)^{6}\gnl^3 $ & 432\\
$P_h^{Z^2 l^2}$ & $A_S^2 \cP_{Z}^{[1,1;2]}$ & $\left(\frac{3}{5}\right)^{8}\gnl^4$ & 648\\
\hline
$P_h^{P}$ & $\cP_{P}^{[1,1;1,1]}$ & $\left(\frac{3}{5}\right)^{4}\fnl^4$ & 32\\
$P_h^{PH}$ & $\cP_{P}^{[1,2;1,1]}$ & $\left(\frac{3}{5}\right)^{6}\fnl^2 \gnl^2$ & 288\\
$P_h^{PR}$ & $\cP_{P}^{[2,2;1,1]}$ & $\left(\frac{3}{5}\right)^{8}\gnl^4$ & 648\\
$P_h^{PC}$ & $\cP_{P}^{[1,1;2,1]}$ & $\left(\frac{3}{5}\right)^{6}\fnl^2 \gnl^2$ & 288\\
$P_h^{PP}$ & $\cP_{P}^{[1,1;2,2]}$ & $\left(\frac{3}{5}\right)^{8}\gnl^4$ & 648\\
\hline
$P_h^{N}$ & $\cP_{N}^{[1,1,1,1]}$ & $\left(\frac{3}{5}\right)^{4}\fnl^4$ & 16\\
$P_h^{NH}$ & $\cP_{N}^{[2,1,1,1]}$ & $\left(\frac{3}{5}\right)^{6}\fnl^2 \gnl^2$ & 288\\
$P_h^{NR}$ & $\cP_{N}^{[2,2,1,1]}$ & $\left(\frac{3}{5}\right)^{8}\gnl^4$ & 648\\
\hline
$P_h^{CZ}$ & $\cP_{CZ}^{[1]}$ & $\left(\frac{3}{5}\right)^{4}\fnl^2 \gnl$ & 192\\
$P_h^{CZl}$ & $A_S \cP_{CZ}^{[1]}$ & $\left(\frac{3}{5}\right)^{6}\fnl^2 \gnl^2$ & 576\\
$P_h^{CZH}$ & $\cP_{CZ}^{[2]}$ & $\left(\frac{3}{5}\right)^{6}\gnl^3$ & 864\\
$P_h^{CZHl}$ & $A_S \cP_{CZ}^{[2]}$ & $\left(\frac{3}{5}\right)^{8}\gnl^4$ & 2592\\
\hline
$P_h^{PZ}$ & $\cP_{PZ}$ & $\left(\frac{3}{5}\right)^{6}\fnl^2 \gnl^2$ & 576\\
\hline
$P_h^{NC}$ & $\cP_{NC}$ & $\left(\frac{3}{5}\right)^{6}\fnl^2 \gnl^2$ & 288 \\
\hline
$P_h^{PN}$ & $\cP_{PN}$ & $\left(\frac{3}{5}\right)^{8}\gnl^4$ & 1296\\
\hline
\end{tabular}
\caption{Table for illustration of the contribution from Diagram-$X$, as shown in \cref{fig:Feynman_Diagrams}, to the power spectrum in Eq.~\eqref{eq:Ph-zeta}.  \label{tab:PX}}
\end{table}

    For a given Feynman-like diagram, we can immediately determine its contribution to the power spectrum in Eq.~(\ref{eq:Ph-zeta}) using the Feynman-like rules. Let's consider a diagram labeled with $X$ ($X \in {G, Gl, ..., PN}$), denoted as Diagram-$X$. Its contribution can be formally expressed as 
    \begin{equation}\label{eq:phxsai} 
    P_h^X = \text{Integral} \times \text{Vertices} \times \text{S.F.}\ , 
    \end{equation} 
    where we provide the details of the ``Integral'', ``Vertices'', and ``S.F.'' in Table~\ref{tab:PX}. The ``Integral'' represents the integrals involving the propagators and $h$-vertices. We have classified these integrals into 9 families, which will be explained in Subsection \ref{subsec:PXcalc} along with the introduction of $\cP_X^{[...]}$. The ``Vertices'' term involves the powers of $\fnl$ and $\gnl$ associated with the Gaussian-vertex, $\fnl$-vertex, and $\gnl$-vertex. Finally, the ``S.F.'' stands for the symmetric factor.

\subsection{Formulae for Energy-Density Fraction Spectrum}

By following Eq.~(\ref{eq:Omegabar-h}), we can determine the contribution from the Feynman-like diagram labeled with $X$ to the energy-density fraction spectrum as 
\begin{equation}\label{eq:Omegabar-X} 
    \bar{\Omega}_\uGW^X
    = \frac{q^5}{96\pi^2 \cH^2} 
        \overbar{P_{h}^X}\ .
\end{equation}
After performing the oscillation average, we can rewrite it as a series of numerically favorable integrals, as shown in Appendix~\ref{sec:num-int}. These integrals can be computed numerically using the publicly available \texttt{vegas} package \cite{Lepage:2020tgj}.

\begin{table}
\renewcommand\arraystretch{1.6}
    \centering
    \begin{tabular}{c|c|c}
        \hline
        $\bar{\Omega}_\uGW^{(a,b)}$& $\left(\frac{3}{5}\right)^{2(a+b)}\fnl^{2a}\gnl^b A_S^{a+b+2}$& Diagram-X\\
        \hline
        (0,0) & $A_S^2$                    & $G$ \\
        \hline
        (0,1) & $\left(\frac{3}{5}\right)^{2} \gnl A_S^3$           & $Gl$ \\
        \hline
        (1,0) & $\left(\frac{3}{5}\right)^{2} \fnl^2 A_S^3$         & $H$, $C$, $Z$ \\
        \hline
        (0,2) & $\left(\frac{3}{5}\right)^{4} \gnl^2 A_S^4$       & $Gl^H$, $Gl^C$, $Gl^Z$, $H^2$, $C^2$, $Z^2$ \\
        \hline
        (1,1) & $\left(\frac{3}{5}\right)^{4} \fnl^2 \gnl A_S^4$  & $Hl$, $Cl$, $Zl$, $CH$, $ZH$, $CZ$\\ 
        \hline
        (2,0) & $\left(\frac{3}{5}\right)^{4} \fnl^4 A_S^4$       & $R$, $P$, $N$\\ 
        \hline
        (0,3) & $\left(\frac{3}{5}\right)^{6} \gnl^3 A_S^5$       & $Gl^3$, $H^2 l$, $C^2 l$, $Z^2 l$, $CZH$\\ 
        \hline
        \multirow{2}{*}{(1,2)} & \multirow{2}{*}{$\left(\frac{3}{5}\right)^{6} \fnl^2 \gnl^2 A_S^5$}& $Hl^2$, $Cl^2$, $Zl^2$, $CHl$, $ZHl$, $CZl$, $RH$,\\
              &                      &$CR$, $ZR$, $PH$, $PC$, $PZ$, $NH$, $NC$ \\ 
              \hline
        (0,4) & $\left(\frac{3}{5}\right)^{8} \gnl^4 A_S^6$       & $Gl^4$, $H^2 l^2$, $R^2$, $C^2 l^2$, $Z^2 l^2$, $PR$, $PP$, $NR$, $CZHl$, $PN$ \\ 
        \hline
    \end{tabular}
    \caption{Table for illustration of the category classification. The Feynman-like diagrams of each category have the same powers in $\fnl$ and $\gnl$. }
    \label{tab:order-FD}
\end{table}

Based on the information provided in the ``Vertices'' column of Table~\ref{tab:PX}, we classify the total of 49 Feynman-like diagrams into 9 categories, as shown in Table~\ref{tab:order-FD}. The diagrams within each category have the same powers in $\fnl$ and $\gnl$, as well as the same powers in $A_{S}$. Consequently, we divide the energy-density fraction spectrum into 9 components, denoted as $\bar{\Omega}_\uGW^{(a,b)}$ with powers given by $(3/5)^{2(a+b)}\fnl^{2a}\gnl^b A_S^{a+b}$. To be more specific, these components can be expressed as follows 
\begin{subequations}\label{eqs:Omegabar-order} 
\begin{eqnarray}
    \bar{\Omega}_\uGW^{(0,0)}  
    &=& \bar{\Omega}_\uGW^G  \ ,\\
    \bar{\Omega}_\uGW^{(0,1)}  
    &=& \bar{\Omega}_\uGW^{Gl}  \ ,\\
    \bar{\Omega}_\uGW^{(1,0)}  
    &=& \bar{\Omega}_\uGW^{H}   + \bar{\Omega}_\uGW^{C}   + \bar{\Omega}_\uGW^{Z}  \ ,\\
    \bar{\Omega}_\uGW^{(0,2)}  
    &=& \bar{\Omega}_\uGW^{Gl^H}   + \bar{\Omega}_\uGW^{Gl^C}   + \bar{\Omega}_\uGW^{Gl^Z}   + \bar{\Omega}_\uGW^{H^2}   + \bar{\Omega}_\uGW^{C^2}   + \bar{\Omega}_\uGW^{Z^2}  \ ,\\
    \bar{\Omega}_\uGW^{(1,1)}  
    &=& \bar{\Omega}_\uGW^{Hl}   + \bar{\Omega}_\uGW^{Cl}   + \bar{\Omega}_\uGW^{Zl}   + \bar{\Omega}_\uGW^{CH}   + \bar{\Omega}_\uGW^{ZH}   + \bar{\Omega}_\uGW^{CZ}  \ ,\\
    \bar{\Omega}_\uGW^{(2,0)}  
    &=& \bar{\Omega}_\uGW^{R}   + \bar{\Omega}_\uGW^{P}   + \bar{\Omega}_\uGW^{N}  \ ,\\
    \bar{\Omega}_\uGW^{(0,3)}  
    &=& \bar{\Omega}_\uGW^{Gl^3}   + \bar{\Omega}_\uGW^{H^2 l}   + \bar{\Omega}_\uGW^{C^2 l}   + \bar{\Omega}_\uGW^{Z^2 l}   + \bar{\Omega}_\uGW^{CZH}   \ ,\\
    \bar{\Omega}_\uGW^{(1,2)}  
    &=& \bar{\Omega}_\uGW^{Hl^2}   + \bar{\Omega}_\uGW^{Cl^2}   + \bar{\Omega}_\uGW^{Zl^2}   + \bar{\Omega}_\uGW^{CHl}   + \bar{\Omega}_\uGW^{ZHl}  + \bar{\Omega}_\uGW^{CZl}   + \bar{\Omega}_\uGW^{RH}  + \bar{\Omega}_\uGW^{CR}     + \bar{\Omega}_\uGW^{ZR} \nonumber\\
     &&
     + \bar{\Omega}_\uGW^{PH}
    + \bar{\Omega}_\uGW^{PC}   + \bar{\Omega}_\uGW^{PZ}   + \bar{\Omega}_\uGW^{NH}   + \bar{\Omega}_\uGW^{NC}  \ ,\\
    \bar{\Omega}_\uGW^{(0,4)}   
    &=& \bar{\Omega}_\uGW^{Gl^4}   + \bar{\Omega}_\uGW^{H^2 l^2}   + \bar{\Omega}_\uGW^{R^2}   + \bar{\Omega}_\uGW^{C^2 l^2}   + \bar{\Omega}_\uGW^{Z^2 l^2}   + \bar{\Omega}_\uGW^{PR}   + \bar{\Omega}_\uGW^{PP} \nonumber\\
     && + \bar{\Omega}_\uGW^{NR}   + \bar{\Omega}_\uGW^{CZH l}   + \bar{\Omega}_\uGW^{PN}   \ .
\end{eqnarray}
\end{subequations}
As a result, the total energy-density spectrum can be expressed as the sum of $\bar{\Omega}_\uGW^{(a,b)}$, i.e., 
\begin{equation}\label{eq:Omegabar-total}
    \bar{\Omega}_\uGW = \bar{\Omega}_\uGW^{(0,0)} + \bar{\Omega}_\uGW^{(0,1)} + \bar{\Omega}_\uGW^{(1,0)} + \bar{\Omega}_\uGW^{(0,2)} + \bar{\Omega}_\uGW^{(1,1)} + \bar{\Omega}_\uGW^{(2,0)} + \bar{\Omega}_\uGW^{(0,3)} + \bar{\Omega}_\uGW^{(1,2)} + \bar{\Omega}_\uGW^{(0,4)}\ .
\end{equation}
In particular, it should be noted that $\bar\Omega_\uGW^{(0,0)}$ represents the energy-density fraction spectrum of \acp{SIGW} under the assumption of Gaussianity, as demonstrated by the semi-analytic calculation in Refs.~\cite{Espinosa:2018eve,Kohri:2018awv}. 
There are a lot of works considering the local-type primordial non-Gaussianties. 
Specifically, Ref.~\cite{Nakama:2016gzw,Cai:2018dig,Zhang:2021rqs} focused on the $\fnl$ contributions denoted as $G$, $H$, and $R$.
Ref.~\cite{Ragavendra:2020sop} conducted a similar analysis, considering a specific model and scale-dependent non-Gaussianity.
Ref.~\cite{Unal:2018yaa} expanded the analysis to include other $\fnl$ contributions, except for the one labeled as $Z$, which was first considered in Ref.~\cite{Atal:2021jyo}.
However, confusion arose between the contributions labeled as ``walnut'' in Ref.~\cite{Unal:2018yaa} and Ref.~\cite{Atal:2021jyo}, respectively, due to an incomplete definition of Feynman-like rules.
Specifically, the contribution referred to as ``walnut'' in Ref.~\cite{Unal:2018yaa} corresponds to $C$, while in Ref.~\cite{Atal:2021jyo}, it corresponds to $Z$.
The complete analysis, including the formulation of Feynman-like rules and diagrams, was originally provided by Ref.~\cite{Adshead:2021hnm} and reexamined in Refs.~\cite{Ragavendra:2021qdu,Li:2023qua}. 
Furthermore, Refs.~\cite{Nakama:2016gzw,Yuan:2020iwf} evaluated contributions from the non-Gaussianity of the $\gnl$ order to the energy-density fraction spectrum of \acp{SIGW}, via focusing on terms belonging to the $G$-like contributions. 
In a recent work (arXiv:2308.07155v1), the analysis was extended to include more diagrams. Our current study goes further by considering two additional diagrams, specifically referred to as $CZH$ and $CZHl$.  
In Ref.~\cite{Abe:2022xur}, the Feynman-like diagrammatic technique was employed to investigate the contributions arising from series-expandable non-Gaussianity, via calculating the energy-density fraction spectrum up to $\cO(A_S^4)$ order.

\subsection{Nine Families of Integrals}\label{subsec:PXcalc}

In Table \ref{tab:PX} of Subsection~\ref{subsec:frd}, we mentioned that the classification of integrals into 9 families facilitates subsequent calculations. In this subsection, we will provide a comprehensive explanation of the criteria used for this classification and introduce $\cP_X^{[...]}$. Quick readers can skip this subsection for now and return to it later if they need more details.

\begin{figure}
    \centering
    \includegraphics[width =1 \columnwidth]{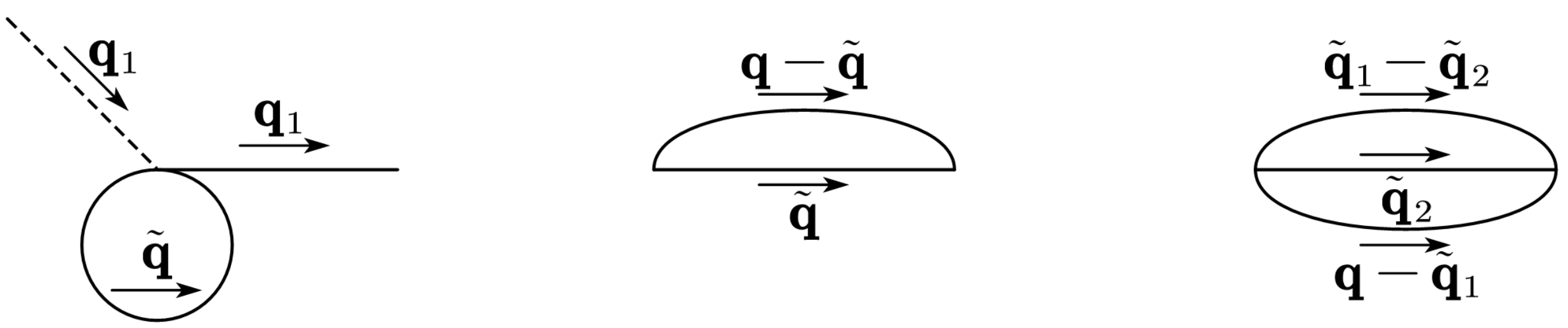}
    \caption{Loop integrals precomputed in advance. }
    \label{fig:FD-loop}
\end{figure}

Let's begin by clarifying two conventions that will simplify our subsequent calculations. The first convention pertains to the loop integral associated with an individual $\gnl$-vertex, as depicted in the left panel of Figure~\ref{fig:FD-loop}. The contribution to $\bar{\Omega}_{\uGW}$ from this loop integral is solely determined by $\gnl$ and $A_S$, specifically, 
\begin{equation}
    3 \times \frac{9}{25}\gnl \int\frac{\ud^3 \tilde{\bq}}{(2\pi)^3} P^{[1]} (\tilde{q}) = 3 \times \left(\frac{3}{5}\right)^{2} \gnl \, A_S \ ,
\end{equation}
In this expression, the number $3$ represents the symmetric factor, $(3/5)^{2}\gnl$ arises from the $\gnl$-vertex, and $A_S$ is associated with the loop integral.
The second convention concerns the multi-propagators between two different vertices, as depicted in the middle and right panels of Figure~\ref{fig:FD-loop}. By performing the integration over loop momenta in advance, we can reduce the dimensions of the multi-variate integration. For convenience, we introduce quantities of the form 
\begin{subequations}\label{eqs:P123-def}
\begin{eqnarray}
    P^{[1]} (q) &=& P_{gS} (q)\ ,\\
    P^{[2]} (q) &=& \int \frac{\ud^3 \tilde{\bq}}{(2 \pi)^3} P^{[1]} (\tilde{q}) P^{[1]} (\abs{\bq - \tilde{\bq}})\ , \\
    P^{[3]} (q) &=& \int \frac{\ud^3 \tilde{\bq}}{(2 \pi)^3} P^{[2]} (\tilde{q}) P^{[1]} (\abs{\bq - \tilde{\bq}})\ .
\end{eqnarray}
\end{subequations}

\begin{table}
\renewcommand\arraystretch{1.6}
    \centering
    \begin{tabular}{c|c}
        \hline
        Family    & Diagram-X\\
        \hline
        \multirow{2}{*}{$G$-like}& $G$, $Gl$, $Gl^H$, $Gl^C$, $Gl^Z$, $Gl^3$, $Gl^4$, \\
                  & $H$, $Hl$, $Hl^2$, $H^2$, $H^2 l$,$H^2 l^2$, $R$, $RH$, $R^2$\\
        \hline
        $C$-like  & $C$, $Cl$, $Cl^2$, $CH$, $CHl$, $CR$, $C^2$, $C^2 l$, $C^2 l^2$ \\
        \hline
        $Z$-like  & $Z$, $Zl$, $Zl^2$, $ZH$, $ZHl$, $ZR$, $Z^2$, $Z^2 l$, $Z^2 l^2$ \\
        \hline
        $P$-like  & $P$, $PH$, $PR$, $PC$, $PP$\\
        \hline
        $N$-like  & $N$, $NH$, $NR$\\ 
        \hline
        $CZ$-like & $CZ$, $CZl$, $CZH$, $CZHl$\\
        \hline
        $PZ$-like &  $PZ$  \\ 
        \hline
        $NC$-like & $NC$  \\ 
        \hline
        $PN$-like & $PN$\\
        \hline
    \end{tabular}
    \caption{Table for illustration of the family classification. The integrals associated with the Feynman-like diagrams of each family share the same form.  }
    \label{tab:FD_class}
\end{table}

By clarifying the aforementioned two conventions, we can immediately determine the integrals modulo these two types of loop integrals. Based on this, we classify them into a total of 9 families, namely ``$G$-like'', ``$C$-like'', ``$Z$-like'', ``$P$-like'', ``$N$-like'', ``$CZ$-like'', ``$PZ$-like'', ``$NC$-like'' and ``$PN$-like'' families. The panels in the first two rows of Figure~\ref{fig:Feynman_Diagrams} belong to the ``G-like'' family, while the panels in each subsequent row belong to the same family. To be more specific, we summarize these families in Table~\ref{tab:FD_class}. One advantage of our classification is that the integrals associated with the same family share the same form, except for their subscripts $^{[...]}$, as demonstrated below.

The integrals belonging to the ``$G$-like'' family can be expressed explicitly as follows 
\begin{eqnarray}\label{eq:G-like} 
    \cP_{G}^{[\alpha,\beta]} (\eta,q) 
    &=& 2^4 \int \frac{\ud^3 \bq_1}{(2\pi)^{3}}\,  P^{[\alpha]} (q_1) P^{[\beta]} (\abs{\bq-\bq_1}) \sum_{\lambda = +,\times} Q_{\lambda}^2 (\bq, \bq_1) 
        \hat{I}^2 (\abs{\bq - \bq_1}, q_1, \eta) \ ,
\end{eqnarray}
where we take $\alpha$ and $\beta$ to be values from 1 to 3, with $\alpha$ less than or equal to $\beta$. The integrals associated with the ``$C$-like'' family can be expressed as follows 
\begin{eqnarray}\label{eq:C-like} 
    \cP_{C}^{[\alpha,\beta;\gamma]} (\eta,q) 
    &=& 2^4 \int \frac{\ud^3 \bq_1}{(2\pi)^{3}} \frac{\ud^3 \bq_2}{(2\pi)^{3}} \,
        P^{[\alpha]} (q_2) P^{[\beta]} (\abs{\bq-\bq_2}) P^{[\gamma]} (\abs{\bq_1-\bq_2}) \nonumber\\
        &&\hphantom{\  2^4} 
        \times \sum_{\lambda = +,\times} Q_{\lambda}(\bq, \bq_1) \hat{I} (\abs{\bq - \bq_1}, q_1, \eta) 
        Q_{\lambda}(\bq, \bq_2) \hat{I} (\abs{\bq - \bq_2}, q_2, \eta) \ , 
\end{eqnarray}
where we take $\alpha$, $\beta$, and $\gamma$ to be values from 1 to 2, with $\alpha$ less than or equal to $\beta$. However, $\beta$ and $\gamma$ cannot both be equal to 2 simultaneously. The integrals associated with the ``$Z$-like'' family can be expressed as follows 
\begin{eqnarray}\label{eq:Z-like} 
    \cP_{Z}^{[\alpha,\beta;\gamma]} (\eta,q) 
    &=& 2^4 \int \frac{\ud^3 \bq_1}{(2\pi)^{3}} \frac{\ud^3 \bq_2}{(2\pi)^{3}} \,
        P^{[\alpha]} (q_2) P^{[\beta]} (\abs{\bq-\bq_1}) P^{[\gamma]} (\abs{\bq_1-\bq_2}) \nonumber\\
        &&\hphantom{\  2^4} 
        \times \sum_{\lambda = +,\times} Q_{\lambda}(\bq, \bq_1) \hat{I} (\abs{\bq - \bq_1}, q_1, \eta) 
        Q_{\lambda}(\bq, \bq_2) \hat{I} (\abs{\bq - \bq_2}, q_2, \eta) \ , 
\end{eqnarray}
where we consider $\alpha$, $\beta$, and $\gamma$ as values from 1 to 2, with $\alpha$ less than or equal to $\beta$. However, it is not possible for both $\beta$ and $\gamma$ to be equal to 2 at the same time. The integrals associated with the ``$P$-like'' family can be expressed as follows
\begin{eqnarray}\label{eq:P-like}  
    \cP_{P}^{[\alpha,\beta;\gamma,\delta]} (\eta,q) 
    &=& 2^4 \int \frac{\ud^3 \bq_1}{(2\pi)^{3}} \frac{\ud^3 \bq_2}{(2\pi)^{3}} \frac{\ud^3 \bq_3}{(2\pi)^{3}}\, 
        P^{[\alpha]} (q_3) P^{[\beta]} (\abs{\bq-\bq_3}) P^{[\gamma]} (\abs{\bq_1-\bq_3}) \nonumber\\
        &&\hphantom{2^4 \int \frac{\ud^3 \bq_1}{(2\pi)^{3}} \frac{\ud^3 \bq_2}{(2\pi)^{3}} \frac{\ud^3 \bq_3}{(2\pi)^{3}}} 
        \times P^{[\delta]} (\abs{\bq_2-\bq_3}) \\
        &&\hphantom{\  2^4} 
        \times \sum_{\lambda = +,\times} Q_{\lambda}(\bq, \bq_1) \hat{I} (\abs{\bq - \bq_1}, q_1, \eta) 
        Q_{\lambda}(\bq, \bq_2) \hat{I} (\abs{\bq - \bq_2}, q_2, \eta) \ , \nonumber
\end{eqnarray}
where we consider $\alpha$, $\beta$, $\gamma$, and $\delta$ as values from 1 to 2, with $\alpha$ less than or equal to $\beta$, and $\gamma$ greater than or equal to $\delta$. However, it is not possible for both $\beta$ and $\gamma$ to be equal to 2 at the same time. The integrals associated with the ``$N$-like'' family can be expressed as follows
\begin{eqnarray}\label{eq:N-like}  
    \cP_{N}^{[\alpha,\beta;1,1]} (\eta,q) 
    &=& 2^4 \int \frac{\ud^3 \bq_1}{(2\pi)^{3}} \frac{\ud^3 \bq_2}{(2\pi)^{3}} \frac{\ud^3 \bq_3}{(2\pi)^{3}} \, 
        P^{[\alpha]} (\abs{\bq-\bq_3}) P^{[\beta]} (\abs{\bq_1+\bq_2-\bq_3}) \nonumber\\
        &&\hphantom{2^4 \int \frac{\ud^3 \bq_1}{(2\pi)^{3}} \frac{\ud^3 \bq_2}{(2\pi)^{3}} \frac{\ud^3 \bq_3}{(2\pi)^{3}}}
        \times P^{[1]} (\abs{\bq_1-\bq_3}) P^{[1]} (\abs{\bq_2-\bq_3}) \\
        &&\hphantom{\  2^4} 
        \times \sum_{\lambda = +,\times} Q_{\lambda}(\bq, \bq_1) \hat{I} (\abs{\bq - \bq_1}, q_1, \eta) 
        Q_{\lambda}(\bq, \bq_2) \hat{I} (\abs{\bq - \bq_2}, q_2, \eta) \ , \nonumber
\end{eqnarray}
where we consider $\alpha$ and $\beta$ as values from 1 to 2, with $\alpha$ greater than or equal to $\beta$. The integrals associated with the ``$CZ$-like'' family can be expressed as follows
\begin{eqnarray}\label{eq:CZ-like}
    \cP_{CZ}^{[\alpha]} (\eta,q) 
    &=& 2^4 \int \frac{\ud^3 \bq_1}{(2\pi)^{3}} \frac{\ud^3 \bq_2}{(2\pi)^{3}} \frac{\ud^3 \bq_3}{(2\pi)^{3}}\, 
        P^{[1]} (q_2) P^{[\alpha]} (\abs{\bq-\bq_3}) P^{[1]} (\abs{\bq_1-\bq_3}) \nonumber\\
        &&\hphantom{2^4 \int \frac{\ud^3 \bq_1}{(2\pi)^{3}} \frac{\ud^3 \bq_2}{(2\pi)^{3}} \frac{\ud^3 \bq_3}{(2\pi)^{3}}}
        \times P^{[1]} (\abs{\bq_2-\bq_3}) \\
        &&\hphantom{\  2^4} 
        \times \sum_{\lambda = +,\times} Q_{\lambda}(\bq, \bq_1) \hat{I} (\abs{\bq - \bq_1}, q_1, \eta) 
        Q_{\lambda}(\bq, \bq_2) \hat{I} (\abs{\bq - \bq_2}, q_2, \eta) \ , \nonumber
\end{eqnarray}
where we consider $\alpha$ to be either 1 or 2. 
Since each of the ``$PZ$-like'', ``$NC$-like'', and ``$PN$-like'' families consists of only one diagram, we omit their superscripts for brevity. The associated integrals can be expressed as follows
\begin{eqnarray}
    \cP_{PZ} (\eta,q) 
    &=& 2^4 \int \frac{\ud^3 \bq_1}{(2\pi)^{3}} \frac{\ud^3 \bq_2}{(2\pi)^{3}} \frac{\ud^3 \bq_3}{(2\pi)^{3}} \frac{\ud^3 \bq_4}{(2\pi)^{3}} \, 
        P^{[1]} (q_4) P^{[1]} (\abs{\bq-\bq_3}) P^{[1]} (\abs{\bq_1-\bq_3}) \nonumber\\
        &&\hphantom{2^4 \int \frac{\ud^3 \bq_1}{(2\pi)^{3}} \frac{\ud^3 \bq_2}{(2\pi)^{3}} \frac{\ud^3 \bq_3}{(2\pi)^{3}} \frac{\ud^3 \bq_4}{(2\pi)^{3}}}
        \times P^{[1]} (\abs{\bq_2-\bq_4}) P^{[1]} (\abs{\bq_3-\bq_4}) \label{eq:PZ-like}\\
        &&\hphantom{\  2^4} 
        \times \sum_{\lambda = +,\times} Q_{\lambda}(\bq, \bq_1) \hat{I} (\abs{\bq - \bq_1}, q_1, \eta) 
        Q_{\lambda}(\bq, \bq_2) \hat{I} (\abs{\bq - \bq_2}, q_2, \eta) \ , \nonumber\\
    \cP_{NC} (\eta,q) 
    &=& 2^4 \int \frac{\ud^3 \bq_1}{(2\pi)^{3}} \frac{\ud^3 \bq_2}{(2\pi)^{3}} \frac{\ud^3 \bq_3}{(2\pi)^{3}} \frac{\ud^3 \bq_4}{(2\pi)^{3}} \, 
        P^{[1]} (q_3) P^{[1]} (\abs{\bq-\bq_4}) P^{[1]} (\abs{\bq_2-\bq_3}) \nonumber\\
        &&\hphantom{2^4 \int \frac{\ud^3 \bq_1}{(2\pi)^{3}} \frac{\ud^3 \bq_2}{(2\pi)^{3}} \frac{\ud^3 \bq_3}{(2\pi)^{3}} \frac{\ud^3 \bq_4}{(2\pi)^{3}}}
        \times P^{[1]} (\abs{\bq_2-\bq_4}) P^{[1]} (\abs{\bq_1 + \bq_3-\bq_4}) \label{eq:NC-like}\\
        &&\hphantom{\  2^4} 
        \times \sum_{\lambda = +,\times} Q_{\lambda}(\bq, \bq_1) \hat{I} (\abs{\bq - \bq_1}, q_1, \eta) 
        Q_{\lambda}(\bq, \bq_2) \hat{I} (\abs{\bq - \bq_2}, q_2, \eta) \ , \nonumber\\
    \cP_{PN} (\eta,q) 
    &=& 2^4 \int \frac{\ud^3 \bq_1}{(2\pi)^{3}} \frac{\ud^3 \bq_2}{(2\pi)^{3}} \frac{\ud^3 \bq_3}{(2\pi)^{3}} \frac{\ud^3 \bq_4}{(2\pi)^{3}} \frac{\ud^3 \bq_5}{(2\pi)^{3}} \, 
        P^{[1]} (q_3) P^{[1]} (\abs{\bq-\bq_4}) P^{[1]} (\abs{\bq_2-\bq_5}) \nonumber\\
        &&\hphantom{2^4 \int \frac{\ud^3 \bq_1}{(2\pi)^{3}} \frac{\ud^3 \bq_2}{(2\pi)^{3}} }
        \times P^{[1]} (\abs{\bq_3-\bq_5}) P^{[1]} (\abs{\bq_4-\bq_5}) P^{[1]} (\abs{\bq_1 + \bq_3-\bq_4}) \label{eq:PN-like}\\
        &&\hphantom{\  2^4} 
        \times \sum_{\lambda = +,\times} Q_{\lambda}(\bq, \bq_1) \hat{I} (\abs{\bq - \bq_1}, q_1, \eta) 
        Q_{\lambda}(\bq, \bq_2) \hat{I} (\abs{\bq - \bq_2}, q_2, \eta) \ .\nonumber
\end{eqnarray}

\subsection{Numerical Results}

\begin{figure}[htbp]
    \centering
    \includegraphics[width =1 \columnwidth]{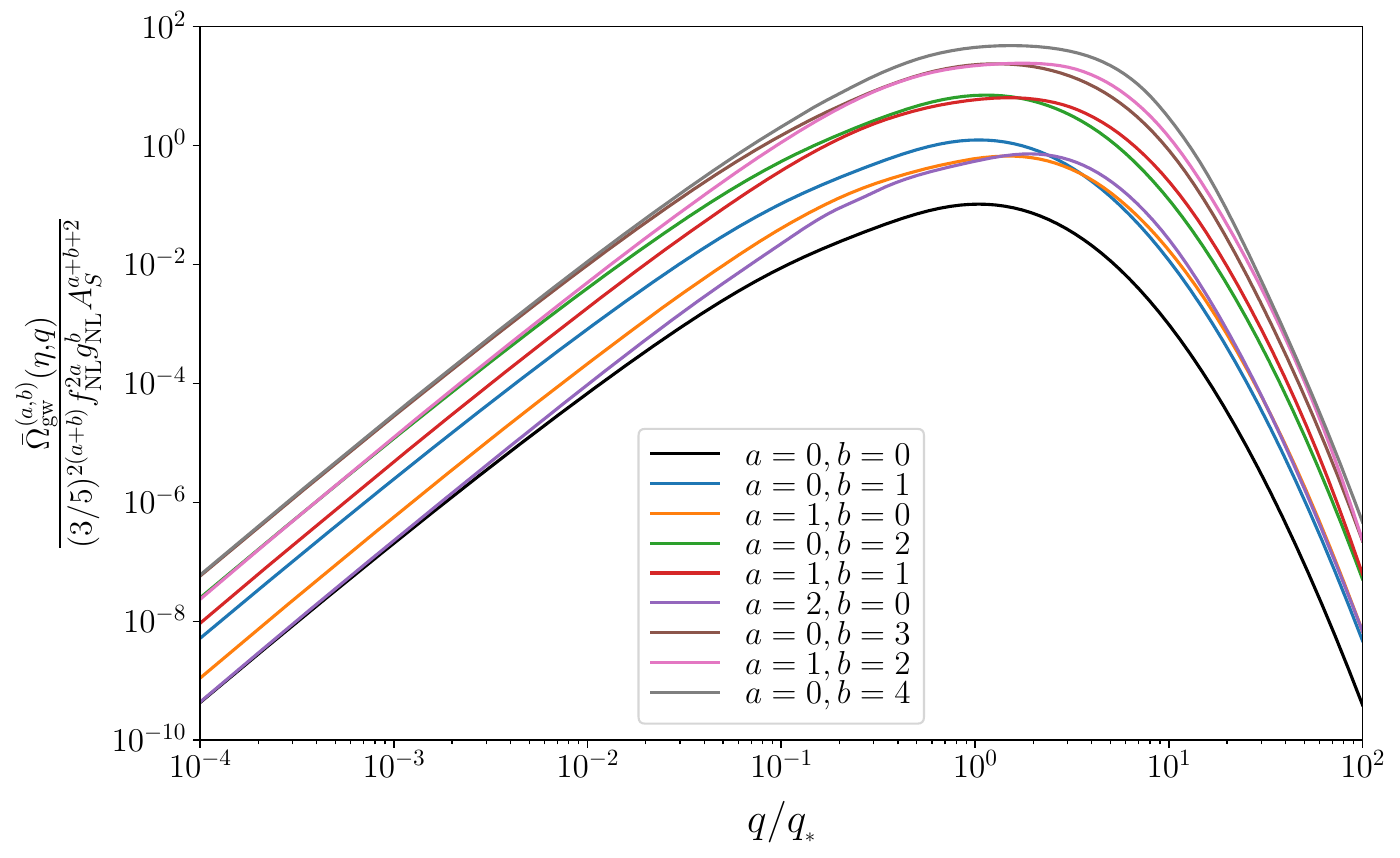}
    \caption{Unscaled (or equivalently, $A_S=1$, $3\fnl/5=1$ and $9\gnl/25=1$) components of the energy-density fraction spectra of \acp{SIGW} in powers of the primordial non-Gaussian parameters $\fnl$ and $\gnl$. }\label{fig:Unscaled_Omegabar}
\end{figure}

In this analysis, we highlight the significant effects of $\fnl$ and $\gnl$ on the energy-density fraction spectrum of \acp{SIGW}. Additionally, we illustrate the presence of significant degeneracies in the model parameters for the spectrum, which pose challenges in accurately probing $\fnl$ and $\gnl$ using the spectrum alone.

To evaluate Eq.~(\ref{eqs:Omegabar-order}) numerically for the \acp{SIGW} generated during the radiation-dominated phase of the early Universe, we set $\sigma=1$. The results of the unscaled energy-density fraction spectra $[(3/5)^{2(a+b)}\fnl^{2a}\gnl^{b}A_{S}^{a+b+2}]^{-1}\bar{\Omega}_{\uGW}^{(a,b)}(\eta,q)$ are shown in Figure~\ref{fig:Unscaled_Omegabar}. By scaling these unscaled spectra and summing them up, we obtain the total spectrum $\bar{\Omega}_{\uGW}(\eta,q)$ as $\bar{\Omega}_{\uGW}(\eta,q)=\sum_{(a,b)}\bar{\Omega}_{\uGW}^{(a,b)}(\eta,q)$, as given in Eq.~(\ref{eq:Omegabar-total}). It is worth noting that generalizations for studying other values of $\sigma$ and different epochs of the early Universe can be carried out relatively easily.

We are also interested in studying the physical energy-density fraction spectrum of \acp{SIGW} in the present Universe, specifically at the conformal time $\eta_0$. This spectrum is related to the total spectrum at the production time $\eta$ in the following manner \cite{Wang:2019kaf}
\begin{eqnarray}\label{eq:Omega0}
    h^{2} \bar{\Omega}_{\uGW,0} (\nu) 
    & = & h^{2} \Omega_{\mathrm{rad}, 0} 
        \left[\frac{g_{*,\rho}(T)}{g_{*,\rho}(T_\mathrm{eq})} \right]
        \left[\frac{g_{*,s}(T_\mathrm{eq})}{g_{*,s}(T)} \right]^{4/3} \bar{\Omega}_\uGW (\eta,q) \ ,
\end{eqnarray}
where $h$ represents the dimensionless Hubble constant, and $h^{2}\Omega_{\mathrm{rad},0}=4.2 \times 10^{-5}$ denotes the physical energy-density fraction of radiation in the present Universe \cite{Planck:2018vyg}. The temperature $T$ (and $T_\mathrm{eq}$) corresponds to the cosmic temperature at the production time (and the epoch of matter-radiation equality), while $\nu$ refers to the gravitational-wave frequency. Furthermore, it should be noted that $\nu$ is determined by $T$, meaning that there is a relationship between the two, i.e., \cite{Zhao:2022kvz}
\begin{eqnarray}
    \frac{\nu}{\mathrm{nHz}} 
    = 26.5 \left(\frac{T}{\mathrm{GeV}}\right)\left(\frac{g_{*,\rho}(T)}{106.75}\right)^{1/2}\left(\frac{g_{*,s}(T)}{106.75}\right)^{-1/3}\ ,
\end{eqnarray} 
where $g_{,\rho}$ and $g_{,s}$ are quantities associated with the number of relativistic species, and they are tabulated functions of the temperature $T$. These tabulated functions can be found in Ref.~\cite{Saikawa:2018rcs}.

\begin{figure}[htbp]
    \centering
    \includegraphics[width =1 \columnwidth]{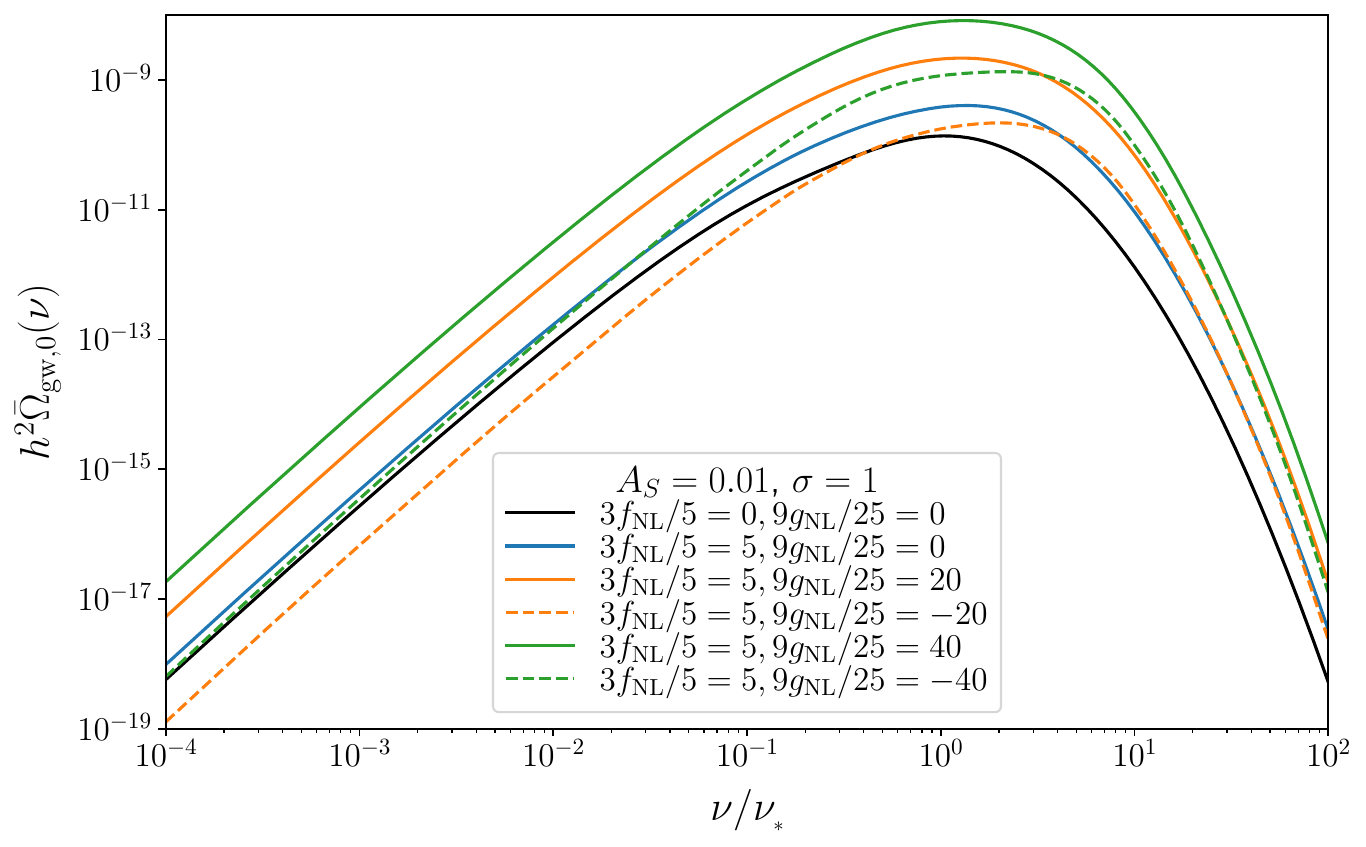}
    \caption{Energy-density fraction spectra of \acp{SIGW} in the present Universe. }\label{fig:Total_Omegabar}
\end{figure}

Figure~\ref{fig:Total_Omegabar} illustrates the contributions of both $\fnl$ and $\gnl$ to the energy-density fraction spectrum of \acp{SIGW} in the present Universe. By combining Figure~\ref{fig:Unscaled_Omegabar} with Figure~\ref{fig:Total_Omegabar}, we observe that the contributions from $\fnl$ are always positive, while those from $\gnl$ depend not only on its sign but also on its amplitude. This behavior arises because the unscaled energy-density fraction spectrum $\bar{\Omega}_{\uGW}^{(a,b)}$ is explicitly proportional to $\fnl^{2a}$ and $\gnl^{b}$, where $a=0,1,2$ and $b=0,1,2,3,4$. It is evident that there is a sign degeneracy for $\fnl$ in $\bar{\Omega}_{\uGW}^{(a,b)}$, but not for $\gnl$. When $\gnl$ is positive (represented by solid curves), the contributions from it to $\bar{\Omega}_{\uGW}^{(a,b)}$ are always positive. On the other hand, when $\gnl$ is negative (indicated by dashed curves), the contributions are positive for even values of $b$ and negative for odd values of $b$. Furthermore, the contributions to $\bar{\Omega}_{\uGW}$ also depend on the absolute value of $\gnl$ (more precisely, on the combinations $\fnl^{2a}\gnl^{b}A_{S}^{a+b+2}$).

The aforementioned observations are visually demonstrated in Figure~\ref{fig:Total_Omegabar}. When comparing the scenario with Gaussianity, we observe that the inclusion of $\fnl$ and positive $\gnl$ consistently enhances the overall energy-density fraction spectrum  for a fixed value of $A_S$. Conversely,  a small negative $\gnl$ can suppress this spectrum, especially in the low-frequency range, compared to a positive one. In fact, the suppression can be substantial to the extent that the suppressed spectrum falls below the level of the Gaussian spectrum.

\begin{figure*}[htbp]
    \centering
    \includegraphics[width = 1 \textwidth]{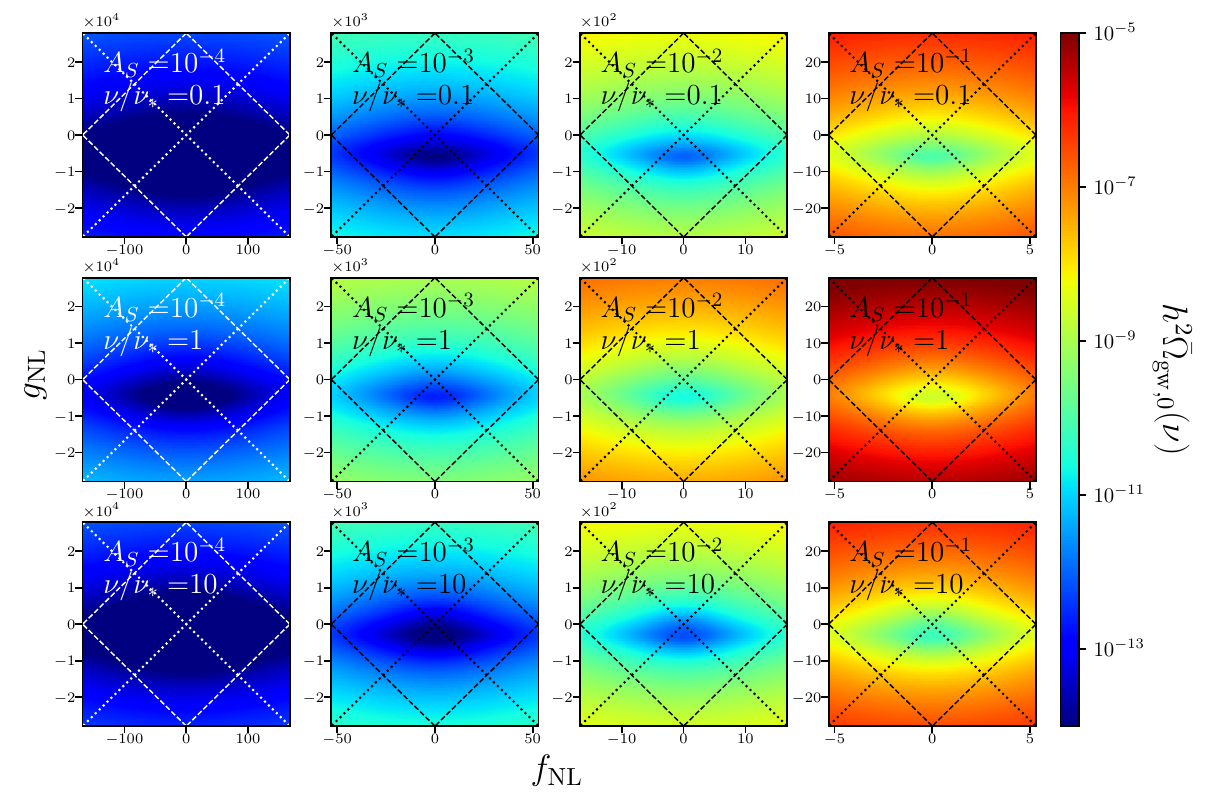}
    \caption{ The present-day energy-density fraction spectrum of SIGWs with respect to the primordial non-Gaussian parameters $\fnl$ and $\gnl$. Dotted lines represent $\gnl = \pm 5 \fnl / (3 \sqrt{A_S})$, which signify that the contribution from $\fnl$ to $\zeta$ is equal to that from $\gnl$. Dashed lines represent $3|\fnl|\sqrt{A_S}/5 + 9|\gnl|A_S/25 = 1$, which signify that the contribution from non-Gaussian terms to $\zeta$ is equal to that from the Gaussian term. }\label{fig:Omega_f-g}
\end{figure*}


To comprehensively illustrate the dependence of the \ac{SIGW} energy-density fraction spectrum on $\fnl$ and $\gnl$, we present an array of contour plots in Figure~\ref{fig:Omega_f-g}. 
The contour plots depict the energy-density fraction spectrum in the present Universe with $\fnl$ and $\gnl$ as variables. 
The value of $A_S$ remains constant within each column, and $\nu/\nu_\ast$ is fixed within each row. 
The same position on each panel corresponds to specific values of $\fnl\sqrt{A_S}$ and $\gnl A_S$.
Each panel is divided into four triangular regions by dotted lines. 
The left and right regions represent the contribution of the $\gnl$ term being smaller than that of the $\fnl$ term, ensuring perturbativity, while the top and bottom regions represent the opposite. 
The square regions surrounded by dashed lines represent the contributions of the primordial non-Gaussian terms being greater than that of the Gaussian term.
In each panel, it is evident that $h^2 \bar{\Omega}_{\uGW,0} (\nu)$ is more sensitive to $9 \gnl A_S / 25$ than $3 \fnl\sqrt{A_S} / 5$. 
Additionally, increasing $A_S$ consistently enhances the spectral amplitudes as expected. 
Although large non-Gaussianity typically leads to large spectral amplitudes in each panel, the position of minimal amplitude consistently appears in the regions of zero $\fnl$ and small negative $\gnl$.

\begin{figure}[htbp]
    \centering
    \includegraphics[width =1 \columnwidth]{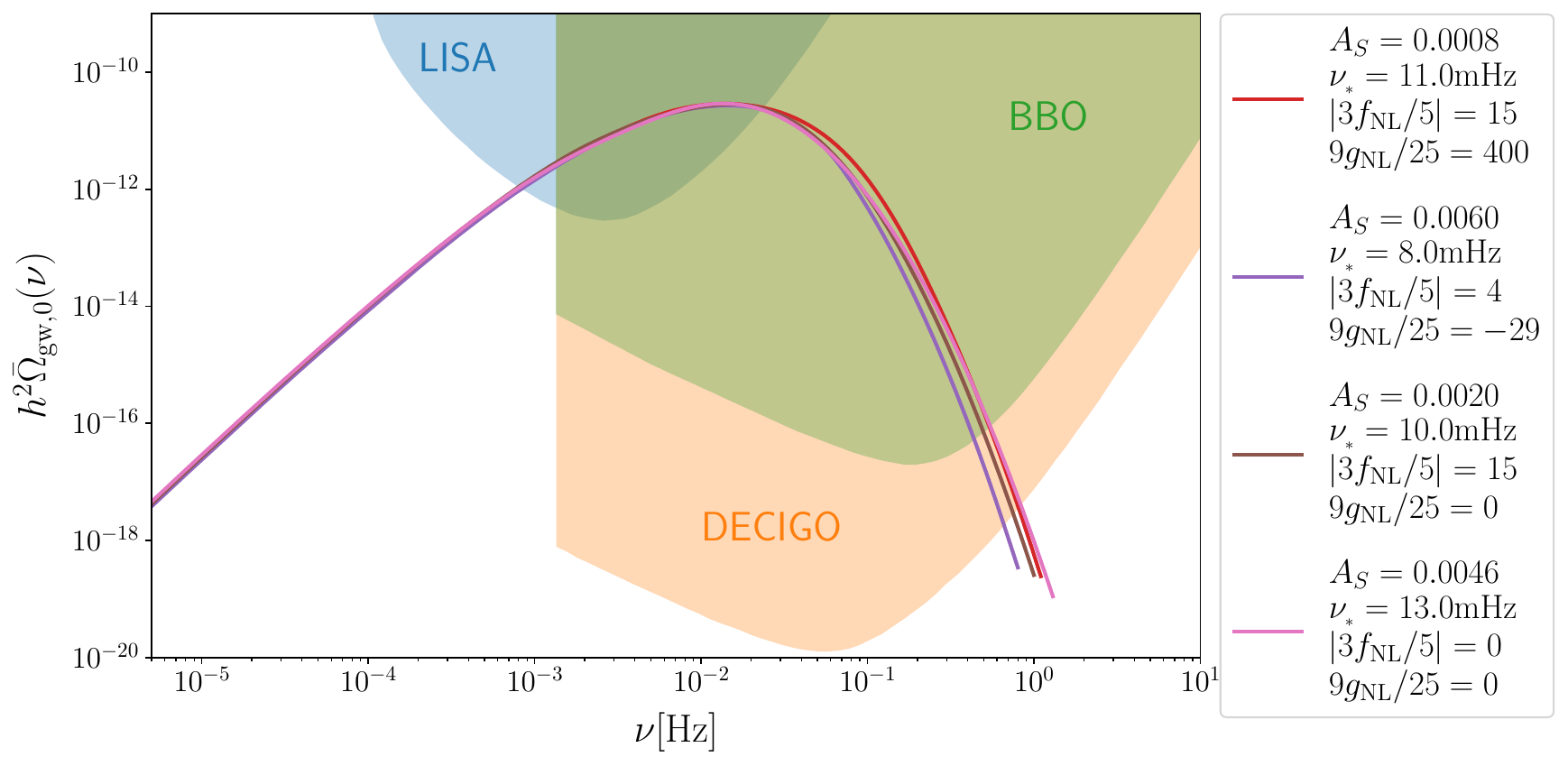}
    \caption{Energy-density fraction spectra of \ac{SIGW} in the present Universe compared with the sensitivity curves of \acp{LISA} (blue shaded region) \cite{Baker:2019nia,Smith:2019wny}, \acp{DECIGO} (orange shaded region) \cite{Seto:2001qf,Kawamura:2020pcg}, and \acp{BBO} (green shaded region) \cite{Crowder:2005nr,Smith:2016jqs}. 
    For all the cases, $\sigma=1$ is assumed for the spectra.
    }\label{fig:Omega_degeneracy}
\end{figure}

Our theoretical predictions are expected to be tested using space-borne gravitational-wave detectors and detector networks such as Taiji \cite{Hu:2017mde,Ren:2023yec}, Tianqin \cite{TianQin:2015yph,TianQin:2020hid,Zhou:2023rop}, \ac{LISA} \cite{LISACosmologyWorkingGroup:2022jok,Robson:2018ifk}, \ac{BBO} \cite{Crowder:2005nr,Harry:2006fi}, \ac{DECIGO} \cite{Sato:2017dkf,Kawamura:2020pcg}, and others. In Figure~\ref{fig:Omega_degeneracy}, we compare a range of theoretical results with the sensitivity curves of the latter three detectors. We consider parameter regimes that may be relevant to scenarios involving \acp{PBH} \cite{Hawking:1971ei,Carr:2020gox}. For \acp{PBH} within mass ranges detectable by space-borne detectors for \acp{SIGW} \cite{Saito:2009jt,Inomata:2018epa,Zhao:2022kvz}, they are considered as potential candidates for cold dark matter. Therefore, it is crucial to investigate the significant impact of primordial non-Gaussianity on the abundance of \acp{PBH} \cite{Byrnes:2012yx,Young:2013oia,Franciolini:2018vbk,Inomata:2020xad,Nakama:2016gzw,Yoo:2018kvb,Young:2014ana,Ferrante:2022mui}. From Figure~\ref{fig:Omega_degeneracy}, we observe that these detectors have the potential to shed light on this question in the future.

However, it is important to note that there exist significant degeneracies among the model parameters affecting the energy-density fraction spectra. In Figure~\ref{fig:Omega_degeneracy}, the four spectra appear to closely overlap, yet they correspond to entirely different parameter combinations. Breaking these degeneracies is crucial, and therefore, we propose a new approach to address this issue in Section~\ref{sec:aps}.

\section{Angular power spectrum}\label{sec:aps}

In this section, we extend our analysis to include the effects of large-scale inhomogeneities, which were previously neglected in Section~\ref{sec:ogw}. For the first time, we comprehensively investigate the inhomogeneities and anisotropies in \acp{SIGW} by incorporating both $\fnl$ and $\gnl$. We develop systematic formulas to calculate the angular power spectrum, allowing us to study the spatial variations of \acp{SIGW} on large scales. Similar to the observations of temperature anisotropies in \ac{CMB}, the study of \acp{SIGW} as a form of \ac{GWB} is expected to reveal inhomogeneities and anisotropies. These inhomogeneities can provide valuable insights into the nature of our Universe, similar to how the \ac{CMB} has led to precise measurements of six fundamental parameters in the concordance model of cosmology.

In the existing literature, there have been three relevant works addressing the same topic. In Ref.~\cite{Bartolo:2019zvb}, the authors examined the angular power spectrum of \acp{SIGW} by incorporating $\fnl$ for the first time. Subsequently, we conducted the first comprehensive analysis of the angular power spectrum by incorporating $\fnl$ in our previous works \cite{Li:2023qua, Wang:2023ost}. Compared to the former work, the latter studies considered all contributions of $\fnl$. In this current work, we extend our previous investigations by including all contributions of $\gnl$ in the analysis of the angular power spectrum of \acp{SIGW}.

The presence of large-scale inhomogeneities in \acp{SIGW} can be attributed to two main factors, as discussed in Refs.~\cite{Bartolo:2019zvb, Li:2023qua, Wang:2023ost}. Firstly, the initial inhomogeneities arise due to the coupling between long-wavelength and short-wavelength modes. The long-wavelength modes redistribute the energy density of gravitational waves induced by the short-wavelength modes, resulting in the formation of initial inhomogeneities on large scales. Secondly, similar to the analysis of \ac{CMB}, inhomogeneities can also arise from propagation effects caused by the presence of long-wavelength gravitational potentials, such as the \ac{SW} effect \cite{Sachs:1967er}. In our analysis, we consider both of these factors. It is important to note that our evaluation adopts the cosmological principle, which assumes statistical homogeneity and isotropy on large scales.

Similar to the measurement of temperature fluctuations in relic photons \cite{Seljak:1996is}, the large-scale inhomogeneities in \acp{SIGW} within an observed region centered at position $\bx$ and conformal time $\eta'$ can be quantified using the density contrast $\delta_{\uGW}$. This density contrast is defined as follows 
\begin{equation}\label{eq:delta-def}
    \delta_\uGW (\eta',\bx,\bq) = 4\pi \frac{\omega_\uGW (\eta',\bx,\bq)}{\bar{\Omega}_\uGW (\eta',q)} - 1 \ .
\end{equation}
By averaging over a large number of such regions, the resulting averaged energy-density fraction spectrum can be expressed as 
\begin{equation}\label{eq:bgdomega}
    \bar{\Omega}_{\uGW}(\eta',q) = 4\pi \langle \omega_{\uGW}(\eta',\bx,\bq) \rangle \ ,
\end{equation}
which yields results that are identical to those obtained in Section~\ref{sec:ogw}. Similarly, we find that $\langle\delta_\uGW (\eta',\bx,\bq)\rangle = 0$, indicating that, on average, the density contrast is zero. The energy-density full spectrum $\omega_{\uGW}(\eta',\bx,\bq)$ is defined in relation to the energy density $\rho_\uGW (\eta',\bx)$ as follows \cite{Maggiore:1999vm} 
\begin{equation}\label{eq:omega-def}
    \rho_\uGW (\eta',\bx) = \rho_\uc(\eta') \int \ud \ln q \, \ud^{2} \hat{\bq} \, \omega_\uGW (\eta',\bx,\bq) \ ,
\end{equation}
where the symbol $\hat{\bq}$ represents the unit directional vector, given by $\hat{\bq}=\bq/q$. Additionally, the energy density of the inhomogeneous \acp{SIGW} is defined as 
\begin{equation}
\rho_\uGW (\eta,\bx) = \frac{\mpl^2 }{16a^{2}(\eta)} \, \overbar{\partial_l h_{ij}(\eta,\bx) \partial_l h_{ij}(\eta,\bx)} \ .
\end{equation}
Hence, we can express the initial energy-density full spectrum $\omega_\uGW(\eta,\bx,\bq)$ in the following manner 
\begin{equation}\label{eq:omega-h}
     \omega_\uGW(\eta,\bx,\bq) 
     = - \frac{q^3}{48 \cH^2} \int \frac{\ud^3 \bk}{(2\pi)^{3}} e^{i\bk\cdot\bx} 
        \left[\left(\bk-\bq\right) \cdot \bq \right] \, 
        \overbar{h_{ij}(\eta,\bk-\bq)
        h_{ij}(\eta,\bq)}\ ,
\end{equation}
which depends not only on the magnitude of $q$ but also on the unit directional vector $\hat{\bq}$. In the given equations, $\bq$ represents the comoving momentum of \acp{SIGW} associated with short wavelengths, while $\bk$ is associated with a Fourier mode of the inhomogeneities in \acp{SIGW}, corresponding to long wavelengths. Hence, in the subsequent analysis, we will consider the regime where $q \gtrsim \cH^{-1} \gg k$.

The evolution of the density contrast is described by the Boltzmann equation \cite{Dodelson:2003ft}. In the context of relic photons, this equation has been solved using the line-of-sight approach \cite{Seljak:1996is}. More recently, the same approach has been applied to the study of general \acp{GWB} \cite{Contaldi:2016koz}. Subsequently, it has been used to investigate two-point and three-point correlators for cosmological \acp{GWB} \cite{Bartolo:2019oiq,Bartolo:2019yeu}. In the context of anisotropies in \acp{SIGW}, the authors of Refs.~\cite{Bartolo:2019zvb,Li:2023qua,Wang:2023ost} have explored the significant influence of $\fnl$. Other relevant works can be found in Refs.~\cite{ValbusaDallArmi:2020ifo,Dimastrogiovanni:2021mfs,LISACosmologyWorkingGroup:2022kbp,LISACosmologyWorkingGroup:2022jok,Unal:2020mts,Malhotra:2020ket,Carr:2020gox,Cui:2023dlo,Malhotra:2022ply,ValbusaDallArmi:2023nqn}.

Based on the line-of-sight solution to the Boltzmann equation, as derived in Appendix~\ref{sec:Boltz}, we demonstrate that the present density contrast $\delta_{\uGW,0}(\bq)=\delta_{\uGW}(\eta_{0},\bx_{0},\bq)$ incorporates both the initial inhomogeneities and the effects of propagation, i.e.,  \cite{Bartolo:2019zvb,Li:2023qua,Wang:2023ost}
\begin{equation}\label{eq:delta-0}
\delta_{\uGW,0}(\bq) = \delta_\uGW (\eta,\bx,\bq) + \left[4-n_{\uGW} (\nu)\right] \Phi (\eta, \bx)\ ,
\end{equation}
where the value of $q$ is given by the equation $q = 2\pi \nu$, with $\nu$ representing the frequency, and the determination of the index of the energy-density fraction spectrum is based on
\begin{equation}\label{eq:ngw-def}
    n_{\uGW} (\nu) = \frac{\partial\ln \bar{\Omega}_{\uGW,0} (\nu)}{\partial\ln \nu} \simeq \frac{\partial\ln \bar{\Omega}_\uGW (\eta,q)}{\partial\ln q}\Big|_{q=2\pi\nu} \ .
\end{equation}
The first term on the right-hand side of Eq.~(\ref{eq:delta-0}) represents the initial inhomogeneities, while the second term corresponds to the \ac{SW} effect \cite{Sachs:1967er}. 
Furthermore, the \ac{ISW} effect \cite{Sachs:1967er}, as shown in Ref.~\cite{Bartolo:2019zvb}, is found to be less significant compared to the \ac{SW} effect.
In this study, we disregard the \ac{ISW} effect, but it can be easily accounted for if necessary.
When considering the long-wavelength modes that reentered the Hubble horizon during matter domination, we characterize the \ac{SW} effect using the Bardeen potential on large scales, i.e.,
\begin{equation}\label{eq:SWe}
    \Phi (\eta,\bx) = \frac{3}{5} \int \frac{\ud^{3}\bk}{(2\pi)^{3/2}} e^{i\bk\cdot\bx} \zeta_{gL}(\bk) \ ,
\end{equation}
where $\zeta_{gL}$ represents the primordial curvature perturbations of long-wavelength as introduced in Eq.~(\ref{eq:shortlong}).
One of the remaining tasks is to assess the initial inhomogeneities $\delta_{\uGW}(\eta,\bx,\bq)$.
This will be addressed in the upcoming subsection.

Now, it is imperative to provide some insights into the physical implications of Eq.~(\ref{eq:delta-0}), or more broadly, Eq.~(\ref{eq:deltaGW-CGW}).
Firstly, let us consider the propagation effects, namely the \ac{SW} and \ac{ISW} effects.
It has been established that these effects are identical for both the \ac{CMB} and \acp{GWB}, as massless photons and gravitons follow the same perturbed geodesics \cite{Contaldi:2016koz,Bartolo:2019oiq,Bartolo:2019yeu}.
Moreover, it is crucial to note that the second comment pertains to the initial inhomogeneities, which differ for the \ac{CMB} and \acp{GWB}.
If any initial inhomogeneities existed in the \ac{CMB}, they have been completely wiped out due to frequent scatterings between photons and electrons in the early Universe \cite{Dodelson:2003ft}. 
In contrast, the initial inhomogeneities in \acp{GWB} persist, as the early Universe is transparent to \acp{GW} \cite{Bartolo:2018igk,Flauger:2019cam}.
Consequently, \acp{GWB} have the potential to serve as powerful probes of early-Universe physics.

The reduced angular power spectrum is widely used to describe the statistical properties of the anisotropies in \acp{SIGW}. This spectrum is defined by the two-point correlator of the present density contrast, namely,
\begin{equation}\label{eq:Ctilde-def}
    \langle\delta_{\uGW,0,\ell m}(2\pi\nu) \delta_{\uGW,0,\ell' m'}^\ast(2\pi\nu)\rangle
    = \delta_{\ell \ell'} \delta_{mm'} \tilde{C}_\ell (\nu)\ ,
\end{equation}
where $\delta_{\text{GW},0}(\mathbf{q})$, defined in Eq.~(\ref{eq:delta-0}), is expanded using spherical harmonics, that is,
\begin{equation}\label{eq:shsai}
    \delta_{\uGW,0}(\bq) = \sum_{\ell m} \delta_{\uGW,0,\ell m}(q) Y_{\ell m}(\hat{\bq})\ . 
\end{equation} 
By inserting Eq.~(\ref{eq:delta-0}) into Eq.~(\ref{eq:shsai}) and subsequently Eq.~(\ref{eq:Ctilde-def}), the reduced angular power spectrum for \acp{SIGW} can be readily obtained.

Practically, evaluating the terms that involve only the propagation effects, such as the \ac{SW} effect shown in Eq.~(\ref{eq:SWe}), is somewhat straightforward. However, when we consider the initial inhomogeneities, which are related to Eq.~(\ref{eq:omega-h}), we need to evaluate the multi-point correlator $\langle h^{4} \rangle \sim \langle \zeta^{8} \rangle$. This involves, at most, the 24-point correlator of the form $\langle \zeta_{g}^{24} \rangle$.
Expanding these correlators straightforwardly using Wick's theorem proves to be particularly challenging. Therefore, we also adopt a Feynman-like diagrammatic technique to simplify the evaluation in the following.

\subsection{Feynman-like Rules and Diagrams}

In this section, we exclusively concentrate on the initial inhomogeneities at large scales, arising from the spatial modulation of energy density caused by the long-wavelength modes.

By substituting Eqs.~(\ref{eq:Fnl-Gnl-def}) and (\ref{eq:shortlong}) into Eq.~(\ref{eq:omega-h}), we can approximately express the effect of long-wavelength modulation as follows
\begin{equation}\label{eq:ogwexpand}
    \omega_{\uGW}(\eta,\bx,\bq) \sim \langle\zeta^4\rangle_{\bx} \sim \langle\zeta_{S}^4\rangle_{\bx} + \mathcal{O}(\zeta_{gL})  \fnl \langle\zeta_{gS}\zeta_{S}^3\rangle_{\bx} +  \mathcal{O}(\zeta_{gL})  \gnl \langle\zeta_{gS}^{2}\zeta_{S}^3\rangle_{\bx} \ , 
\end{equation}
where $\zeta_{S}$ represents the short-wavelength modes associated with the primordial non-Gaussianity, and the subscript $_\bx$ denotes an ensemble average within the observed region centered at $\bx$.
The term $\langle\zeta_{S}^4\rangle_{\bx}$ represents the monopole, denoted as $\bar{\Omega}(\eta,q)$, while the last two terms capture the effects of long-wavelength modulation, indicating that $\delta_{\uGW}(\eta,\bx,\bq)\sim\zeta_{gL}$ as defined in Eq.~(\ref{eq:delta-def}).
In this analysis, we have neglected higher-order terms in $\zeta_{gL}$ due to the negligible dimensionless power spectrum of long-wavelength modes compared to that of short-wavelength modes, specifically $\Delta_L^2 \ll \Delta_S^2$.
Therefore, at the linear order in $\Delta_L^2\sim\langle\zeta_{gL}^{2}\rangle$, we obtain the angular power spectrum defined in Eq.~(\ref{eq:Ctilde-def}) as follows
\begin{equation}\label{eq:deltaomega}
\langle\delta_{\uGW}(\eta,\bx,\bq)\delta_{\uGW}(\eta,\bx',\bq')\rangle \sim \langle\omega_\uGW (\eta,\bx,\bq)\omega_\uGW (\eta,\bx',\bq')\rangle^{\mathcal{O}(\Delta_{L}^{2})} \sim  \Delta_L^2 \langle...\rangle_{\bx} \langle...\rangle_{\bx'} \ ,
\end{equation} 
where $\langle...\rangle_{\bx}$ represents the combination of angle brackets in the last two terms of Eq.~(\ref{eq:ogwexpand}).
In fact, it is more convenient to use $\langle\omega_\uGW (\eta,\bx,\bq)\omega_\uGW (\eta,\bx',\bq')\rangle$ when establishing the Feynman-like rules.

\begin{figure}
    \centering
    \includegraphics[width =1 \columnwidth]{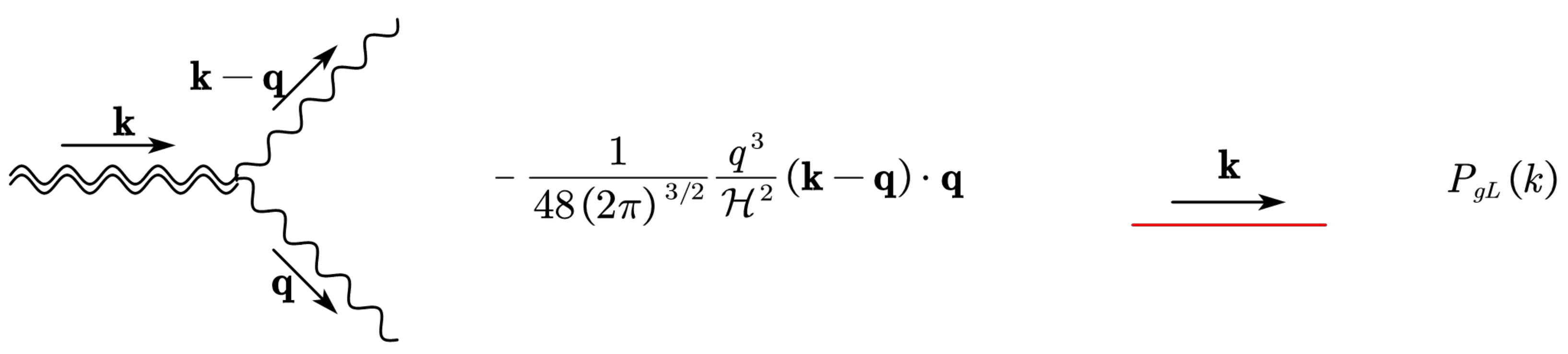}
    \caption{The Feynman-like rules complementary to those of Figure~\ref{fig:FD_Rules}. Left panel represents the $\omega$-vertex, for which double wavy lines denote $\omega_{\uGW}$. Right panel shows the primordial power spectrum of long-wavelength modes in red solid line. 
    }\label{fig:FD_Rules_add}
\end{figure}

Analogous to the Feynman-like diagrammatic technique utilized in Section~\ref{sec:ogw}, we reformulate the expression $\langle\omega_\uGW (\eta,\bx,\bq)\omega_\uGW (\eta,\bx',\bq')\rangle$ as a series of Feynman-like diagrams.
Previous studies that have employed this technique to analyze the angular power spectrum for \acp{SIGW} can be found in Refs~\cite{Bartolo:2019zvb,Li:2023qua,Wang:2023ost}.
By substituting Eqs.~(\ref{eq:omega-h}) into the two-point correlator and incorporating Eqs.~(\ref{eq:Fnl-Gnl-def}) and (\ref{eq:shortlong}), we can explicitly introduce two additional Feynman-like rules, as summarized in Figure~\ref{fig:FD_Rules_add}.
The left panel represents the $\omega$-vertex, while the right panel depicts the propagator of the long-wavelength modes.
These supplementary Feynman-like rules complement those shown in Figure~\ref{fig:FD_Rules} in Section~\ref{sec:ogw}.
It is important to note that the Feynman-like rules for the vertices in Figure~\ref{fig:FD_Rules} remain unchanged, irrespective of the colors of the solid lines. This indicates that these vertices are independent of the propagators of the short-wavelength and long-wavelength modes.

\begin{figure}
    \centering
    \includegraphics[width =1 \columnwidth]{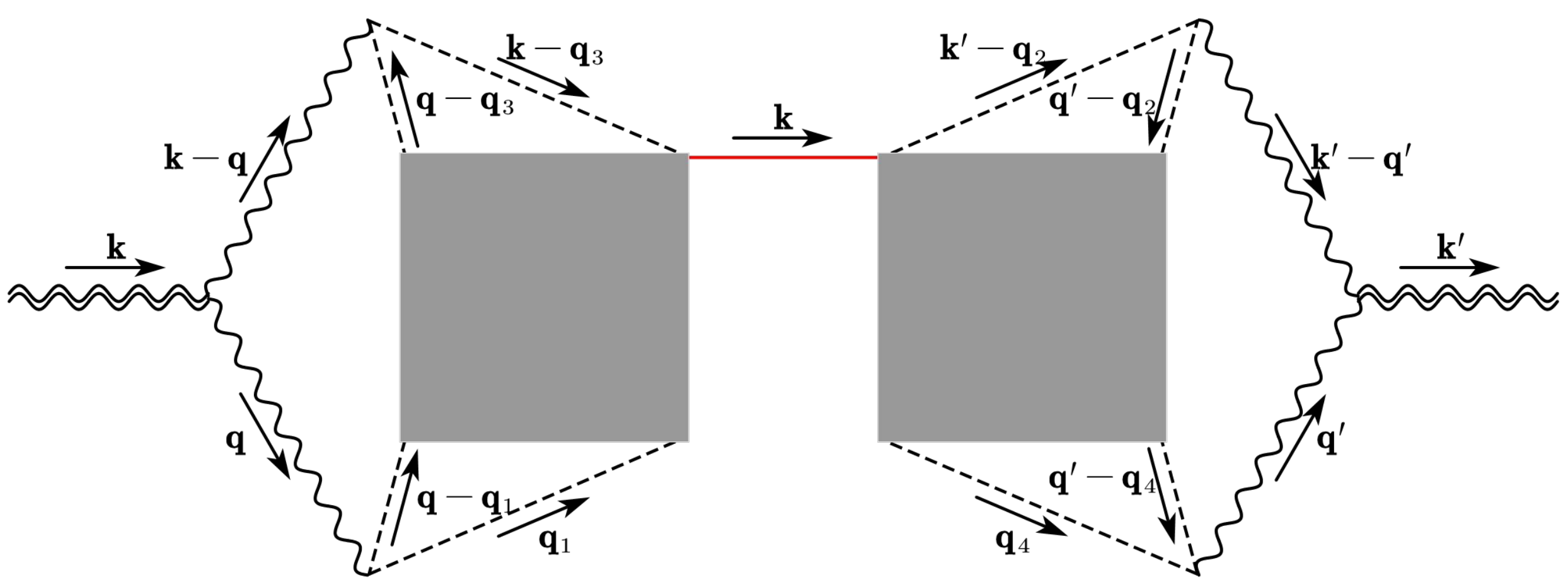}
    \caption{The Feynman-like diagram of the angular power spectrum of \acp{SIGW}. The shadow squares should be replaced with the panels in \cref{fig:Feynman_Diagrams}. }\label{fig:FD_Cl_Frame}
\end{figure}

Corresponding to Eq.~(\ref{eq:deltaomega}), the Feynman-like diagrams are fully depicted in Figure~\ref{fig:FD_Cl_Frame} and Figure~\ref{fig:Feynman_Diagrams}.
To obtain an individual diagram, we need to replace the two shaded squares in Figure~\ref{fig:FD_Cl_Frame} with the panels from Figure~\ref{fig:Feynman_Diagrams}, respectively.
Here, we have neglected the disconnected diagrams, which correspond to the monopole $\bar{\Omega}_{\uGW}(\eta,q)$, and the diagrams of higher order in $\Delta_{L}^{2}$.
For a given Feynman-like diagram, there is a single ``non-Gaussian bridge'', represented by the red line in Figure~\ref{fig:FD_Cl_Frame}, that connects the initial inhomogeneities located at two different positions separated by a large distance.
The contribution of this diagram to the two-point correlator is proportional to $\bar{\Omega}_{\uGW}^{X}\langle\zeta_{gL}\zeta_{gL}\rangle\bar{\Omega}_{\uGW}^{X'}$.
This can be equivalently expressed as a correlator between $\bar{\Omega}_{\uGW}^{X}\zeta_{gL}$ and $\bar{\Omega}_{\uGW}^{X'}\zeta_{gL}$, which represent the contributions of long-wavelength modulation to $\omega_{\uGW}(\eta,\bx,\bq)$.
By summing over all the Feynman-like diagrams, we obtain the total effect of long-wavelength modulation on $\omega_{\uGW}(\eta,\bx,\bq)$, which is analogous to the last two terms in Eq.~(\ref{eq:ogwexpand}).

Precisely speaking, we can reformulate Eq.~(\ref{eq:ogwexpand}) into a novel expression of the following form
\begin{equation}\label{eq:omega-result}
    \omega_{\uGW}(\eta,\bx,\bq) = \frac{\bar{\Omega}_{\uGW}(\eta,q)}{4\pi} + \frac{\Omega_{\mathrm{ng}}(\eta,q)}{4\pi} \int \frac{\ud^{3}\bk}{(2\pi)^{3/2}} e^{i\bk\cdot\bx} \zeta_{gL}(\bk)\ , 
\end{equation}
where for the sake of convenience, we introduce a new quantity $\Omega_{\mathrm{ng}}(\eta,q)$ that follows the form
\begin{eqnarray}\label{eq:Ong-def}
    \Omega_\ung (\eta,q) 
    &=& \frac{6\fnl}{5} \left(4 \bar{\Omega}_\uGW^{(0,0)} + 3 \bar{\Omega}_\uGW^{(0,1)} + 2 \bar{\Omega}_\uGW^{(1,0)} + 2 \bar{\Omega}_\uGW^{(0,2)} +  \bar{\Omega}_\uGW^{(1,1)} +  \bar{\Omega}_\uGW^{(0,3)}\right)\nonumber\\
    && + \frac{9 \gnl}{5 \fnl} \left(2 \bar{\Omega}_\uGW^{(1,0)} + 2 \bar{\Omega}_\uGW^{(1,1)} + 4 \bar{\Omega}_\uGW^{(2,0)} + 2  \bar{\Omega}_\uGW^{(1,2)}\right)\ .
\end{eqnarray} 
The next crucial step is to derive the aforementioned equation.

\begin{figure}
    \centering
    \includegraphics[width =0.35 \columnwidth]{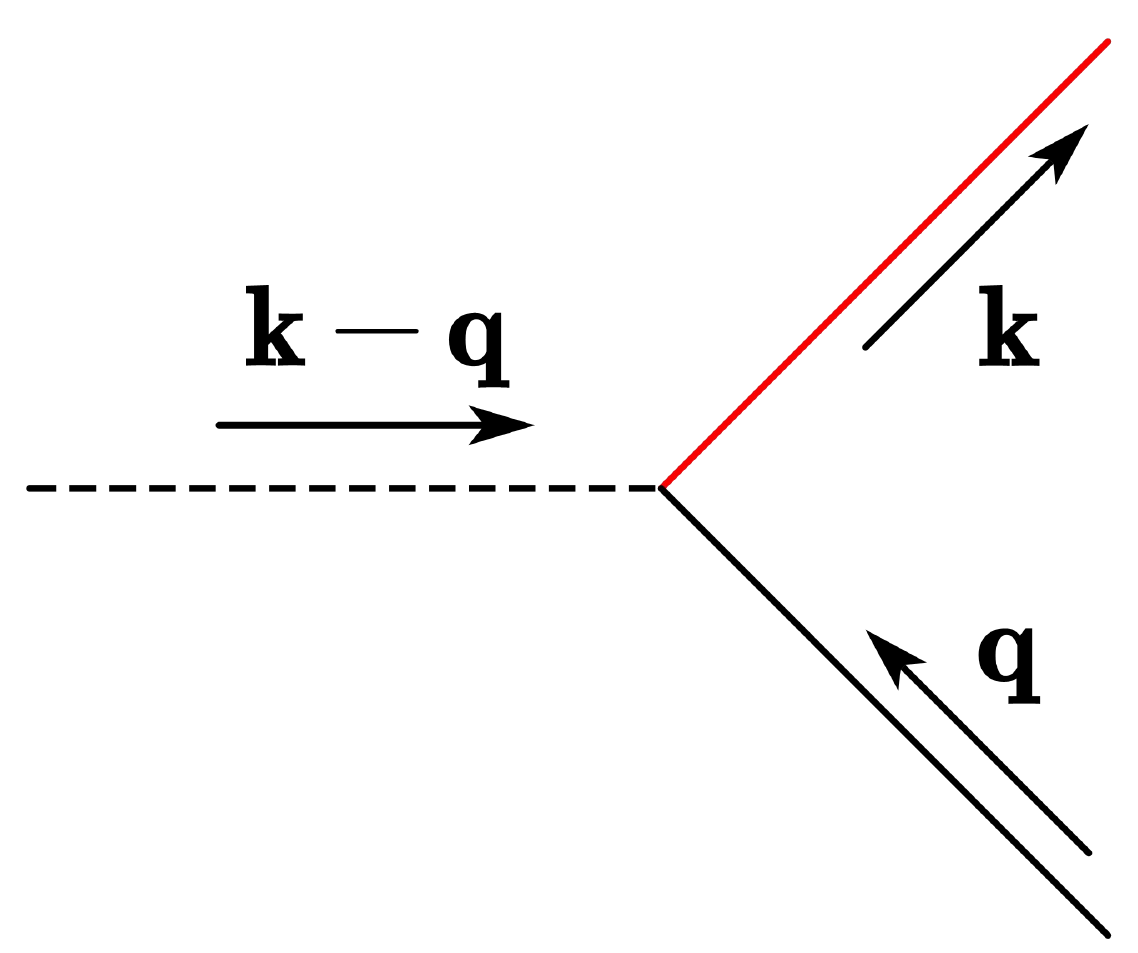}
    \hfil
    \includegraphics[width =0.4 \columnwidth]{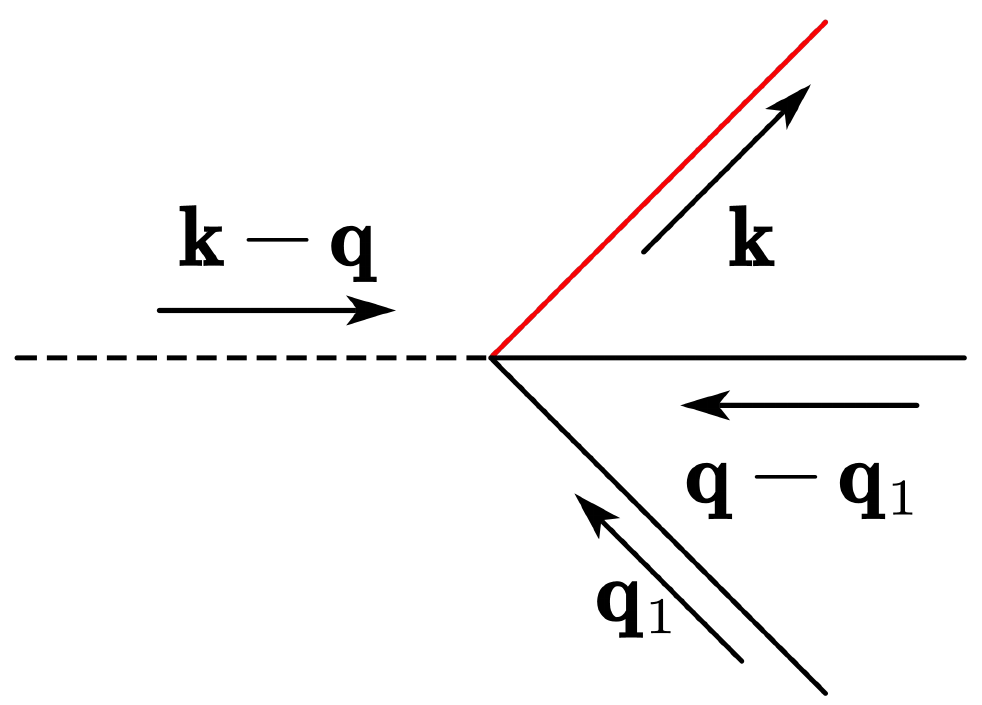}
    \caption{Vertices involved in the non-Gaussian bridge. 
    }\label{fig:v-transform}
\end{figure}

To simplify the process, we adopt a diagrammatic approach to derive Eq.~(\ref{eq:Ong-def}).
This approach is equivalent to the traditional expansion outlined in Eq.~(\ref{eq:ogwexpand}), utilizing Wick's theorem.
As illustrated in the left panel of Figure~\ref{fig:v-transform}, the inclusion of the non-Gaussian bridge leads to the transformation of the Gaussian vertex into the $\fnl$-vertex.
This transformation also results in a doubling of the symmetric factor, leading to an additional factor of $2$.
In this case, the contribution from Diagram-$X$ to $\Omega_{\ung}$ is given by $2 N^X_\mathrm{Gau}[(3\fnl/5)/1] \bar\Omega_\uGW^X$, where $N^X_\mathrm{Gau}$ represents the number of Gaussian vertices.
As shown in the right panel of Figure~\ref{fig:v-transform}, the inclusion of the non-Gaussian bridge converts the $\fnl$-vertex into the $\gnl$-vertex.
This transformation also triples the symmetric factor, resulting in an additional factor of $3$.
In this scenario, the contribution from Diagram-$X$ to $\Omega_{\ung}$ is given by $3 N^X_{\fnl}[(9\gnl/25)/(3\fnl/5)] \bar\Omega_\uGW^X$, where $N^X_{\fnl}$ represents the number of $\fnl$-vertices.
The count of various vertices is determined by the category associated with Diagram-X, which implies $N^X_\mathrm{Gau} = 4 - 2a -b$ and $N^X_{\fnl} = 2a$.
By employing Table~\ref{tab:order-FD} as a dictionary, we can readily obtain Eq.~\eqref{eq:Ong-def}.
It is worth noting that the ten diagrams of the $(0,4)$-category do not contribute to $\Omega_{\mathrm{ng}}$, as we are considering non-Gaussianity up to the order of $\gnl$. This suggests a decoupling of modes with short-wavelength and long-wavelength for these diagrams.
Please note that we have revised the definition of $\Omega_\ung$ compared to our previous works \cite{Li:2023qua,Wang:2023ost}, where we only considered $\fnl$.
When $\gnl$ is zero, our new definition of $\Omega_\ung$ reverts back to the old definition if we divide the new definition by $(3\fnl/5)$.

\subsection{Formulae for Angular Power Spectrum}

Based on the aforementioned findings, it is now necessary to derive the equation for the angular power spectrum of \acp{SIGW}, as defined in Eq.~(\ref{eq:Ctilde-def}). By substituting Eq.~(\ref{eq:omega-result}) into Eq.~\eqref{eq:delta-def}, we obtain the initial density contrast, denoted as $ \delta(\bm{k})$, which can be expressed as
\begin{equation}\label{eq:delta-result}
\delta_{\uGW}(\eta,\bx,\bq) = \frac{\Omega_{\mathrm{ng}}(\eta,q)}{\bar{\Omega}_{\uGW}(\eta,q)} \int \frac{\ud^{3}\bk}{(2\pi)^{3/2}} e^{i\bk\cdot\bx} \zeta_{gL}(\bk)\ .  
\end{equation}
Subsequently, by further substituting Eq.~(\ref{eq:SWe}) and Eq.~(\ref{eq:delta-result}) into Eq.~(\ref{eq:delta-0}), we can derive the present density contrast, i.e., 
\begin{equation}\label{eq:delta-0-result}
\delta_{\uGW,0}(\bq) = \biggl\{
            \frac{\Omega_{\mathrm{ng}} (\eta,2\pi\nu)}{\bar{\Omega}_\uGW (\eta,2\pi\nu)}
            + \frac{3}{5} \bigl[4 - n_{\uGW} (\nu)\bigr]
        \biggr\} 
        \int \frac{\ud^{3}\bk}{(2\pi)^{3/2}} e^{i\bk\cdot\bx} \zeta_{gL}(\bk)\ ,
\end{equation}
where the expression for $\Omega_{\mathrm{ng}}$ is given in Eq.~(\ref{eq:Ong-def}).
Ultimately, by substituting Eq.~(\ref{eq:delta-0-result}) into Eq.~(\ref{eq:shsai}) and subsequently Eq.~(\ref{eq:Ctilde-def}), we derive the fundamental equation of this study for the reduced angular power spectrum, i.e.,
\begin{equation}\label{eq:reduced-APS}
    \tilde{C}_\ell (\nu) 
    = \frac{2\pi\Delta^2_L}{\ell (\ell+1)} 
        \left\{
            \frac{\Omega_{\mathrm{ng}} (\eta,2\pi\nu)}{\bar{\Omega}_\uGW (\eta,2\pi\nu)}
            + \frac{3}{5} \left[4 - n_{\uGW} (\nu)\right]
        \right\}^2\ ,
\end{equation}
which can be further reformulated as the angular power spectrum, i.e.,
\begin{equation}\label{eq:cellsai}
C_{\ell}(\nu) =  \left[\frac{\bar{\Omega}_{\uGW,0}(\nu)}{4\pi}\right]^{2} \tilde{C}_{\ell}(\nu)  \ .
\end{equation}
Eq.~\eqref{eq:reduced-APS} demonstrates that the angular power spectrum is influenced by the initial inhomogeneities, propagation effects, and their intersections.
Similar to the analysis of \ac{CMB}, we can define the root-mean-squared (rms) energy density for the anisotropies in \acp{SIGW} as $\left[\ell(\ell+1)C_{\ell}(\nu) / 2\pi\right]^{1/2}$.
This quantity is independent of the angular multipoles $\ell$, but varies with the frequencies of \ac{SIGW}.

We would like to provide further comments regarding the multipole dependence of the reduced angular power spectrum, as illustrated by Eq.~(\ref{eq:reduced-APS}). It is worth noting that we have observed a distinctive multipole dependence, where $C_{\ell}\propto[\ell(\ell+1)]^{-1}$, for \acp{SIGW}. This behavior has also been reported in the literature \cite{Bartolo:2019zvb, Li:2023qua, Wang:2023ost, Dimastrogiovanni:2022eir}.
In contrast, for \acp{GWB} generated by \acp{BBH}, the multipole dependence has been found to be roughly proportional to $(\ell+1/2)^{-1}$ \cite{Cusin:2018rsq, Cusin:2017fwz, Cusin:2019jhg, Cusin:2019jpv, Jenkins:2018kxc, Jenkins:2018uac, Jenkins:2019nks, Contaldi:2016koz, Wang:2021djr, Mukherjee:2019oma, Bavera:2021wmw, Bellomo:2021mer}. On the other hand, for a \ac{GWB} originating from cosmic string loops, the multipole dependence has been shown to be proportional to $\ell^{0}$ \cite{Jenkins:2018nty, Kuroyanagi:2016ugi, Olmez:2011cg}. 
Therefore, the multipole dependence plays a crucial role in differentiating \acp{SIGW} from these other sources of \acp{GWB} (for reviews, see Ref.~\cite{LISACosmologyWorkingGroup:2022kbp}). It is important to note that there are other cosmological sources of gravitational waves, such as inflation \cite{Adshead:2020bji, Dimastrogiovanni:2021mfs, Dimastrogiovanni:2019bfl, Jeong:2012df, ValbusaDallArmi:2023nqn}, domain walls \cite{Liu:2020mru}, first-order phase transitions \cite{Li:2022svl,Li:2021iva, Domcke:2020xmn, Jinno:2021ury, Geller:2018mwu, Kumar:2021ffi, Racco:2022bwj}, and preheating \cite{Bethke:2013aba, Bethke:2013vca}, which predict angular power spectra with a similar $\tilde{C}_{\ell}\sim\ell^{-2}$ dependence as \acp{SIGW}.
To distinguish \acp{SIGW} from these sources, cross-correlation studies between \acp{GWB} and various observables, such as \ac{CMB} \cite{Dimastrogiovanni:2021mfs, Cusin:2018rsq, Ricciardone:2021kel, Malhotra:2020ket, Braglia:2021fxn, Capurri:2021prz, Dimastrogiovanni:2022eir,Galloni:2022rgg,Ding:2023xeg,Cyr:2023pgw}, \ac{LSS} \cite{Cusin:2018rsq, Canas-Herrera:2019npr, Alonso:2020mva, Yang:2020usq, Yang:2023eqi, Libanore:2023ovr, Bosi:2023amu, Balaudo:2022znx}, and 21cm lines \cite{Scelfo:2021fqe, Seto:2005tq}, have been proposed.
Furthermore, it is worth mentioning that apart from the information contained in the characteristic energy density $\bar{\Omega}_{\uGW}$, there is additional information embedded in the frequency dependence of $C_{\ell}$. This frequency dependence could potentially serve as a valuable tool for further distinguishing between different sources of \acp{GW} \cite{Li:2023qua, Dimastrogiovanni:2022eir, Cui:2023dlo, ValbusaDallArmi:2023ydl}. We will now proceed to demonstrate this frequency dependence in the following discussion.

\subsection{Numerical Results}

We present numerical results illustrating the impact of both $\fnl$ and $\gnl$ on the angular power spectrum of \acp{SIGW}. By studying this spectrum, we are able to reexamine the issue of degeneracies in the model parameters, which can be effectively resolved.

\begin{figure}[htbp]
    \centering
    \includegraphics[width =1 \columnwidth]{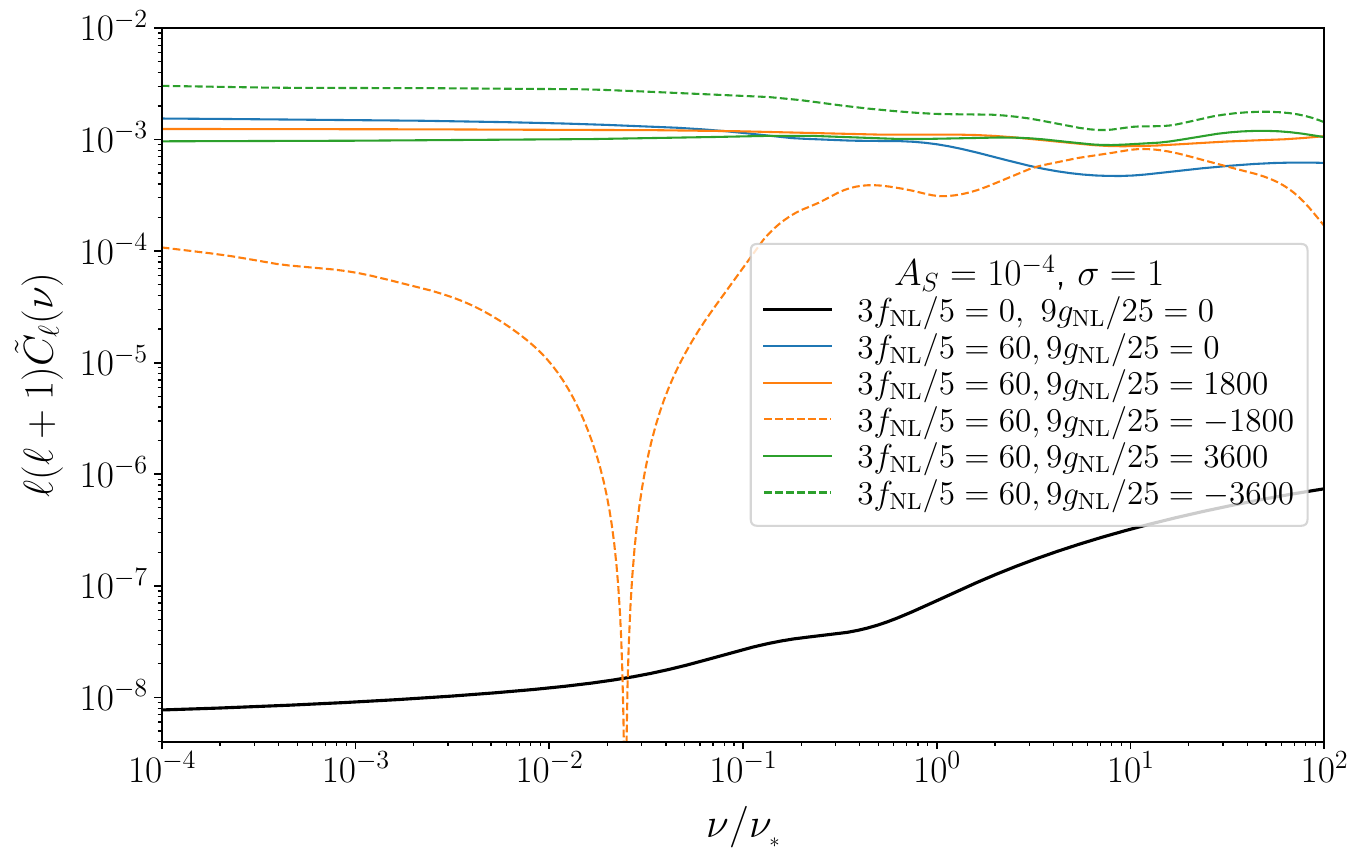}
    \caption{Reduced angular power spectrum for the anisotropies in SIGWs.  }\label{fig:Cl_tilde}
\end{figure}

During the radiation-dominated era of the early Universe, we investigate the reduced angular power spectrum of \acp{SIGW}. This spectrum is defined in Eq.~(\ref{eq:reduced-APS}). 
We keep $A_{S}=10^{-4}$ and $\sigma=1$ fixed, while varying the values of $\fnl$ and $\gnl$. 
Throughout this work, we let $\sigma=1$ for the simplicity. Variations of $\sigma$ may change the following numerical results, but do not change our conclusions. 
To visually represent our numerical results, we present the plot of $\ell(\ell+1)\tilde{C}_{\ell}(\nu)$ as a function of $\nu/\nu_\ast$ in Figure~\ref{fig:Cl_tilde}.
The black solid curve corresponds to the results in the Gaussian case, where the contributions are solely due to the propagation effects. 
The colored curves indicate the additional contributions from the primordial non-Gaussianity. 
We observe that such non-Gaussianity can generate significant anisotropies, with the spectral amplitude reaching values as high as $\sim10^{-3}$. 
Notably, this amplitude already exceeds that of the \acp{GWB} generated by inspiralling \acp{BBH} in the \ac{LISA} band, where $\tilde{C}_{\ell}\lesssim\mathcal{O}(10^{-4})$ \cite{Wang:2021djr,Cusin:2019jhg,Capurri:2022lze}. 
This suggests that the anticipated signal is comparable to these astrophysical foregrounds. 
Furthermore, we find that the resulting enhancement, when compared to the Gaussian case, is five orders of magnitude. 
In this study, we specifically focus on the parameter $\gnl$, as the parameter $\fnl$ has already been extensively investigated in our previous works \cite{Li:2023qua,Wang:2023ost}. 
To illustrate our findings, we fix $3\fnl/5=60$ and vary $\gnl$. 
We observe that the sign of $\gnl$ can significantly impact both the spectral amplitude and the spectral index.

\begin{figure}[htbp]
    \centering
    \includegraphics[width =1 \columnwidth]{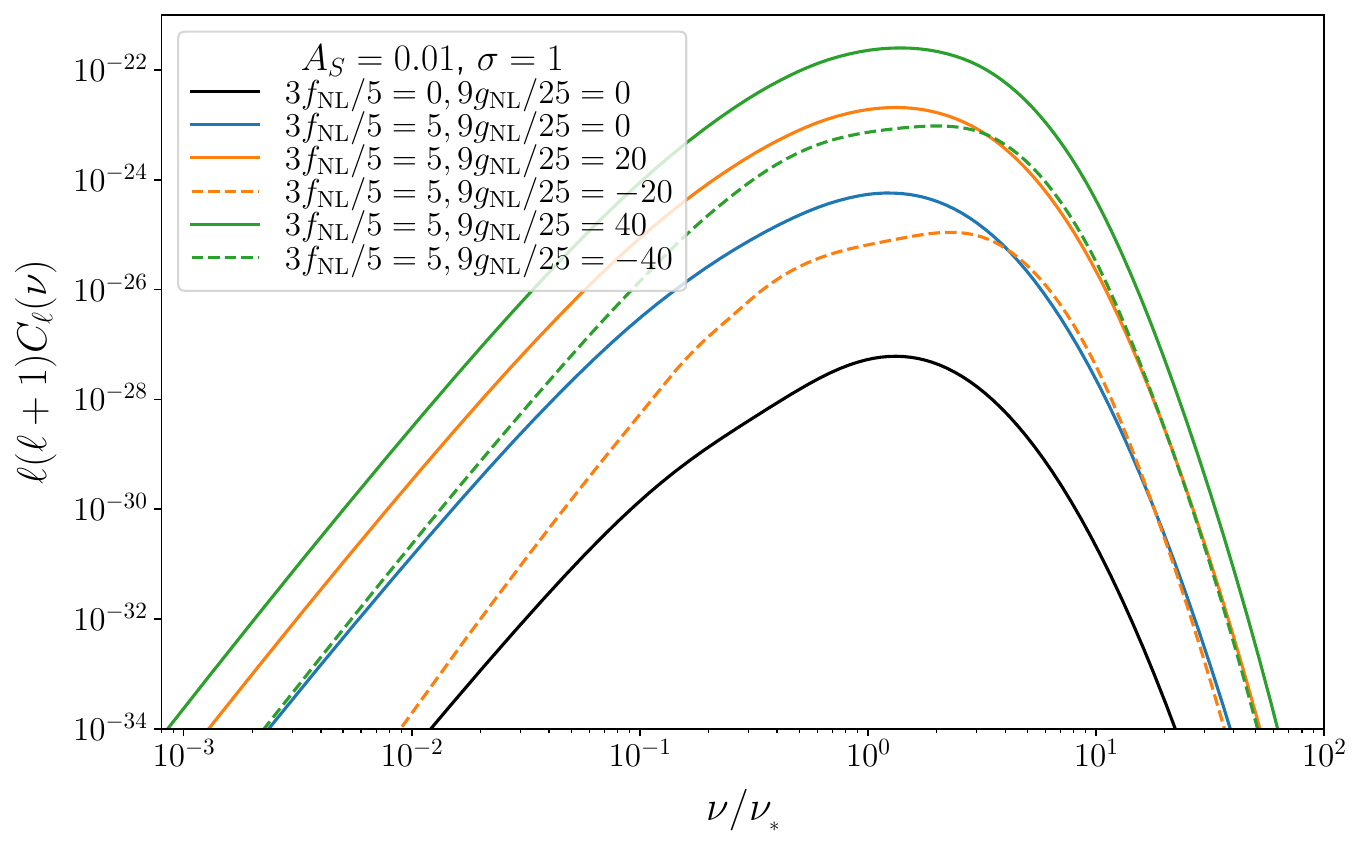}
    \caption{Angular power spectrum for the anisotropies in SIGWs. We use the same sets of model parameter as those of Figure~\ref{fig:Total_Omegabar}.  
     }\label{fig:Cl}
\end{figure}

We assess the angular power spectrum, as defined in Eq.~(\ref{eq:cellsai}), and present the numerical outcomes of $\ell(\ell+1)C_{\ell}(\nu)$ in relation to $\nu$ in Figure~\ref{fig:Cl}. We employ the same set of model parameters as depicted in Figure~\ref{fig:Total_Omegabar}.
When compared to the Gaussian scenario, the presence of primordial non-Gaussianity consistently amplifies the spectral amplitude. In this analysis, we specifically focus on $\gnl$ while keeping $3\fnl/5=5$ fixed. As demonstrated in Figure~\ref{fig:Cl}, the spectral amplitude for positive values of $\gnl$ consistently surpasses that for negative values of $\gnl$.
Furthermore, in comparison to the case of $\gnl=0$, the spectral amplitude is suppressed (enhanced) when $\gnl$ is negative but possesses a small (large) absolute value. When $\gnl$ is positive, the spectral amplitude is always enhanced relative to the $\gnl=0$ scenario.
Moreover, the degree of enhancement becomes more pronounced as the positive $\gnl$ value increases.

\begin{figure}[htbp]
    \centering
    \includegraphics[width =1 \columnwidth]{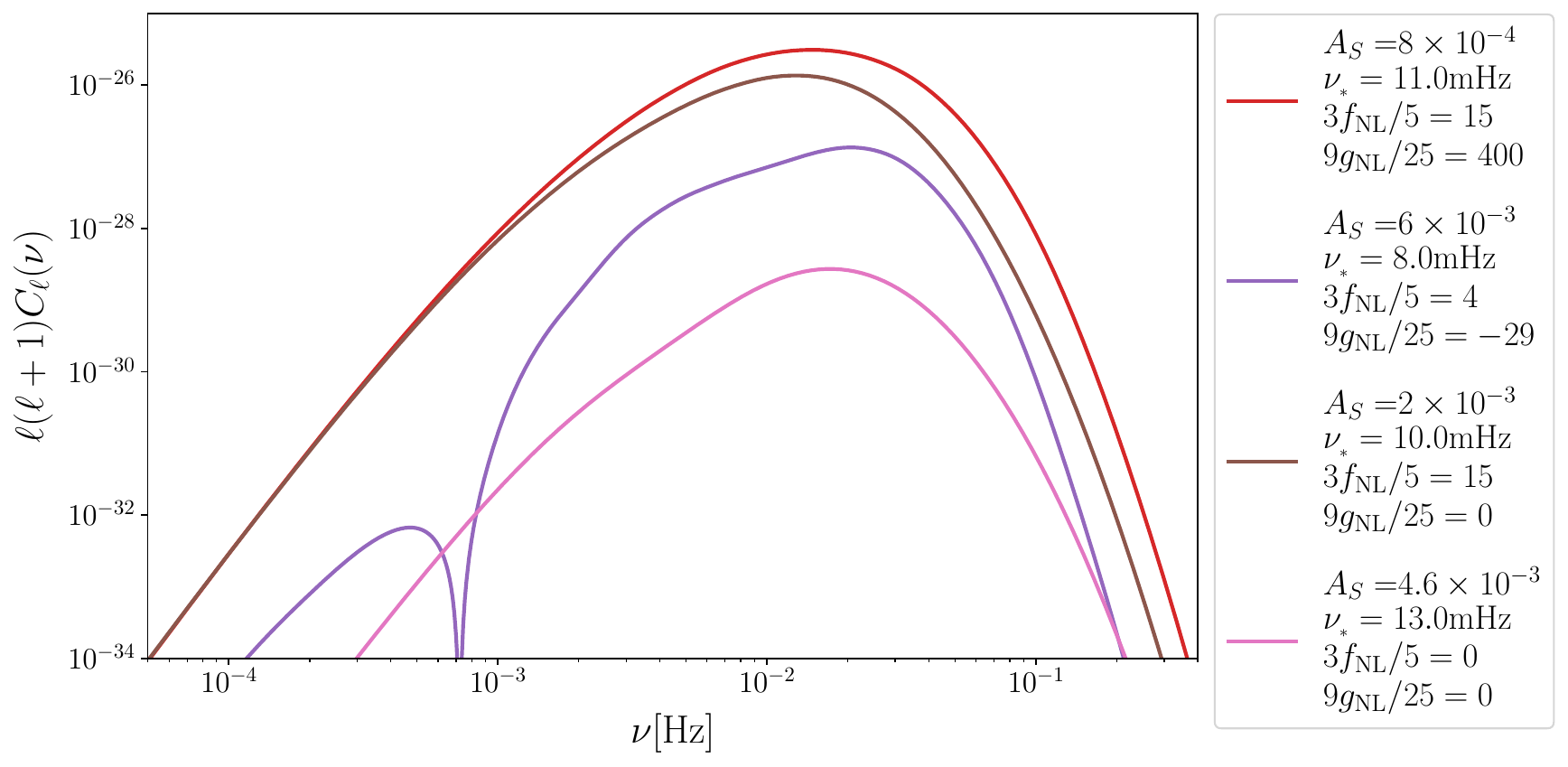}
    \caption{Degeneracies in the model parameters are broken by the angular power spectrum. Here, we adopt the same model parameters as those of \cref{fig:Omega_degeneracy}, where they are shown to be degenerate for the energy-density fraction spectrum. }\label{fig:Cl_degeneracy_broken}
\end{figure}

As demonstrated in Refs.~\cite{Li:2023qua,Wang:2023ost}, the degeneracies in the model parameters of the energy-density fraction spectrum can be overcome by considering the angular power spectrum.
Figure~\ref{fig:Omega_degeneracy} illustrates these degeneracies, as the energy-density fraction spectra appear nearly identical for four different sets of model parameters.
However, Figure~\ref{fig:Cl_degeneracy_broken} displays the corresponding angular power spectra for the same sets of model parameters, revealing noticeable differences in both spectral amplitudes and profiles.
Consequently, these previously mentioned degeneracies are explicitly broken.
Thus, the angular power spectrum can be employed to determine the model parameters that cannot be individually determined based solely on the energy-density fraction spectrum.

\begin{figure}[htbp]
    \centering
    \includegraphics[width =1 \columnwidth]{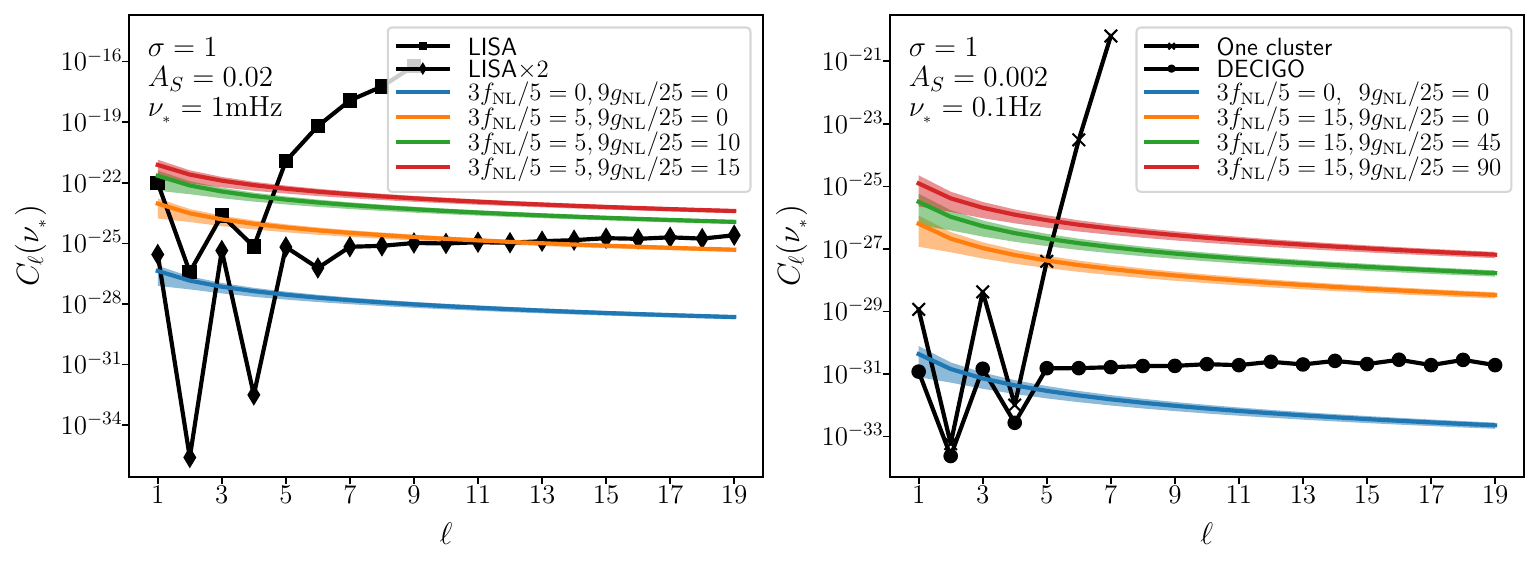}
    \caption{Comparison between the theoretical angular power spectra and the noise angular power spectra of \ac{LISA} at the 1 milli-Hz band (left panel) \cite{Capurri:2022lze} and \ac{DECIGO} at the deci-Hz band (right panel) \cite{Capurri:2022lze}. Shaded regions represent the uncertainties (68\% confidence level) due to the cosmic variance, i.e., $\Delta C_{\ell}/C_{\ell}=\sqrt{2/(2\ell+1)}$.   }
    \label{fig:Cl_sensitivity}
\end{figure}

Our theoretical predictions regarding the angular power spectrum can potentially be tested using future \ac{GW} detectors. In Figure~\ref{fig:Cl_sensitivity}, we compare the sensitivity curves of the \ac{LISA} and \ac{DECIGO} detectors \cite{Capurri:2022lze} with the expected angular power spectra for values of $A_{S}$ that are relevant in the context of scenarios involving the formation of \acp{PBH} (see reviews, e.g., in Ref.~\cite{Carr:2020gox}).
The frequency ranges of the \acp{SIGW} corresponding to these detectors correspond to the mass ranges of \acp{PBH}, within which \acp{PBH} could potentially account for the abundance of cold dark matter in the present Universe.
Consistent with our previous work \cite{Li:2023qua}, we find that detector networks exhibit superior sensitivities compared to individual detectors.

\begin{figure}[htbp]
    \centering
    \includegraphics[width =1 \textwidth]{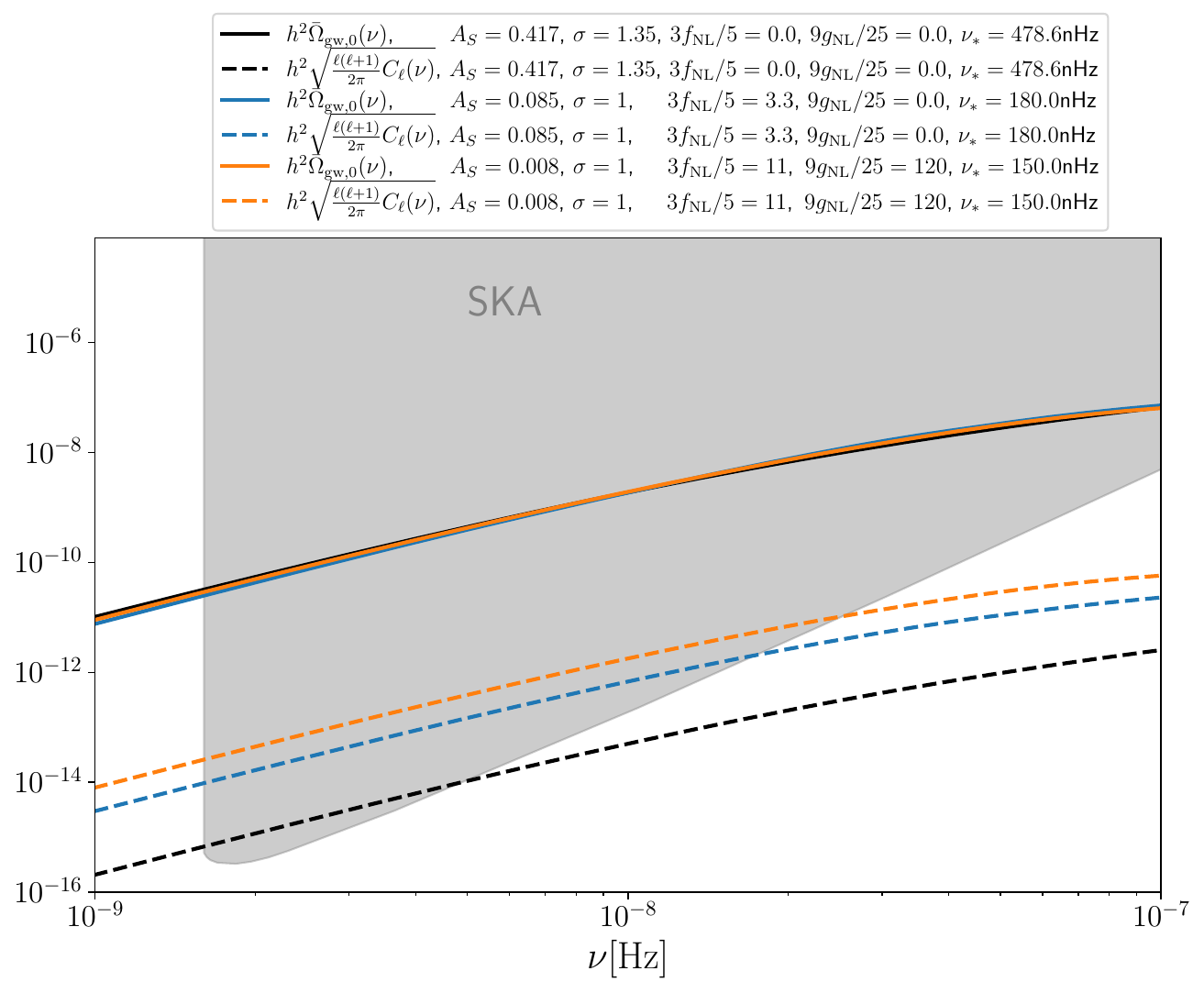} 
    \caption{The root-mean-squared energy density for the anisotropies in SIGWs in the PTA band. Energy-density fraction spectra are displayed to compare with the NANOGrav 15-year best-fit result. Shaded region can be explored with SKA. }\label{fig:PTA-SKA}
\end{figure}

We further investigate the angular power spectrum of \acp{SIGW} within the frequency band of \ac{PTA}. In Figure~\ref{fig:PTA-SKA}, the black solid curve represents the model parameters derived from the \ac{NG15} data release \cite{NANOGrav:2023hvm}, which only considered the Gaussian case. It is noteworthy that this curve overlaps with the other two colored solid curves, which include the effects of primordial non-Gaussianity. This implies that all three sets of model parameters can be used for interpreting the observed signal \cite{Xu:2023wog,EPTA:2023fyk,NANOGrav:2023gor,Reardon:2023gzh}.
To obtain a more precise estimation of the model parameters, we calculate the rms energy densities of \acp{SIGW} for these three sets of parameters and compare them (dashed curves) with the sensitivity curve of the Square Kilometre Array (\ac{SKA}) \cite{Schmitz:2020syl} in Figure~\ref{fig:PTA-SKA}. Notably, there are noticeable differences in the spectral amplitudes of the rms energy densities, suggesting the potential for determining the underlying theory and model parameters accurately. Similar results can also be found in Ref.~\cite{Wang:2023ost}.

\begin{figure*}[htbp]
    \centering
    \includegraphics[width = 1 \textwidth]{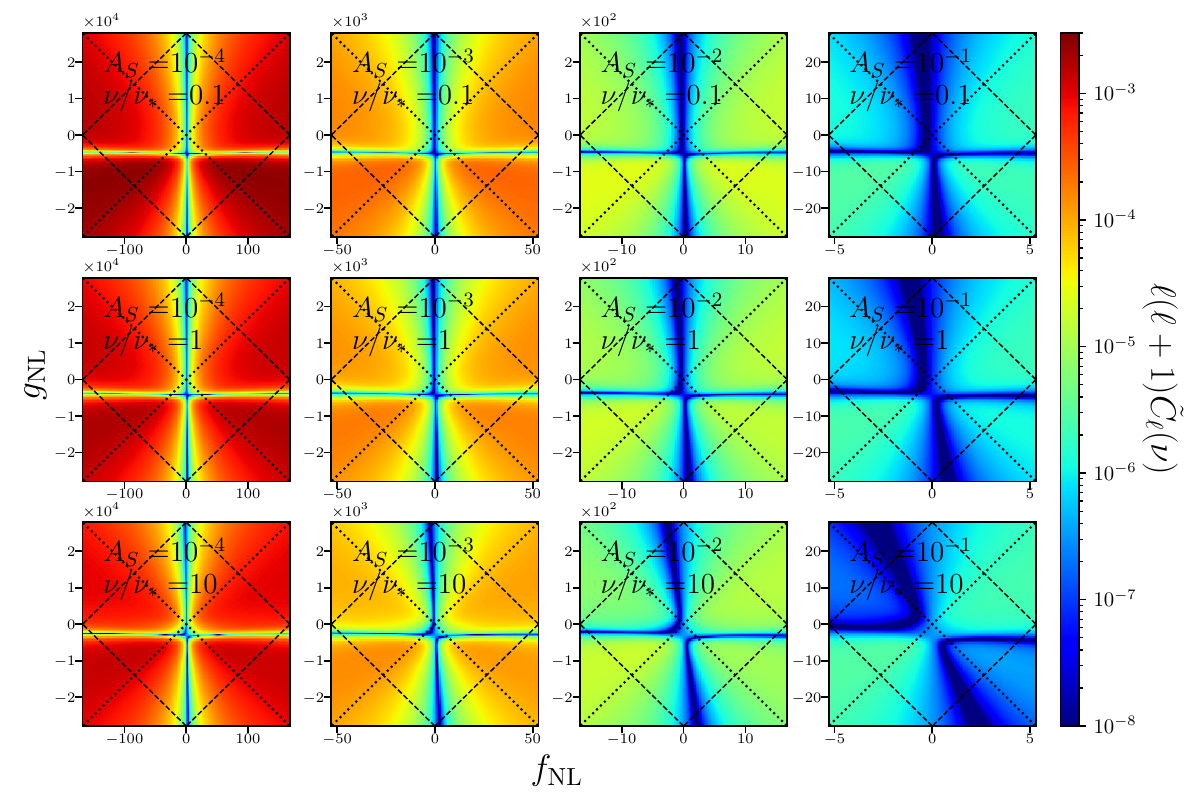}
    \caption{ Reduced angular power spectrum of SIGWs with respect to the primordial non-Gaussian parameters $\fnl$ and $\gnl$. The conventions for the dotted and dashed lines are the same as those in Figure~\ref{fig:Omega_f-g}.  }\label{fig:Clt_f-g}
\end{figure*}


We also present an array of contour plots for $\ell (\ell+1) \tilde{C}_\ell (\nu)$ in Figure~\ref{fig:Clt_f-g}, arranged identically to Figure~\ref{fig:Omega_f-g}. 
The dependence of the \ac{SIGW} angular power spectrum on $\fnl$ and $\gnl$ can be easily observed from this array. 
It is noted that the magnitude of the reduced angular power spectrum decreases with an increase in $A_S$. 
For the case of $A_S \leq 10^{-3}$, when $\fnl$ takes a nearly-zero value regardless of the value of $\gnl$, or when $\gnl$ takes a small value with minus sign regardless of the value of $\fnl$, $\tilde{C}_\ell (\nu)$ reaches the minimum value of 0, which implies that the anisotropies in \ac{SIGW} vanish at the corresponding frequency band.
Moreover, each panel for such a value of $A_S$ reveals that the \ac{SIGW} anisotropies are more significant for a large negative $\gnl$ than a positive one. 
Otherwise, if $A_S$ takes a larger value, the magnitude of $\tilde{C}_\ell (\nu)$ becomes larger when the signs of $\fnl$ and $\gnl$ are the same, especially in the high-frequency range.

\begin{figure}
    \centering
    \includegraphics[width = 1 \columnwidth]{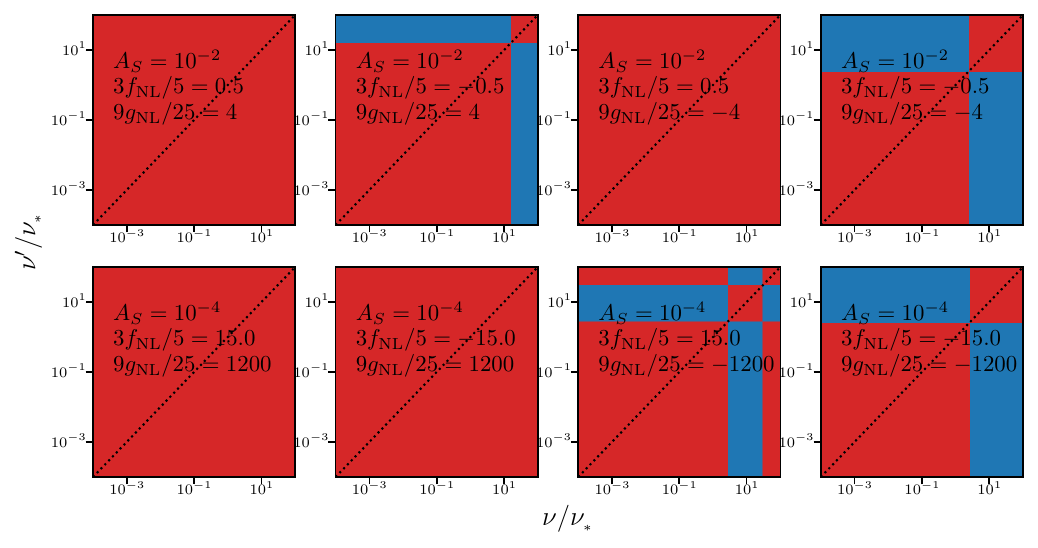}
    \caption{Correlation factor in red color for $r_{\ell}(\nu,\nu')=+1$ and blue color for $r_{\ell}(\nu,\nu')=-1$.  }\label{fig:rellpm}
\end{figure}

Additional information about the anisotropies in \acp{SIGW} may be concealed within the cross-correlated angular power spectrum. This spectrum, defined as $\langle\delta_{\uGW,0,\ell m}(2\pi\nu) \delta_{\uGW,0,\ell' m'}^\ast(2\pi\nu')\rangle = \delta_{\ell \ell'} \delta_{mm'} \tilde{C}_{\ell} (\nu,\nu')$, describes the correlation between two different frequency bands. However, this definition is somewhat redundant. Therefore, we propose a redefinition of the cross-correlation as a correlation factor, $r_{\ell} (\nu,\nu')={\tilde{C}_{\ell}(\nu,\nu')}/{\sqrt{\tilde{C}_{\ell}(\nu)\tilde{C}_{\ell}(\nu')}}$, where $\pm1$ represents the sign depending on the frequency bands.
In Figure~\ref{fig:rellpm}, we illustrate the impact of primordial non-Gaussianity on $r_{\ell}(\nu,\nu')$ in the $\nu-\nu'$ plane. The red shaded regions indicate $r_{\ell}=+1$, while the blue shaded regions represent $r_{\ell}=-1$.

\section{Discussion and Conclusion}
\label{sec:conc}

In this study, we have undertaken the first complete analysis of the energy-density fraction spectrum and the angular power spectrum of \acp{SIGW}. Notably, we have incorporated the local-type primordial non-Gaussianity parameterized by $\fnl$ and $\gnl$ simultaneously, marking a significant advancement in our understanding of these phenomena.

By utilizing a Feynman-like diagrammatic technique, we have delved into the noteworthy effects of $\fnl$ and $\gnl$ on the energy-density fraction spectrum of \acp{SIGW}. Through a thorough consideration of all relevant Feynman-like diagrams, we have established a comprehensive dictionary that correlates each diagram with its corresponding contribution to this spectrum.
For the first time, we have successfully derived semi-analytic formulas for the energy-density fraction spectrum by systematically evaluating the contributions from these diagrams in a hierarchical manner.
Our numerical analysis reveals that non-Gaussianity has a significant impact on both the amplitude and shape of the \acp{SIGW} spectrum, contingent upon the specific levels and functional forms of the non-Gaussianity. This arises from the fact that the non-Gaussianity of the primordial curvature perturbations introduces additional contributions to the energy-density fraction spectrum.

We have anticipated that the energy-density fraction spectrum of \acp{SIGW} can be experimentally tested in the future using space-borne \ac{GW} detector networks and \ac{PTA} experiments.
In particular, we have shown that the recent strong evidence for \ac{GWB} reported by \ac{PTA} collaborations can be explained by this theory.
Furthermore, we have demonstrated the significant degeneracies in the model parameters, specifically the non-Gaussian parameters $\fnl$ and $\gnl$, which pose notable challenges in determining these parameters.

For the first time, we have conducted a comprehensive analysis of the inhomogeneities and anisotropies in \acp{SIGW}, incorporating both the $\fnl$ and $\gnl$ effects.
The presence of primordial non-Gaussianity can induce anisotropies in \acp{SIGW}.
In this study, we have taken into account both the initial inhomogeneities and the propagation effects.
Interestingly, we have found that the propagation effects in \acp{SIGW} are identical to those observed in \ac{CMB}.
However, while the initial inhomogeneities vanish for the \ac{CMB}, they persist for \acp{SIGW} and reach the Earth in the present day.
To explore the properties of the initial inhomogeneities, we employed a Feynman-like diagrammatic technique, which involved dealing with a multitude of complex diagrams.
Nevertheless, we have developed systematic approaches to simplify our analytic evaluation.
Specifically, we have identified all the Feynman-like diagrams that allow for the modulation of energy densities.
In other words, we have focused on the diagrams that involve couplings between short-wavelength and long-wavelength modes when evaluating the initial inhomogeneities.

We have successfully derived the formulae for the angular power spectrum of \acp{SIGW}. From our analysis, we have discovered a universal characteristic of this spectrum, namely, the multipole dependence of the form $C_{\ell}\propto[\ell(\ell+1)]^{-1}$. 
This multipole dependence has the potential to distinguish \acp{SIGW} from other sources of \acp{GWB}, which may exhibit distinct multipole dependencies.
Our numerical results have revealed significant influences of $\fnl$ and $\gnl$ on the angular power spectrum. In particular, we have observed enhancements of the spectral amplitude by several orders of magnitude. Furthermore, we have found that large anisotropies in \acp{SIGW} can be generated.
This suggests promising prospects for detection and potential differentiation from other sources of \acp{GWB}.
The sign of $\gnl$ has been shown to have a significant impact on both the spectral amplitude and index. Moreover, we have demonstrated that the angular power spectrum can explicitly break the degeneracies in the model parameters. This exciting result highlights the potential of the angular power spectrum as a powerful probe of the early-Universe physics that would otherwise remain unexplored.

We anticipate that the angular power spectrum predictions will undergo testing through space-borne \ac{GW} detector networks \cite{Baker:2019nia,Smith:2019wny,Hu:2017mde,Wang:2021njt,Ren:2023yec,TianQin:2015yph,TianQin:2020hid,Zhou:2023rop,Seto:2001qf,Kawamura:2020pcg,Crowder:2005nr,Smith:2016jqs,Capurri:2022lze} and \ac{PTA} detectors \cite{Hobbs:2009yy,Demorest:2012bv,Kramer:2013kea,Manchester:2012za,Sesana:2008mz,Thrane:2013oya,Janssen:2014dka,2009IEEEP..97.1482D,Weltman:2018zrl,Moore:2014lga}. Such tests would not only provide new insights into the early-Universe physics but also into the nature of dark matter. For example, \acp{SIGW} of the \ac{LISA} band could be a potential probe of scenarios involving asteroid-mass \acp{PBH}, which have the potential to account for all cold dark matter. Future measurements of the angular power spectrum in the \ac{PTA} band, such as those from the \ac{SKA} program, are expected to reveal the origins of the recent \ac{PTA} signal. This would be particularly promising in terms of identifying the underlying theory, such as the theory of \acp{SIGW}, and its relevant parameters.

Our investigation of \acp{SIGW} also has significant implications for exploring scenarios involving \acp{PBH}, which are hypothesized to be a possible candidate for cold dark matter and a potential source of the individual \ac{GW} events observed by the \acl{LVK} \cite{LIGOScientific:2018mvr,LIGOScientific:2020ibl,LIGOScientific:2021djp,Wang:2022nml,He:2023yvl}.
In fact, the presence of primordial non-Gaussianity has a notable impact on the abundance and mass distribution of \acp{PBH}, as the threshold for their formation can be influenced by the levels and specific forms of these non-Gaussianity.
Furthermore, due to the coupling between short-wavelength and long-wavelength modes, they can induce a more clustered distribution of \acp{PBH} compared to the Gaussian case.
This clustering effect could enhance the rate of binary \ac{PBH} formation, resulting in a higher local merger rate that can potentially explain the observed individual \ac{GW} events.

Overall, the existence of primordial non-Gaussianity can have significant implications for both the energy-density fraction spectrum and the angular power spectrum of \acp{SIGW}. Conversely, a thorough examination of the two spectra can provide valuable insights into the primordial curvature perturbations, the nature of \ac{PBH} dark matter, and the physics of the early Universe, particularly the inflation models.
Moreover, by employing the Feynman-like diagrammatic technique, it becomes straightforward to verify the correctness of the findings presented in this paper and make comparisons with results from other studies.

Finally, we highlight that our research approach can be easily extended to study contributions of primordial non-Gaussianity to the angular bispectrum and trispectrum of \acp{SIGW}, which have been comprehensively analyzed in Ref.~\cite{Li:2024zwx}. 
Furthermore, it can be straightforwardly generalized to investigate impacts of higher-order non-Gaussianity on \acp{SIGW}, which are left to our future works.

\acknowledgments

We acknowledge Profs. Bin Gong, Tao Liu, Yi Wang, and Mr. Yan-Heng Yu for helpful discussion. 
S.W. and J.P.L. are partially supported by the National Natural Science Foundation of China (Grant No. 12175243), the National Key R\&D Program of China No. 2023YFC2206403, the Science Research Grants from the China Manned Space Project with No. CMS-CSST-2021-B01, and the Key Research Program of the Chinese Academy of Sciences (Grant No. XDPB15). Z.C.Z. is supported by the National Key Research and Development Program of China Grant No. 2021YFC2203001 and the National Natural Science Foundation of China (Grant NO. 12005016). K.K. is supported by KAKENHI Grant No. JP22H05270.

\appendix
\section{Equation of Motion}\label{sec:basic} 

We consider the perturbed spatially-flat \ac{FRW} spacetime in the conformal Newtonian gauge.
The metric thus reads 
\begin{equation}\label{metric} 
    \ud s^2 
    =  a^2(\eta) \left\{ 
            - (1 + 2 \Phi) \ud \eta^2 
            + \left[ (1 -2 \Phi) \delta_{ij} + \frac{1}{2} h_{ij} \right] \ud x^i \ud x^j 
    \right\}\ ,
\end{equation}
where $\Phi(\eta,\bx)$  denotes the linear scalar perturbations, and $h_{ij}(\eta,\bx)$ stands for the transverse-traceless tensor perturbations induced by $\Phi(\eta,\bx)$, i.e., the \acp{SIGW}. 
The conformal Hubble parameter $\cH (\eta)$ is defined as $\cH (\eta) = \partial_{\eta} a / a$, which equals to $1 / \eta$ during radiation domination especially.

The evolution of $\Phi(\eta,\mathbf{q})$ is governed by a master equation derived from the first-order Einstein's equation. 
In the absence of entropy perturbations, the explicit form of the master equation reads \cite{Maggiore:2018sht} 
\begin{equation}\label{eq:master}
    \partial_\eta^2 \Phi + 3\cH \left(1+c_s^2\right) \partial_\eta \Phi + 3\left(c_s^2 -w\right) \cH^2\Phi + c_s^2 q^2 \Phi  = 0\ ,
\end{equation}
where $c_s^2$ denotes the speed of sound. 
For the above linear differential equation, $\Phi(\eta,\bq)$ can be connected to the primordial (comoving) curvature perturbations $\zeta(\bq)$ through the scalar transfer function $T(q\eta)$, i.e.,  
\begin{equation}\label{eq:T-zeta-def}
    \Phi(\eta, \bq)
        = \left(\frac{3+3w}{5+3w}\right) T(q \eta) \zeta(\bq)\ ,
\end{equation}
where $w$ is the equation-of-state parameter of the Universe. 
In the case of radiation domination with $w=c_s^2=1/3$, the master equation can be solved to obtain the transfer function $T(q\eta)$. 
By substituting \cref{eq:T-zeta-def} into \cref{eq:master}, we get the solution for $T(q\eta)$, i.e.,  
\begin{equation}\label{eq:T-RD}
    T (x) = \frac{9}{x^2} 
            \left[
                \frac{\sin (x/\sqrt{3})}{x/\sqrt{3}}
                -\cos (x/\sqrt{3})
            \right] \ ,
\end{equation}
where we denote $x= q\eta$ for brevity.

The evolution of $h_{ij}(\eta,\bq)$ is governed by an equation of motion, which arises from the spatial components of Einstein's equation at second order, i.e.,  \cite{Ananda:2006af,Baumann:2007zm} 
\begin{equation}\label{eq:SIGW-motion} 
    \partial_\eta^2 h_{\lambda}(\eta, \bq)
    + 2\cH \partial_\eta h_{\lambda}(\eta,\bq) 
    +q^2 h_{\lambda}(\eta,\bq) 
    = 4 S_{\lambda}(\eta,\bq)\ , 
\end{equation} 
where $S_{\lambda}(\eta,\bq)$ represents the source term quadratic in the scalar perturbations $\Phi$, i.e.,   
\begin{eqnarray}\label{eq:source} 
    \cS_\lambda(\eta,\bq) 
    & = & \int\frac{\ud^3 \bq_a}{(2\pi)^{3/2}}
            \epsilon_{ij}^{\lambda}(\bq) q_{a}^i q_{a}^j
            \Big\{
                2 \Phi (\eta,\bq - \bq_a)\Phi(\eta,\bq_a)
            \\
            && 
                +\frac{4}{3(1+w)\cH^2}
                \left[
                    \partial_\eta \Phi(\eta,\bq -\bq_a)
                    + \cH\Phi(\eta,\bq -\bq_a)
                \right]
                \left[
                    \partial_\eta \Phi(\eta,\bq_a) 
                    + \cH\Phi(\eta,\bq_a)
                \right]
            \Big\}\ .\nonumber
\end{eqnarray} 
Following the Green' function method \cite{Espinosa:2018eve,Kohri:2018awv}, the solution to \cref{eq:SIGW-motion} is formally given by   
\begin{equation}\label{eq:h-G} 
    a(\eta) h_\lambda(\eta, \bq) 
    = 4 \int^{\eta}_{} \ud \eta' \,
        G_\bq(\eta,\eta') a(\eta') \cS_\lambda(\eta', \bq)\ . 
\end{equation}
Here, the Green's function $G_\bq(\eta,\eta')$ satisfies the differential equation of the form   
\begin{equation}\label{eq:Green}
    \partial_{\eta}^2 G_\bq(\eta,\eta') 
        +  \left[q^2-\frac{\partial_\eta^2 a(\eta)}{a(\eta)}\right] G_\bq(\eta, \eta')
        = \delta(\eta - \eta')\ ,
\end{equation}
for which the solution during radiation domination is given by 
\begin{equation}\label{eq:G-RD}
    G_\bq (\eta,\eta')
        = \Theta (\eta -\eta') 
        \frac{\sin [q(\eta - \eta')]}{q}\ ,
\end{equation}
with $\Theta(x)$ being the Heaviside function with variable $x$.

To get \cref{eq:h}, we need to reformulate $\mathcal{S}_{\lambda}(\eta,\bq)$ of \cref{eq:h-G}. 
Following Eq.~(\ref{eq:source}), we immediately get  
\begin{equation}\label{eq:S}
    \cS_\lambda(\eta, \bq)
        = \int \frac{\ud^3 \bq_a}{(2\pi)^{3/2}} q^2 Q_{\lambda}(\bq,\bq_a)
            F(\abs{\bq-\bq_a}, q_a, \eta)
            \zeta(\bq_a)
            \zeta(\bq-\bq_a)\ ,
\end{equation}
where \cref{eq:T-zeta-def} has been used. 
Here, we have introduced the projection factor of the form   
\begin{equation}\label{eq:Qsai}
    Q_{\lambda}(\bq, \bq_i) = \epsilon_{ij}^{\lambda}(\bq) \frac{q_{a}^i q_{a}^j}{q^2}
    = \frac{\sin^2 \theta}{\sqrt{2}}
     \times
        \begin{cases}
            \cos(2\phi_i) &\lambda = + \\
            \sin(2\phi_i) &\lambda = \times 
        \end{cases}\ , 
\end{equation}
where $\theta$ denotes the separation angle between $\mathbf{q}$ and $\mathbf{q}_i$, while $\phi_i$ is the azimuthal angle of $\mathbf{q}_i$ when $\mathbf{q}$ is aligned with the $\mathbf{z}$ axis. 
We also have introduced a functional of the form  
\begin{eqnarray}
    F(p_a,q_a,\eta)
    & = & \frac{4}{3} T(p_a \eta) T(q_a \eta)
            + \frac{4\eta^2}{9} \partial_\eta T(p_a \eta) \partial_\eta T (q_a \eta)
            + \frac{4\eta}{9} \partial_\eta \left[
                T(p_a \eta) T(q_a \eta) 
            \right]\ ,
\end{eqnarray}
where we denote $p_{a}=|\bq-\bq_{a}|$. 
By substituting \cref{eq:S} into \cref{eq:h-G}, and introducing a kernel function of the form
\begin{equation}\label{eq:I-def}
    \hat{I}(\abs{\bq - \bq_a},q_a,\eta)
    =\int^{\eta}_{} \ud \eta' \,
        q^2 G_\bq(\eta, \eta')
        \frac{a(\eta')}{a(\eta)} 
        F(\abs{\bq - \bq_a}, q_a, \eta')\ ,
\end{equation}
we can thus express $h_{\lambda}$ in terms of $Q_{\lambda}$, $\hat{I}$, and $\zeta$, namely, 
\begin{eqnarray}
    h_\lambda(\eta, \bq) 
    &=& 4 \int^{\eta}_{} \ud \eta' \,
        G_\bq(\eta,\eta') \frac{a(\eta')}{a(\eta)} 
        \int \frac{\ud^3 \bq_a}{(2\pi)^{3/2}} q^2 Q_{\lambda}(\bq,\bq_a)
            F(\abs{\bq-\bq_a}, q_a, \eta)
            \zeta(\bq_a)
            \zeta(\bq-\bq_a)\nonumber\\
    &=& 4 \int \frac{\ud^3 \bq_a}{(2\pi)^{3/2}} 
        \zeta(\bq_a) \zeta(\bq-\bq_a) Q_{\lambda}(\bq,\bq_a) 
        \hat{I} (\abs{\bq - \bq_a},q,\eta)\ .
\end{eqnarray}
This is exactly the expression of \cref{eq:h}.

\section{Numerical integrals}\label{sec:num-int}

To evaluate $\bar\Omega_\uGW^X (\eta,q)$ in \cref{eq:Omegabar-X}, we seek to rephrase the corresponding integrals in \cref{sec:ogw} in a form that is more amenable to numerical calculations. 
Here, we calculate the unscaled components of energy-density fraction spectrum, which are still labeled with $\bar\Omega_\uGW^X (\eta,q)$ in this Appendix. 
This is equivalent to fixing the model parameters, i.e., $A_S = 1$, $3\fnl/5 = 1$, and $9\gnl/25 = 1$. 
Following \cref{tab:order-FD}, the eventual results can be straightforwardly recovered via multiplying the unscaled results by $(3/5)^{2(a+b)}\fnl^{2a}\gnl^b A_S^{a+b+2}$, where $a$ and $b$ denote the powers of these components in $\fnl$ and $\gnl$, respectively.

By substituting Eqs.~(\ref{eq:G-like} -- \ref{eq:PN-like}) into \cref{tab:PX} and further substituting the results into \cref{eq:phxsai}, 
we can express \cref{eq:Omegabar-X} as follows 
\begin{eqnarray}\label{eq:Omega-formal}
    \frac{\bar{\Omega}_\uGW^X (\eta, q)}{\text{S.F.}}
    &=& \frac{q^5 \eta^2}{16\pi^2} 
        \Bigg\{\prod_{i}\int \frac{\ud^3 \bq_i}{(2\pi)^{3}}\Bigg\} \Bigg\{\prod P^{[...]}\Bigg\}\nonumber\\ 
        &&\times\Bigg\{\sum_{\lambda = +,\times} Q_{\lambda}(\bq, \bq_1) Q_{\lambda}(\bq, \bq_j) 
        \overbar{\hat{I} (\abs{\bq - \bq_1}, q_1, \eta) 
        \hat{I} (\abs{\bq - \bq_j}, q_j, \eta)}\Bigg\}\ ,
\end{eqnarray}
where we have $i=1,...,n$ with $n \leq 5$ being the order of integration, $j=1,2$, and $\prod P^{[...]}$ stands for products of power spectra in Eqs.~(\ref{eq:G-like} -- \ref{eq:PN-like}). 
In the following, we evaluate the terms in the three pairs of braces on the right hand side of Eq.~(\ref{eq:Omega-formal}), respectively.

For the first brace, we transform the variables of integration into five sets of new variables $(u_i,v_i)$ with $i=1,2,3,4,5$, namely,
\begin{equation}\label{eq:uv-def}
    v_i = \frac{q_i}{q}\ ,\qquad\qquad\quad
    u_i = \frac{\abs{\bq-\bq_i}}{q}\ .
\end{equation}
We further convert the integration regions into rectangle ones via introducing the variable transformation as follows 
\begin{equation}\label{eq:st-def}
    s_i = u_i - v_i\ ,\qquad\qquad\quad
    t_i = u_i + v_i -1\ .
\end{equation}
Hence, the first brace transforms to the following form 
\begin{equation}\label{eq:int-tf}
    \prod_i \int \frac{\ud^3 \bq_i}{(2\pi)^{3}} = \frac{q^3}{2(2\pi)^{3}} \prod_i \left(\int_0^\infty \ud t_i \int_{-1}^1 \ud s_i\, v_i u_i \int_0^{2\pi} \ud \phi_i\right)\ ,
\end{equation} 
where $\phi_i$ is defined below \cref{eq:Qsai}.

For the second brace, we show the expressions of $P^{[1]}(q)$, $P^{[2]}(q)$, $P^{[3]}(q)$, and the variable transformations concerned with them. 
Based on \cref{eq:Lognormal}, we represent the unscaled $P^{[1]}$ (or equivalently $P_{gS}$) as follows 
\begin{equation}\label{eq:P1}
    P^{[1]} (q) = \frac{2\pi^2}{q^3} \Delta_S^2 (q) 
    = \frac{2\pi^2}{q^3} \frac{1}{\sqrt{2\pi\sigma^2}}\exp\left(-\frac{\ln^2 (q/q_\ast)}{2 \sigma^2}\right)\ .
\end{equation}
We further express the unscaled $P^{[2]}$ and $P^{[3]}$ as follows 
\begin{eqnarray}\label{eqs:P2-P3}
    P^{[2]} (q) &=& \int \frac{\ud^3 \tilde{\bq}_1}{(2 \pi)^3} P^{[1]} (\tilde{q}_1) P^{[1]} (\abs{\bq - \tilde{\bq}_1})
    = \frac{\pi^2}{2 q^3} \int_0^\infty \ud \tilde{t}_1 \int_{-1}^1 \ud \tilde{s}_1\, \frac{\Delta_S^2 (\tilde{v}_1 q) \Delta_S^2 (\tilde{u}_1 q)}{\left(\tilde{v}_1 \tilde{u}_1\right)^2}\ ,\\
    P^{[3]} (q) &=& \int \frac{\ud^3 \tilde{\bq}_1}{(2 \pi)^3} P^{[2]} (\tilde{q}_1) P^{[1]} (\abs{\bq - \tilde{\bq}_1})\nonumber\\
    &=& \frac{\pi^2}{8 q^3} \prod_{i=1,2}\left(\int_0^\infty \ud \tilde{t_i} \int_{-1}^1 \ud \tilde{s_i}\right) \frac{\Delta_S^2 (\tilde{u}_1 q) \Delta_S^2 (\tilde{v}_1 \tilde{v}_2 q) \Delta_S^2 (\tilde{v}_1 \tilde{u}_2 q)}{\left(\tilde{v}_1 \tilde{u}_1 \tilde{v}_2 \tilde{u}_2\right)^2}\ ,
\end{eqnarray}
where we have introduced the transformation of variables $(\tilde{v}_1,\tilde{u}_1)$ and $(\tilde{v}_2,\tilde{u}_2)$ of the form 
\begin{eqnarray}
    \tilde{v}_1 &=& \frac{\tilde{q}_1}{q}\ ,\qquad\qquad\quad
    \tilde{u}_1 = \frac{\abs{\bq-\tilde{\bq}_1}}{q}\ ,\nonumber\\
    \tilde{v}_2 &=& \frac{\tilde{q}_2}{\tilde{q}_1}\ ,\qquad\qquad\quad
    \tilde{u}_2 = \frac{\abs{\tilde{\bq}_1-\tilde{\bq}_2}}{\tilde{q_1}}\ ,\nonumber
\end{eqnarray}
and the transformations from $(\tilde{u}_i, \tilde{v}_i)$ to two sets of new variables $(\tilde{s}_i, \tilde{t}_i)$ with $i=1, 2$ are similar to \cref{eq:st-def}, i.e., $\tilde{s}_i = \tilde{u}_i - \tilde{v}_i\ ,\tilde{t}_i = \tilde{u}_i + \tilde{v}_i -1\ $. 
Since the above integrals involve inner products of $\bq_i$ and $\bq_j$, it is convenient to introduce a new quantity, i.e., 
\begin{eqnarray}\label{eqs:yij-def}
    y_{ij} = \frac{\bq_i \cdot \bq_j}{q^2} 
        &=& \frac{\cos\varphi_{ij}}{4}\sqrt{t_i (t_i + 2) (1 - s_i^2) t_j (t_j + 2) (1 - s_j^2)} 
            \nonumber\\
        &&  + \frac{1}{4}[1 - s_i (t_i + 1)][1 - s_j (t_j + 1)]\ ,
\end{eqnarray}
where we denote $\varphi_{ij}=\phi_i-\phi_j$ for the sake of brevity. 
We further introduce two new quantities of the form  
\begin{eqnarray}\label{eqs:wij-def}
    w_{ij} &=& \frac{\abs{\bq_i-\bq_j}}{q} =  \sqrt{v_i^2 + v_j^2 - 2 y_{ij}}\ ,\\
    w_{ijk} &=& \frac{\abs{\bq_i+\bq_j-\bq_k}}{q} = \sqrt{v_i^2 + v_j^2 + v_k^2 + 2 y_{ij} - 2 y_{ik} - 2 y_{jk}}\ .
\end{eqnarray}
Consequently, all the variables of $P^{[...]}$ in Eqs.~(\ref{eq:G-like} -- \ref{eq:PN-like}) can be expressed in terms of $v_i q$, $u_i q$, $w_{ij} q$ or $w_{ijk} q$.

For the third brace, we first recast the kernel function $\hat{I} \left(\abs{\bq-\bq_i},q_i,\eta\right)$ during radiation domination into $I_{\uRD}(u_i,v_i,x)$ with $x=q\eta$, namely, 
\begin{equation}\label{eq:Ihat-Iuv}
    \hat{I} \left(\abs{\bq-\bq_i},q_i,\eta\right) 
        =  I_{\uRD} \left(\frac{\abs{\bq-\bq_i}}{q},\frac{q_i}{q},q\eta\right)
        = I_{\uRD} (u_i,v_i,x)\ ,
\end{equation}
which can be straightforwardly derived from \cref{eq:I-def}. 
As demonstrated in Refs.~\cite{Kohri:2018awv,Espinosa:2018eve}, the analytical formula for $I(u_i,v_i,x)$ on subhorizon scales, i.e., $x \gg 1$, can be expressed as follows 
\begin{subequations}\label{eq:I-RD} 
\begin{eqnarray}
I_{\uRD} (u_i,v_i,x \gg 1) &=& \frac{1}{x} I_A (u_i,v_i) 
            \left[I_B (u_i,v_i) \sin x - \pi I_C (u_i,v_i) \cos x\right]\ ,\\
I_A (u_i,v_i) &=& \frac{3\left(u_i^2 + v_i^2 - 3\right)}{4 u_i^3 v_i^3}\ , \\
I_B (u_i,v_i) &=& - 4 u_i v_i  
        + \left(u_i^2+v_i^2-3\right) \ln \abs{\frac{3-(u_i + v_i)^2}{3-(u_i - v_i)^2}}\ , \\
I_C (u_i,v_i) &=& \left(u_i^2 + v_i^2 - 3\right)
        \Theta\left(u_i + v_i - \sqrt{3}\right)\ .
\end{eqnarray} 
\end{subequations} 
Moreover, the oscillation average of two kernel functions with the same $x$ is shown as \cite{Adshead:2021hnm}
\begin{eqnarray}\label{eq:I-ave-12}
    && \overbar{I_\uRD (u_i,v_i,x \gg 1) 
        I_\uRD (u_j,v_j,x \gg 1)} \nonumber\\
        & = & \frac{I_A (u_i,v_i) I_A (u_j,v_j)}{2 x^2} 
        \left[
                I_B (u_i,v_i) I_B (u_j,v_j)
                + \pi^2 I_C (u_i,v_i) I_C (u_j,v_j)
            \right]\ .
\end{eqnarray}
Combining the kernel function with the projection factor, we can introduce a new function for simplifying our evaluation, i.e.,  
\begin{equation}\label{eq:J-def}
    J (u_i,v_i,x) 
    = \frac{x}{8}\bigl[(v_i+u_i)^2-1\bigr] \bigl[1-(v_i-u_i)^2\bigr] I_{\uRD} (u_i,v_i,x)\ , 
\end{equation}
where $Q_\lambda (\bq,\bq_i)$ is shown in Eq.~(\ref{eq:Qsai}) and $I_{\uRD}(u_i,v_i,x \gg 1)$ is shown in \cref{eq:I-RD}. 
After explicit computation, we get the following relations 
\begin{equation}\label{eq:QI-J}
    \sum_\lambda Q_{\lambda}^2 (\bq, \bq_1) \overbar{\hat{I}^2 (\abs{\bq - \bq_1}, q_1, \eta)} = \frac{1}{x^2} \overbar{J^2 (u_1,v_1,x)}\ ,
\end{equation} 
\begin{equation}\label{eq:QQII-JJ}
    \sum_\lambda Q_{\lambda}(\bq, \bq_1) Q_{\lambda}(\bq, \bq_2) \overbar{\hat{I} (\abs{\bq - \bq_1}, q_1, \eta) \hat{I} (\abs{\bq - \bq_2}, q_2, \eta)} 
    = \frac{\cos (2\varphi_{12})}{x^2} \overbar{J(u_1,v_1,x)J(u_2,v_2,x)} \ .
\end{equation}

\section{Boltzmann equation}\label{sec:Boltz}

To derive the present density contrast $\delta_{\uGW,0}$ in Eq.~(\ref{eq:delta-0}), we review the Boltzmann equation for gravitons following Refs.~\cite{Contaldi:2016koz,Bartolo:2019oiq,Bartolo:2019yeu}. 
We decompose the distribution function $f(\eta',\bx,\bq)$ for gravitons into a background $\bar{f}(\eta',q)$ and perturbations $\Gamma(\eta',\bx,\bq)$, namely, 
\begin{equation}\label{eqn:Gamma-def}
    f(\eta',\bx,\bq)
        =\bar{f}(\eta',q)
        - q\frac{\partial \bar{f}}{\partial q}\Gamma(\eta',\bx,\bq)\ .
\end{equation}
The evolution of $f(\eta',\bx,\bq)$ follows the Boltzmann equation, i.e., 
\begin{equation}\label{eq:Boltzmann-def} 
\frac{\ud f}{\ud \eta'}=\mathcal{I}(f) + \mathcal{C}(f)\ ,
\end{equation}
where $\mathcal{I}$ denotes the emissivity term and $\mathcal{C}$ stands for the collision term. 
For gravitons, the collision term is negligible, i.e., $\mathcal{C}=0$ \cite{Contaldi:2016koz,Bartolo:2018igk,Flauger:2019cam}. 
For cosmological sources, the emissivity term is treated as the initial condition, implying $\mathcal{I}=0$ \cite{Bartolo:2019oiq,Bartolo:2019yeu}. 
Therefore, the Boltzmann equation is expressed as  
\begin{equation}\label{eq:f-total-derivation}
    \frac{\ud f}{\ud \eta'}
        = \frac{\partial f}{\partial \eta'}
        +\frac{\partial f}{\partial x^i}\frac{\ud x^i}{\ud \eta'}
        +\frac{\partial f}{\partial q}\frac{\ud q}{\ud \eta'} 
        +\frac{\partial f}{\partial n^i}\frac{\ud n^i}{\ud \eta'}=0\ .
\end{equation}
To calculate the Boltzmann equation up to first order, we adopt relations of $\ud x^i / \ud \eta' = n^i$, $\ud q / \ud \eta'= \left(\partial_{\eta'}\Phi - n^i \partial_i \Phi\right) q$, and $\ud n^i / \ud \eta' = 0$ resulted from perturbed geodesics of massless gravitons. 
Hence, Eq.~\eqref{eq:f-total-derivation} can be separated into a background equation and a perturbation equation, i.e., 
\begin{eqnarray}
    \partial_{\eta'} \bar{f} & = & 0\ ,\label{eq:Boltzmannfbar}\\
    \partial_{\eta'}\Gamma + n^i \partial_i \Gamma 
    & = & \partial_{\eta'}\Phi - n^i\partial_i \Phi\ . \label{eq:Boltzmann-1st}
\end{eqnarray}
Eq.~(\ref{eq:Boltzmannfbar}) indicates that the background remains unchanged during process of evolution. 
For the perturbations, we can transform Eq.~(\ref{eq:Boltzmann-1st}) into Fourier space, namely,  
\begin{equation}\label{eq:Boltzmann-k}
    \partial_{\eta'}\Gamma + i k \mu \Gamma 
    = \partial_{\eta'}\Phi - i k \mu \Phi\ , 
\end{equation}
where we denote $k\mu = \bk\cdot\bn$ for simplicity. 
Following Refs.~\cite{Contaldi:2016koz,Bartolo:2019oiq,Bartolo:2019yeu}, we can obtain the line-of-sight solution to \cref{eq:Boltzmann-k} as follows \cite{Bartolo:2019oiq,Bartolo:2019yeu} 
\begin{eqnarray}\label{eq:Gamma-solution}
    \Gamma (\eta_0,\bk,\bq) 
    & = & e^{i k \mu (\eta - \eta_0)} 
        \left[
            \Gamma (\eta, \bk, \bq) + \Phi (\eta, \bk)
        \right]
        + 2 \int_{\eta}^{\eta_0} \ud \eta'\,
            e^{i k \mu (\eta' - \eta_0)}
                \partial_{\eta'} \Phi (\eta',\bk)\ ,
\end{eqnarray}
where we have disregarded the monopole term $\Phi (\eta_0,\bk)$.

The perturbations $\Gamma$ lead to the inhomogeneities of the energy density of \acp{GW}, implying that $\delta_\uGW \propto \Gamma$. 
Exactly, we express the energy density $\rho_\uGW (\eta',\bx)$ in terms of $f(\eta',\bx, \bq)$, i.e., 
\begin{equation}\label{eqn:rho-f}
    \rho_\uGW (\eta',\bx) 
        = \frac{1}{a^4} \int \ud^3 \bq\, q f(\eta',\bx,\bq)\ .
\end{equation} 
Hence, we establish a relation between $f$ and the energy-density full spectrum $\omega_\uGW (\eta',\bx,\bq)$, defined in \cref{eq:omega-def}, as follows  
\begin{equation}\label{eq:f-omega}
    f(\eta',\bx,\bq) 
        = \rho_\uc \left(\frac{a}{q}\right)^4 \omega_\uGW (\eta',\bx,\bq) \ .
\end{equation}
Since $\langle\omega_\uGW (\eta',\bx,q)\rangle = \bar{\Omega}_\uGW (\eta',q) / 4\pi$ is obtained in \cref{eq:bgdomega} and $\bar{f}(\eta',q)$ stands for the background term of $f(\eta',\bx,\bq)$, we get 
\begin{equation}\label{eq:fbar-Omegabar}
    \bar{f}(\eta',q)
        = \frac{\rho_\uc}{4\pi} \left(\frac{a}{q}\right)^4 \bar{\Omega}_\uGW (\eta',q) \ . 
\end{equation}
Consequently, the density contrast, as defined in \cref{eq:delta-def}, can be represented in terms of $\Gamma(\eta',\bx,\bq)$, i.e., \cite{Bartolo:2019oiq,Bartolo:2019yeu} 
\begin{equation}\label{eq:delta-Gamma}
    \delta_\uGW (\eta',\bx,\bq)
        = \left[
            4-\frac{\partial \ln \bar{\Omega}_\uGW (\eta',q)}{\partial\ln q}
        \right] \Gamma (\eta',\bx,\bq)
        = \left[4-n_{\uGW} (\nu)\right] \Gamma (\eta',\bx,\bq)\ ,
\end{equation}
where $n_{\uGW} (\nu)$ is defined in Eq.~\eqref{eq:ngw-def}. 
Considering the line-of-sight $\bx_0-\bx = (\eta_0-\eta)\bn_0$ and combining Eq.~(\ref{eq:Gamma-solution}) with Eq.~\eqref{eq:delta-Gamma}, we immediately obtain a formula for the present density contrast in configuration space, i.e., \cite{Li:2023qua} 
\begin{eqnarray}\label{eq:deltaGW-CGW}
    \delta_{\uGW,0} (\bq) 
    & = & \delta_\uGW (\eta, \bx,\bq) + 
        \left[4 - n_{\uGW}(\nu)\right]\Phi (\eta, \bx)\\
            && + 2 \left[4 - n_{\uGW}(\nu)\right] 
            \int \frac{\ud^3 \bk}{(2\pi)^{3/2}} e^{i\bk\cdot\bx_0}
            \int_{\eta}^{\eta_0} \ud \eta' \,  
            e^{i k \mu (\eta' - \eta_0)}
                \partial_{\eta'} \Phi (\eta',\bk)\ ,\nonumber
\end{eqnarray}
where $\delta_\uGW (\eta, \bx, \bq)$ represents the initial perturbations, $\Phi (\eta, \bx)$ corresponds to the \ac{SW} effect \cite{Sachs:1967er}, and the integral signifies the \ac{ISW} effect \cite{Sachs:1967er}. 
According to Ref.~\cite{Bartolo:2019zvb}, the \ac{ISW} effect is typically of less importance and can be disregarded.
Finally, we obtain the formula for the present density contrast of \acp{SIGW}, as shown in Eq.~\eqref{eq:delta-0}.

\bibliography{biblio}
\bibliographystyle{JHEP}

\end{document}